\documentclass[acmsmall]{acmart}

\usepackage{algorithm}
\usepackage{algpseudocode}
\usepackage{algorithmicx}
\usepackage{capt-of}


\usepackage{amsthm}
\newtheorem{example}{Example}[section]
\newtheorem{problem}{Problem}[section]
\newtheorem{challenge}{Challenge}[section]
\usepackage{amsmath}
\usepackage{booktabs} 
\usepackage{multirow}
\usepackage{makecell}
\usepackage{array}
\usepackage{graphicx}
\usepackage{subcaption}
\AtBeginDocument{%
  \setlength{\textfloatsep}{0pt plus 2pt minus 2pt}
  \setlength{\floatsep}{0pt plus 2pt minus 2pt}
  \setlength{\dbltextfloatsep}{2pt plus 2pt minus 2pt}
  \setlength{\dblfloatsep}{2pt plus 2pt minus 2pt}
}
\usepackage{siunitx}
\usepackage{threeparttable}
\setcopyright{cc}
\setcctype{by}
\acmJournal{PACMMOD}
\acmYear{2026} \acmVolume{4} \acmNumber{1 (SIGMOD)} \acmArticle{75} \acmMonth{2} \acmPrice{}\acmDOI{10.1145/3786689}

\begin{document}

\title{Reqo: A Comprehensive Learning-Based Cost Model for Robust and Explainable Query Optimization}

\author{Baoming Chang}
\orcid{0009-0005-7414-3631}
\affiliation{%
  \institution{University of Ottawa}
  \city{Ottawa}
  \country{Canada}
}
\email{bchan081@uottawa.ca}

\author{Amin Kamali}
\orcid{0000-0003-2176-4088}
\affiliation{%
  \institution{University of Ottawa}
  \city{Ottawa}
  \country{Canada}
}
\email{skama043@uottawa.ca}

\author{Verena Kantere}
\orcid{0000-0002-3586-9406}
\affiliation{%
  \institution{University of Ottawa}
  \city{Ottawa}
  \country{Canada}
}
\email{vkantere@uottawa.ca}


\begin{abstract}
Although machine learning (ML) shows potential in improving query optimization by generating and selecting more efficient plans, ensuring the robustness of learning-based cost models (LCMs) remains challenging. These LCMs currently lack explainability, which undermines user trust and limits the ability to derive insights from their cost predictions to improve plan quality. Accurately converting tree-structured query plans into representations via tree models is also essential, as omitting any details may negatively impact subsequent cost model performance. Additionally, inherent uncertainty in cost estimation leads to inaccurate predictions, resulting in suboptimal plan selection. To address these challenges, we introduce Reqo, a \underline{R}obust and \underline{E}xplainable \underline{Q}uery \underline{O}ptimization cost model that comprehensively enhances three main stages in query optimization: plan generation, plan representation, and plan selection. Reqo integrates three innovations: the first explainability technique for LCMs that quantifies subgraph contributions and produces plan generation hints to enhance candidate plan quality; a novel tree model based on Bidirectional Graph Neural Networks (Bi-GNNs) with a Gated Recurrent Unit (GRU) aggregator to further capture both node-level and structural information and effectively strengthen plan representation; and an uncertainty-aware learning-to-rank cost estimator that adaptively integrates cost estimates with uncertainties to enhance plan selection robustness. Extensive experiments demonstrate that Reqo outperforms state-of-the-art approaches across all three stages.
\end{abstract}



\received{July 2025}
\received[revised]{October 2025}
\received[accepted]{November 2025}

\maketitle

\section{Introduction}

Query optimization is crucial for database management systems (DBMSs), aiming to generate and select the most efficient execution plan. Recently, the database community has begun exploring learning-based optimizers as alternatives to classical methods. However, robustly selecting the optimal candidate plan for a query remains challenging across three sequential stages in learning-based query optimization: plan generation, plan representation, and plan selection.

The worst-case query performance occurs if the optimizer selects the least efficient plan from its candidate plan pool. The black-box nature of LCMs makes their predictions difficult to understand and trust~\cite{lipton2018mythos} and obstructs identifying and optimizing components of candidate plans that significantly increase estimated costs and may recur in future generated plans. Existing LCMs lack explainability and therefore cannot accurately attribute predictions to specific nodes or maintain monotonicity in cost predictions across nested components within the same plan. This limited transparency restricts their practical adoption in trust-critical scenarios and prevents the use of predicted costs for such plan components to improve candidate plan quality and optimizer robustness.

Accurate cost estimation by LCMs is also crucial for robust query optimization. However, their performance ceiling is limited by the quality of plan representation. A query plan is a tree with nodes representing operators to access, join, or aggregate data, and edges indicating parent-child dependencies. LCMs use tree models to aggregate node-level features over the plan tree into a plan representation for cost prediction and subsequent plan selection. Current tree models~\cite{treelstm, treecnn} cannot capture information flow along long paths in large query plans~\cite{queryformer}. Thus, preserving both node-level and structural information when encoding plans into representations is essential, as omitting critical details degrades prediction accuracy and undermines optimizer performance.

Finally, plan selection typically ranks candidate plans by their cost estimates. However, inherent uncertainty often leads to inaccurate cost estimation and causes the optimizer to select suboptimal plans with longer runtimes. Robust query optimization aims to reduce this sensitivity to estimation errors and to avoid simplifying assumptions~\cite{roq}, making it promising to improve optimizer performance. Recent LCMs have increasingly addressed robustness. State-of-the-art approaches~\cite{zhao2021uncertainty, liu2021fauce, chen2023leon, doshi2023kepler, roq} quantify uncertainty using probabilistic ML and apply it with predefined rules for plan selection. These rules are fixed and not learned during training. They cannot adapt to workload shifts or self-refine, often necessitating manual retuning and limiting further robustness improvements.

This paper introduces Reqo to address challenges in these three stages. We present the first explainability technique for LCMs, which quantifies each subgraph’s context-aware contribution to the plan cost prediction, improving transparency and promoting monotonic cost predictions across nested subplans within the same plan. These explanations can then be converted into plan generation hints to improve candidate plan quality. Second, we propose a novel tree model that utilizes Bi-GNN~\cite{dirgnn} and GRU-based aggregation~\cite{gruaggr} to further capture both node and structural information of query plans, providing a stronger basis for downstream representation-based models. Third, we propose a learning-to-rank cost estimator with uncertainty quantification. It adaptively integrates cost estimates and uncertainties for plan selection, significantly improving robustness while maintaining cost estimation accuracy. Reqo integrates all three techniques into a comprehensive cost model, forming a virtuous cycle. Our tree model first generates richer plan representations, enabling more accurate cost estimates and explanations. These explanations yield hints to indirectly improve the worst candidate in plan selection, while also enriching the tree model with subplan-level information. The learning-to-rank cost estimator learns from actual plan comparisons to refine cost and uncertainty estimates as well as their integration for plan selection, and feeds back to benefit the tree model learning. Experimental results demonstrate that this synergy enables Reqo to outperform state-of-the-art approaches across all three stages.

To summarize, our main contributions are:
\vspace{-0.03in}
\begin{itemize}
    \item We introduce explainability into LCMs for the first time by quantifying subgraph contributions to cost predictions.
    \item As an example application, we use explanations to produce plan generation hints that avoid costly operator implementations in specific subplan patterns and improve plan quality.
    \item We propose a novel query plan tree model using Bi-GNNs with a GRU-based aggregator to enhance plan representation.
    \item We design an uncertainty-aware learning-to-rank cost estimator that adaptively integrates estimated cost with quantified uncertainty to improve the robustness of plan selection.
    \item We develop \emph{Reqo}, a comprehensive learning-based cost model that integrates the above three techniques and experimentally demonstrates its significant improvements in all three stages compared to the corresponding state-of-the-art approaches.
\end{itemize}

In the paper, Section~\ref{section_problem_statement} outlines the problem statement, Section~\ref{section_model_overview} details three techniques included in Reqo, Section~\ref{section_model_architecture} describes Reqo’s architecture, Section~\ref{section_experimental_study} presents the experimental study, Section~\ref{section_related_work} reviews related work, and Section~\ref{section_conclusion} concludes the paper.

\section{Problem Statement}
\label{section_problem_statement}

This section outlines the three core challenges in learning-based query optimization. First, we discuss the requirement for plan representations that preserve both node-level features and structural dependencies for downstream tasks (Section~\ref{section2_subsection_query_plan_representation}). Second, we explore the explainability gap between learning-based and classical cost models, and how explanations can be leveraged to improve plan quality (Section~\ref{section2_subsection_explainability}). Finally, we highlight the need to address uncertainty and the limitations of existing approaches for quantifying and leveraging uncertainty in robust plan selection (Section~\ref{section2_subsection_robustness}).

\subsection{Query Plan Representation}
\label{section2_subsection_query_plan_representation}

In learning-based query optimizers, query plan representation learning typically begins by taking physical plans as input, which are encoded using a feature encoder and a tree model to generate plan representations. These representations encapsulate critical information, including operators, parent-child relationships, and underlying data, serving as essential inputs for downstream tasks. Consequently, their quality impacts the performance ceiling for the entire cost model, making the tree model’s output pivotal to the optimization process.
A comparative study on tree models~\cite{bigg} has shown that different tree models can lead to varying cost estimation performance under identical workloads and optimizer settings, highlighting the importance of the tree model choice.

A physical query plan is a tree \( p = (N_p, E_p) \), where each \( n \in N_p \) denotes a node and each edge \( (n_c, n_P) 
\in E_p \) represents an execution dependency from child node $n_c$ to parent node $n_p$. Each node \( n \) is associated with a feature vector \( \mathbf{x}_n \in \mathbb{R}^d \), capturing node-level information such as operator implementation, relations, and predicates. A tree model \( g_{\phi} \), parameterized by \( \phi \), encodes all node features \(X_p = \{\mathbf{x}_n \mid n \in N_p\}\) and edge set \(E_p\) of \(p\) into a fixed-size representation \( r_p \in \mathbb{R}^k \), where $r_p = g_{\phi}(X_p, E_p)$. A downstream cost model \( m_{\theta} \), parameterized by \( \theta \), then utilizes this representation to predict the estimated execution cost \( \hat{y}_p \in \mathbb{R} \), where $\hat{y}_p = m_{\theta}(r_p)$.

\begin{problem}[Effective Plan Representation]
Given a downstream model \( m_{\theta} \), find the optimal parameters \( \phi^* \) for the tree model \( g_{\phi} \), so that the generated representations \( r_p \) maximize the performance of \( m_{\theta} \) by minimizing the expected cost estimation error over query plans in \( P \):
{
\setlength{\abovedisplayskip}{2pt}
\setlength{\belowdisplayskip}{2pt}
\setlength{\abovedisplayshortskip}{2pt}
\setlength{\belowdisplayshortskip}{2pt}
\begin{equation}
\phi^*,\theta^* = \arg\min_{\phi, \theta} \ \mathbb{E}_{(p, y_p) \sim P} \left[ \mathcal{L}\left( m_{\theta}\left( g_{\phi}(X_p, E_p) \right),\ y_p \right) \right]
\end{equation}
}

\noindent where \( \mathcal{L} \) is a loss function and \( y_p\) is the actual execution cost of \(p\).
\label{problem_plan_representation}
\end{problem}

\begin{challenge}
Design a tree model \(g_{\phi}\) that accurately captures both node-level features (e.g., operator implementation, relations, and predicates) and their structural dependencies. Any loss of these details degrades the quality of the input representation, preventing the downstream model \(m_{\theta}\) from reliably assessing the impact of each node and its position in the plan on the total cost. Strengthening \(g_{\phi}\)’s ability to encode complex query plans is therefore essential for learning-based query optimizers to achieve superior downstream performance.
\label{challenge_plan_representation}
\end{challenge}

\subsection{Explainability of LCMs}
\label{section2_subsection_explainability}

In classical query optimizers, cost models typically rely on statistical methods (e.g., histograms) with a transparent, modular structure. The total cost of a query plan is the sum of the costs of its constituent operators. In such models, for a query plan \(p\), the estimated cost of a parent node \(n_p \in p\) is given by the sum of the estimated cost of its children \(n_c\) plus a local cost reflecting the overhead of the operator at \(n_p\):
{
\setlength{\abovedisplayskip}{2pt}
\setlength{\belowdisplayskip}{2pt}
\setlength{\abovedisplayshortskip}{2pt}
\setlength{\belowdisplayshortskip}{2pt}
\begin{equation}
C_{\mathrm{classical}}(n_{p})
\;=\;
c(n_p)
\;+\;
\sum_{n_c \in \mathrm{Children}(n_p)} C_{\mathrm{classical}}(n_c)
\label{eq:classical_node}
\end{equation}
where \( c(n_p) \) is a function that estimates the local cost of operator \( n_p \). Recursively applying Eq.~\ref{eq:classical_node} from leaves to the root yields the accumulated plan's total estimated cost $C_{\mathrm{classical}}(p) = C_{\mathrm{classical}}(n_{root})$.}

This bottom-up aggregation makes cost estimates traceable to each node, allowing developers to tune performance at the operator, subplan, or arbitrary subgraph level with insights into how local cost estimates contribute to the total estimate. In contrast, LCMs take the entire plan as input to predict its cost \( C_{\mathrm{learned}}(p) \) using a parameterized (often black-box) function \( m_{\theta} \) trained on historical query data, which do not explicitly decompose the plan into subgraphs in an explainable manner. Although these models often achieve higher accuracy, they forgo the transparency of classical optimizers and obscure how subgraphs affect the total cost estimate. This lack of transparency hinders the diagnosis of cost estimation errors, complicates fine-grained tuning, and erodes trust when estimates deviate from observed execution times.

To restore the transparency of classical cost models, an LCM should have the ability to quantify the contribution of each subgraph \( sg \subseteq p \) to the total cost \( C_{\mathrm{learned}}(p) \). Such \(sg\)s include operators that can be individually optimized (e.g., joins, scans). Identifying their contributions helps developers locate bottlenecks or highlight operators that may benefit from tuning or rewriting.

\begin{definition}[Explainable LCM]
\label{def:explainability}
Let \( sg \) be any connected subgraph of a plan \( p \). A LCM is considered to be explainable if it quantifies the contribution of \(sg\) to \( C_{\mathrm{learned}}(p) \) via a function $\mathcal{E}$:
{
\setlength{\abovedisplayskip}{2pt}
\setlength{\belowdisplayskip}{2pt}
\setlength{\abovedisplayshortskip}{2pt}
\setlength{\belowdisplayshortskip}{2pt}
\begin{equation}
\mathcal{E} \colon \{(sg, p)\,\mid\,sg\subseteq p\} \to \mathbb{R}
\label{eq:explainer_F}
\end{equation}
Since a black-box model \( m_{\theta} \) does not directly reveal how individual subgraphs contribute to the overall cost, \( \mathcal{E} \) must be integrated with \( m_{\theta} \) to capture the contribution at the subgraph level.}
\end{definition}

\begin{problem}[Accurate Subgraph Contribution Estimation]
The goal of explainability is to accurately estimate each subgraph’s contribution to the entire plan's cost prediction in a learning-based query optimizer. This estimate should closely match the actual contribution \(AC(sg, p)\) calculated from the summed actual execution times \(T\) of all nodes in \(sg \subseteq p\) and in the entire plan \(p\), where \(AC(sg,p) = T_{sg} / T_{p}\). We seek the function \( \mathcal{E}^* \) that estimates the contribution with minimal error:
{
\setlength{\abovedisplayskip}{2pt}
\setlength{\belowdisplayskip}{2pt}
\setlength{\abovedisplayshortskip}{2pt}
\setlength{\belowdisplayshortskip}{2pt}
\begin{equation} \mathcal{E}^* = \arg\min_{\mathcal{E}} \ \mathcal{L}_{\text{Explanation}}\bigl(\mathcal{E}(sg, p), AC(sg,p)\bigr),\ \forall \ sg \subseteq p \end{equation}
\noindent where \( \mathcal{L}_{\text{Explanation}} \) is a loss function that measures the discrepancy between $AC(sg,p)$ and the contribution of \(sg\) to the plan’s cost prediction as quantified by \( \mathcal{E} \).}

\label{problem_subgraph_contribution}
\end{problem}

\begin{figure}
    \centering
    \includegraphics[width=0.775\linewidth]{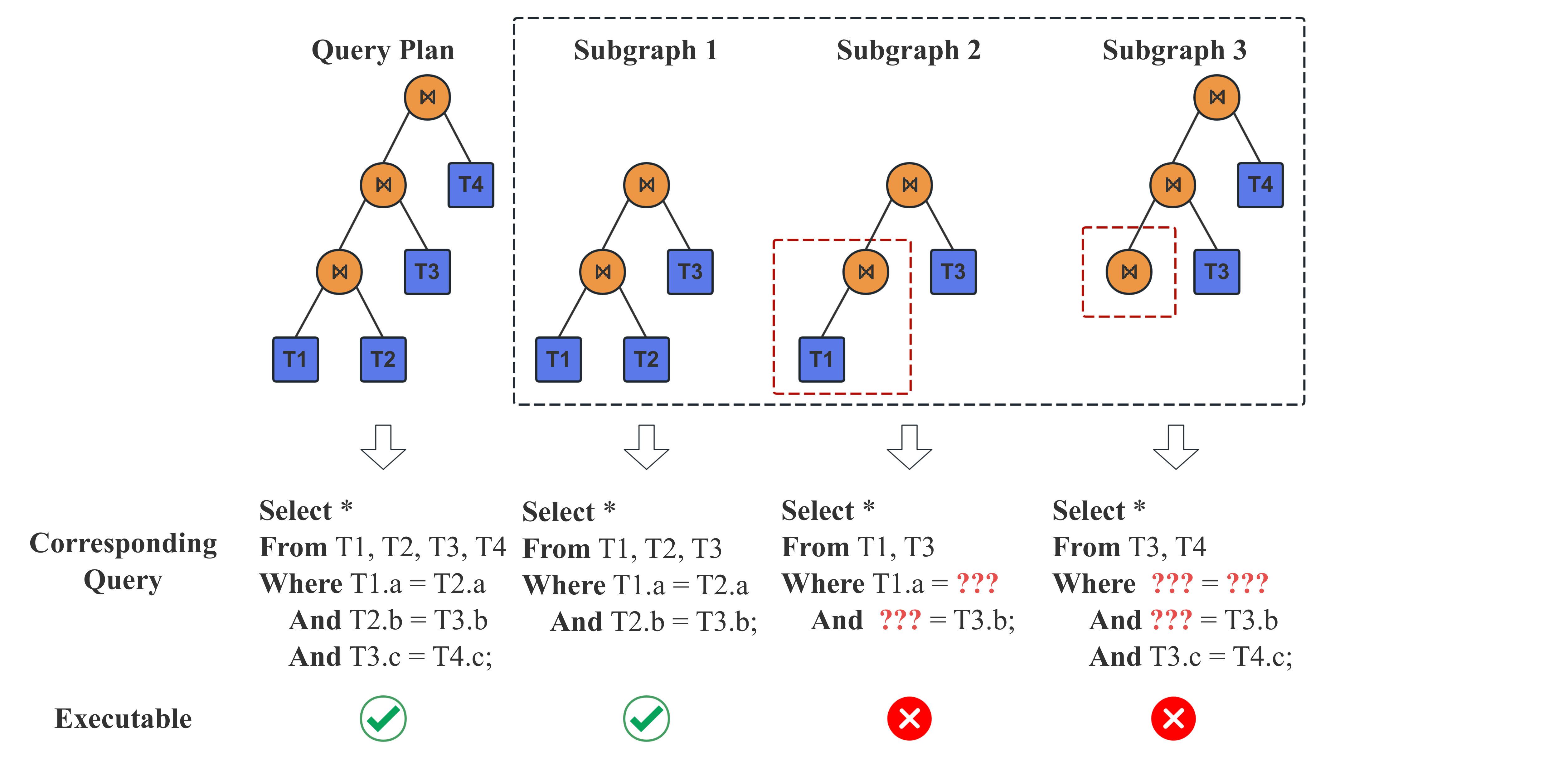}
    \caption{Omitting any leaf nodes during plan subgraph extraction produces non-executable subgraphs, preventing accurate cost estimation}
    \label{query_plan_integrity}
\end{figure}

\begin{challenge}
Design an explanation function \(\mathcal{E}\) that accepts \(sg\)s as input and produces accurate contribution values, where \(sg\)s can be subplans, individual nodes, or any connected subset of nodes in which at least one of its nodes has children in the entire plan that are not included in the subset.
If the contribution of internal plan nodes is required, the corresponding \(sg\) is not a complete subplan. Since it lacks essential information such as leaf relations and downstream operators, its cost contribution becomes difficult for LCMs to estimate, as illustrated in Example~\ref{example_query_plan_integrity}.
Furthermore, \(\mathcal{E}\) must account for the fact that the contribution of an \(sg\) can vary across query plans. \(\mathcal{E}\) should capture both the structure of \(sg\) and its contextual information within the entire plan, such as how \(sg\) is reached and its incoming cardinalities, on which the cost estimation of \(sg\) highly depends.
No existing LCM can accurately explain a given \(sg\) in a plan while accounting for its contextual information. Due to the black-box nature of LCMs, the monotonicity implicitly enforced by classical cost models (a subplan’s cost should not be lower than that of its containing subplans) is not guaranteed, which makes it challenging to infer the contributions of individual nodes or specific \(sg\)s from isolated per-subplan estimates.

\label{challenge_subgraph_contribution}
\end{challenge}

\begin{example}
Subgraphs 2 and 3 in Figure~\ref{query_plan_integrity} illustrate this issue. Since the leaf scan operators for tables \texttt{T1} and \texttt{T2} are omitted, the cost model cannot accurately estimate the join's cost.\qed
\label{example_query_plan_integrity}
\end{example}

If \(\mathcal{E}\) is sufficiently accurate, high-cost subgraphs can be identified prior to query execution, providing opportunities for targeted re-optimization. Leveraging \(\mathcal{E}\)-based explanations in the form of hints to optimize plan generation is feasible, where hints are directives that instruct the optimizer to enforce or disable specific operator implementations rather than rely solely on cost-based decisions during plan generation. Current approaches~\cite{bao, autosteer, fastgres} typically use global disabling hints (GDHints) that restrict specific operator implementations throughout plan generation for a given query. However, when an operator implementation appears multiple times in a plan, global disabling prevents all its occurrences. This can avoid catastrophic cases, but also blocks others that might be the optimal solution, hindering overall plan quality.

\begin{definition}[Context-based Hints for Plan Generation]
A context-based hint generation function is defined as:
{
\setlength{\abovedisplayskip}{3pt}
\setlength{\belowdisplayskip}{3pt}
\setlength{\abovedisplayshortskip}{3pt}
\setlength{\belowdisplayshortskip}{3pt}
\begin{equation}
\pi\colon (\mathcal{O}_n,\,sg_n)\to h\in\{op\in \mathcal{O}_n \colon-1,0,1\}
\end{equation}
where \(\mathcal{O}_n\) is the set of all operator implementations (\(op\)s) considered at node \(n\) during plan generation, and \(sg_n\) denotes the subgraph where \(n \in sg_n\) and provides contextual information (e.g.\ \(n\)'s position in the plan and the relations accessed along the path to \(n\)). Based on \(sg_n\), \(\pi\) generates a hint \(h\) for an \(op \in \mathcal{O}_n\) that enforces (1) or disables (-1) \(op\) at node \(n\), or applies no hint (0), guiding the optimizer to select an appropriate \(op\) for \(n\) in a context-aware manner for more efficient plans.}
\end{definition}

\begin{problem}[Explanation-based Hints for Query Optimization]
Let \(Q\) be a set of queries and let \(H_{\pi}\) denote the union of hints generated by \(\pi\) based on \(sg\)s collected from the plans of \(Q\) and their explanations. At execution time, a subset \(h_q \subseteq H_{\pi}\) is selected for each \(q \in Q\). Let \(\mathrm{T}(q \mid h_q)\) be the runtime of \(q\) using \(h_q\). We seek \(\pi^*\) that generates hints to minimize the average runtime of \(Q\):
{
\setlength{\abovedisplayskip}{2pt}
\setlength{\belowdisplayskip}{2pt}
\setlength{\abovedisplayshortskip}{2pt}
\setlength{\belowdisplayshortskip}{2pt}
\begin{equation}
\pi^* \;=\; \arg\min_{\pi}\; \frac{1}{|Q|}\sum_{q \in Q} \mathrm{T}\bigl(q \mid h_q\bigr),\ h_q \subseteq H_{\pi}
\label{eq_plan_generation_hints}
\end{equation}
}
\label{problem_plan_generation_hints}
\end{problem}

\begin{challenge}
Given a workload of plans for a set of queries, generate hints \(H_{\pi}\) by explanations derived from the workload while incorporating contextual information. Specifically, \(\pi\) must accurately identify meaningful \(sg\) for each node during plan generation, ensuring \(sg\) contains sufficient contexts. \(\pi\) should efficiently recognize the occurrences of \(sg\) in the workload and collect their corresponding explanations produced by \(\mathcal{E}\). Utilizing this information, \(\pi\) decides which \(op\in\mathcal{O}_n\) to enforce, disable or leave unaffected for the node in the input \(sg\). During this process, \(H_{\pi}\) must avoid outlier-driven decisions, ensuring no \(op\) is disabled or enforced solely because it performs badly or well in rare or extreme contexts. For example, an index scan may excel on columns with low row selectivity but underperform on those with high selectivity, so enforcing it does not guarantee better performance.

\label{challenge_plan_generation_hints}
\end{challenge}

\subsection{Robust Learning-based Cost Estimation}
\label{section2_subsection_robustness}

Most cost models prioritize the accuracy of cost estimates in their design, often overlooking inherent uncertainties. In reality, uncertainties in query plan execution are significantly influenced by factors such as structural characteristics, specific operators or predicates, and data properties. Moreover, estimation models themselves may introduce further limitations. In this paper, robustness in cost estimation refers to the ability of a cost model to maintain accurate plan selection despite these inherent uncertainties. The classical problem of optimal plan selection is defined as: given a finite set of candidate execution plans \(\mathcal{P} = \{p_1, p_2, \ldots, p_{|\mathcal{P}|}\}\) and a cost function \(m(p_i)\) that estimates the cost of executing plan \(p_i\) for \(i = 1, \ldots, |\mathcal{P}|\), the goal is to find the optimal plan \(p^*\) such that:
{
\setlength{\abovedisplayskip}{2pt}
\setlength{\belowdisplayskip}{2pt}
\setlength{\abovedisplayshortskip}{2pt}
\setlength{\belowdisplayshortskip}{2pt}
\begin{equation}
    p^* = \arg\min_{p_i \in \mathcal{P}} m(p_i)
\end{equation}
}

This formulation overlooks the inherent inaccuracies in cost estimates, which may lead to selecting suboptimal plans at runtime. Robust query optimization aims to minimize such risks by explicitly modeling and incorporating uncertainties into plan selection. Accordingly, a function \(u(p_i)\) is introduced to quantify the uncertainty in the estimated cost of plan \(p_i\).

\begin{problem}[Robust Plan Selection]
The goal of robust plan selection is to identify the optimal plan \(p^*\) that considers both the estimated cost \(m(p)\) and uncertainty \(u(p)\).
{
\setlength{\abovedisplayskip}{2pt}
\setlength{\belowdisplayskip}{2pt}
\setlength{\abovedisplayshortskip}{2pt}
\setlength{\belowdisplayshortskip}{2pt}
\begin{equation}
    p^* = \arg\min_{p_i \in \mathcal{P}} b\left( m(p_i),\ u(p_i) \right)
\label{eq:problem_robust_plan_selection}
\end{equation}
where \(b\) is a function representing the optimizer's strategy in balancing estimated cost and uncertainty. A learned \(b\) captures nonlinear trade-offs and improves plan selection accuracy. It adapts to workload characteristics and shifts, and can be trained without extensive manual tuning.}
\label{problem_robust_plan_selection}
\end{problem}

\begin{challenge}
Design an adaptive \(b\) that balances \(m(p_i)\) and \(u(p_i)\) based on workload characteristics. Existing approaches~\cite{zhao2021uncertainty, liu2021fauce, chen2023leon, doshi2023kepler, roq} quantify uncertainty but rely on predefined rules to balance \(m(p_i)\) and \(u(p_i)\) in plan selection, which are not learned during their cost model training, inhibiting \(b\)'s adaptation to the workload’s characteristics. Therefore, the function \(b\) should be integrated into the cost model and trained from plan comparison feedback to learn this adaptive balance.
\label{challenge_robust_plan_selection}
\end{challenge}

\section{Model Overview}
\label{section_model_overview}
We propose Reqo, a comprehensive LCM that integrates three novel techniques. First, a novel representation-learning model (Section~\ref{section_tree_model}) based on Bi-GNNs and GRUs preserves both node-level and structural information, producing superior plan representations. Second, to overcome the black-box nature of LCMs, an explainability technique (Section~\ref{section_explainability_technique}) quantifies the contributions of plan subgraphs to cost predictions, thereby promoting transparency and generating hints to improve plan quality. Third, a robust learning-to-rank cost estimator (Section~\ref{learning-to-Rank_uncertainty_quantification}) adaptively quantifies and integrates uncertainty to enhance the robustness of plan selection.

\subsection{A Novel Tree Model for Query Plan Representation Learning}
\label{section_tree_model}
We propose a novel tree model that leverages Bi-GNNs and a GRU-based aggregator to solve Problem~\ref{problem_plan_representation}. This design preserves both node-level and structural information when transforming a physical plan tree into a representation, serving as a powerful \(g_{\phi}\). This richer representation enhances \(m_{\theta}\)'s input quality and provides a stronger foundation for our subsequent techniques. Specifically, our subplan-based explainer (Section~\ref{section_explainability_technique}) uses these representations to accurately explain each subplan’s contribution to the predicted plan cost. Our robust learning-to-rank cost estimator (Section~\ref{learning-to-Rank_uncertainty_quantification}) benefits from these representations to better distinguish candidate plan differences under uncertainty, mitigating misestimation and leading to more robust plan selection.

\subsubsection{Bidirectional GNNs}
\label{subsection_bidirectional_gnn}

\begin{figure}[t!]
\centering
\includegraphics[width=\textwidth]{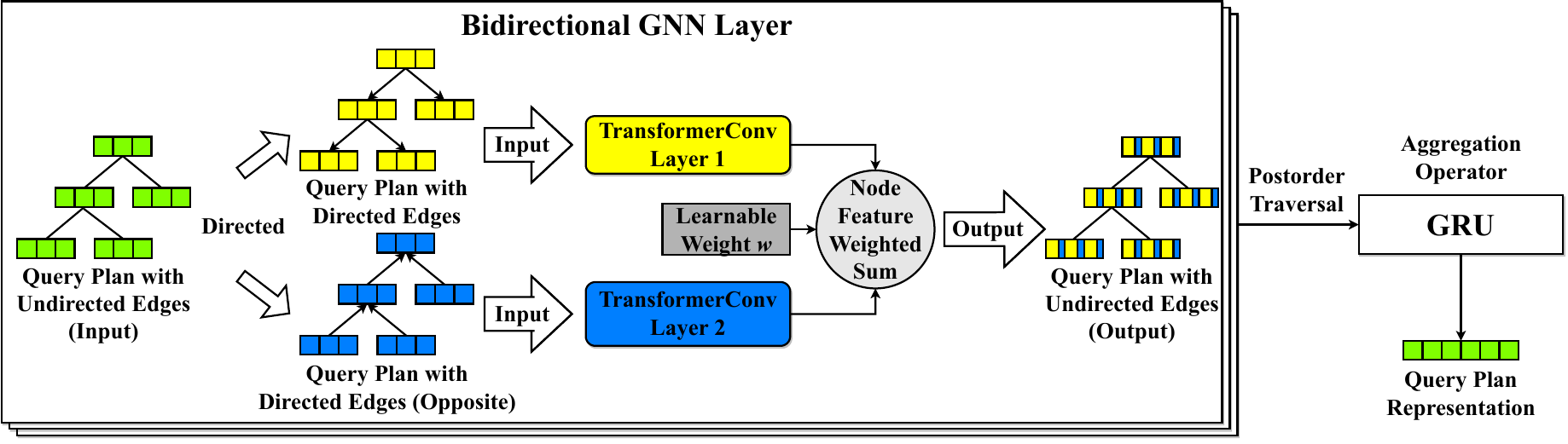}
\caption{Architecture of the proposed tree model using bidirectional GNN with a GRU-based aggregator} \label{fig:bigg}
\end{figure}

To improve the tree model's ability to represent query plans accurately, we employ GNNs due to their proficiency in capturing graph topology~\cite{kipf2016semi}. We innovatively treat each query plan tree as two single-directional graphs with opposite edge directions (parent-to-child and child-to-parent). In each layer shown in Figure~\ref{fig:bigg}, these two graphs are processed independently by TransformerConv~\cite{transformerconv} layers. We then integrate the corresponding node features from the two output learned graphs through a learnable parameter, which makes it possible to transmit information in both directions while still utilizing the direction information of the edges and retaining relevant structural information.
This Bi-GNN design facilitates information flow in both directions, enabling nodes to learn from both sides, unlike using single-directional edges. By treating the query plan tree as two graphs with opposite edge directions, the model preserves dependencies between parent and child nodes compared to using undirected edges. TransformerConv layers with multi-head attention~\cite{selfattention} allow each node to adaptively aggregate neighbor information, enhancing the model's ability to capture the tree's local graph topology and global dependencies. This design benefits from GNNs and addresses the limitations of single-directional and undirected GNNs in learning query plans, markedly improving tree-structured plan representation learning.

\subsubsection{GRU-Based Aggregation Operator}
\label{bidirectional-gnn-aggregation}
Conventional graph aggregation methods often yield inferior performance on query plans, since they typically rely on global pooling and thus ignore the tree's structural information. To address this limitation, we employ GRUs~\cite{gruaggr} to aggregate the Bi-GNN-derived node features after postorder traversal of the plan tree. This traversal approximates the execution order of plan nodes in DBMSs~\cite{saturn}, allowing the model to selectively retain or discard features and to learn how operator order affects cost. Consequently, the model captures node dependencies more accurately and aligns with the actual execution sequence, enabling a graph-level plan embedding while retaining both essential node-level and structural information.

\subsection{An Explainability Technique for LCMs}
\label{section_explainability_technique}
To address Problem~\ref{problem_subgraph_contribution}, we propose a subplan-based explainability technique for LCMs that uses a learning-based explainer to quantify the embedding similarity between each subplan and the entire plan. This explainer allows an LCM to infer each subgraph's contribution to the total cost, promoting transparency similar to that of classical cost models. We further leverage these explanations to generate subplan pattern hints to optimize plan generation and solve Problem~\ref{problem_plan_generation_hints}.

\subsubsection{Learning-based Explainability Technique via Subplan-Plan Embedding Similarity}

To explain the cost contribution of subgraphs (\( sg \)s) in a query plan, a straightforward approach is to feed each \( sg \) directly to the LCM. However, as discussed in Section~\ref{section2_subsection_explainability}, when \(sg\) is not a subplan, the absence of leaf nodes strips away essential information, making its estimated contribution unreliable. Moreover, training LCMs on such non-executable \(sg\)s would inject noise into the training of the base cost model. Therefore, we restrict the training of this explainability technique to subplans only.

\begin{figure}[!t]
    \centering
    \includegraphics[width=0.95\linewidth]{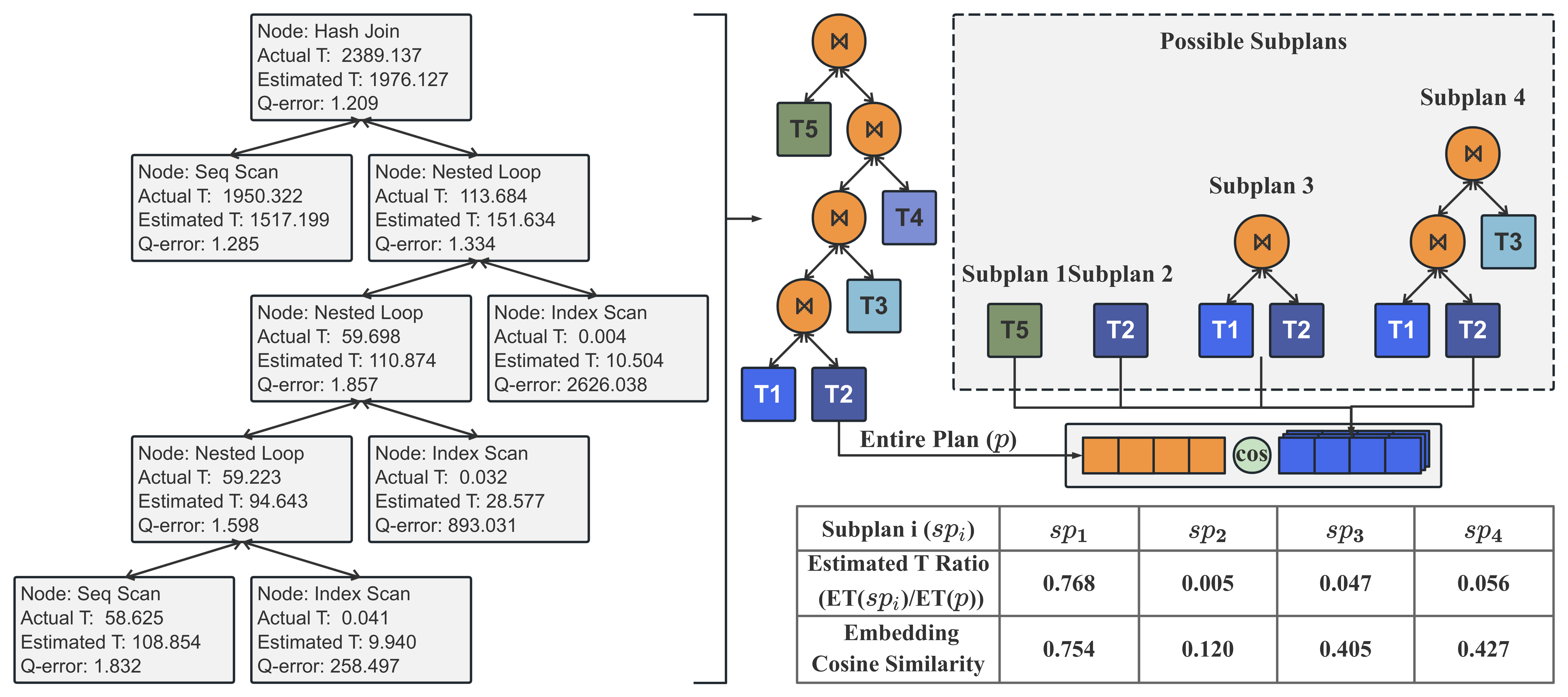}
    \caption{Example of the relationship between the cosine similarity of the entire query plan and its subplan embeddings, and their contributions to the plan-level cost prediction}
    \label{fig_subplan_similarity}
\end{figure}

\begin{example}
Figure~\ref{fig_subplan_similarity} shows a plan whose nodes are annotated with cost estimates obtained by feeding the subplan rooted at each node into an LCM. We extract four subplans to examine the correlation between the embedding cosine similarity of each subplan to the entire plan and its corresponding contribution to the entire plan's estimated cost. For example, subplan 1, although just a leaf, has the highest similarity (0.754) and largest contribution (0.768). In contrast, subplan 2 shows low similarity (0.120) and minimal contribution (0.005). Subplans 3 and 4 exhibit intermediate levels on both. These observations indicate a positive correlation between embedding similarity and subplan contribution.\qed
\label{eg:embedding_similarity}
\end{example}

Given accurate plan-level cost prediction, embeddings of subplans with substantial contributions should exhibit a strong relationship with the entire plan embedding. Without this relationship, the LCM cannot effectively capture costly subplan information from the plan-level embedding, resulting in inaccurate estimates for the entire plan. As shown in Example~\ref{eg:embedding_similarity}, the strength of this relationship reflects subplan contributions and can be approximated by quantifying embedding similarity between each subplan and the entire plan.
Ranking these similarities then identifies the subplans with the largest contribution, where similarity can be measured using functions such as cosine similarity, mutual information (MI)~\cite{kraskov2004estimating}, and learning-based methods.

\begin{figure}
\centering
\includegraphics[width=0.8\textwidth]{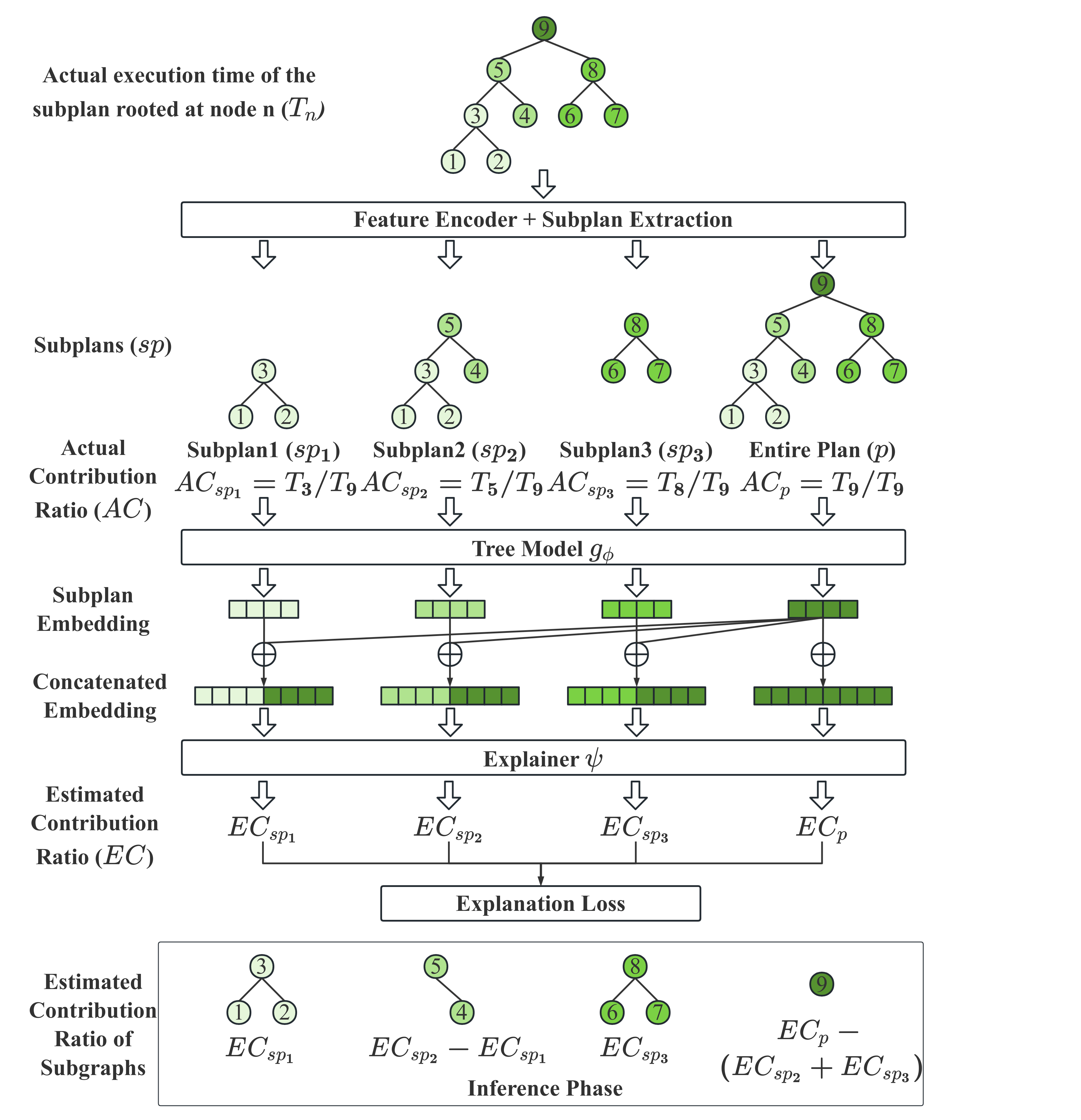}
\caption{The explainability technique for LCMs based on subplan-plan embedding similarity}
\label{fig:explainability_technique}
\end{figure}

Therefore, function \(\mathcal{E}\) (Eq.~\ref{eq:explainer_F}) can be instantiated using a learning-based model to adaptively quantify similarity instead of using fixed mathematical metrics. By concatenating each subplan embedding with the entire plan embedding, the model captures contextual information and estimates the subplan’s contribution to the predicted cost. We therefore propose an explainability technique for LCMs, illustrated in Figure~\ref{fig:explainability_technique}. We first extract all subplans (\(sp\)s) from the entire query plan \(p\) and encode them (including \(p\)) using the same tree model \(g_{\phi}\). During training, a learning-based explainer \(\psi\) automatically learns the contribution of each subplan to cost prediction. In particular, \(\psi\) estimates the contribution \( EC_{sp_k} \in [0,1] \) of each subplan \( sp_k \) by quantifying the similarity between its embedding \(Emb_{sp_k}\) and the embedding of the entire query plan \(Emb_{p}\) as below:
    \begin{equation}
        EC_{sp_k} = \psi(\operatorname{CONCAT}(Emb_{sp_k}, Emb_{p})),\ EC_{sp_k} \in [0,1]
    \end{equation}

The actual contribution ratio \( AC_{sp_k} \) is the ratio of the subplan's actual execution time \(T_{sp_k}\) to the entire plan’s execution time \(T_{p}\):
{
\setlength{\abovedisplayskip}{2pt}
\setlength{\belowdisplayskip}{2pt}
\setlength{\abovedisplayshortskip}{2pt}
\setlength{\belowdisplayshortskip}{2pt}
    \begin{equation}
        AC_{sp_k} = \frac{T_{sp_k}}{T_{p}},\ AC_{sp_k} \in [0,1]
    \end{equation}
}

An explanation loss function is used to minimize the discrepancy between $EC$ and $AC$. Given query plans \(P\) with \(|P|=N\), where each \(p_i\in P\) has \(K_i\) subplans, the explanation loss is:
{
\setlength{\abovedisplayskip}{2pt}
\setlength{\belowdisplayskip}{2pt}
\setlength{\abovedisplayshortskip}{2pt}
\setlength{\belowdisplayshortskip}{2pt}
\begin{equation}
    \mathcal{L}_{\text{Explanation}} = \frac{1}{N} \sum_{i=1}^{N} \left( \frac{\sum_{k=1}^{K_i}  (AC_{sp_{ik}} - EC_{sp_{ik}})^2 }{K_i} \right)
        \label{explanationloss}
\end{equation}
}

Thus, the loss function forces the LCM to estimate each subplan’s contribution while conditioning on its context within the entire plan, improves monotonicity in cost estimates across subplans of different sizes within the same plan, and can be used to infer more accurate explanations for specific nodes or subgraphs. After training, when explanations are not required, \(\psi\) can be disabled to reduce inference time. The explainer \(\psi\) serves as \(\mathcal{E}\) to estimate subplan contributions considering the subplan's contextual information, trained via Eq.~\ref{explanationloss}. These contributions can be leveraged to optimize plan generation. To enable finer-grained optimization, our technique infers contributions for arbitrary subgraphs from subplan contributions, as detailed in the following section.

\subsubsection{Plan Subgraph Contribution Inference via Subplan Explanations}
To enable \(\mathcal{E}\) to accept arbitrary subgraphs and estimate their contributions, we propose a technique that infers any arbitrary subgraph’s contribution from the estimated contributions of subplans within the same plan. Specifically, given a subgraph \(sg \subseteq p\), let \(n_{sgr} \in sg\) be the node closest to the plan root (i.e., with minimal depth). For any subplan \(sp \subseteq p\), let \(EC_{sp}(n)\) be the estimated contribution of the subplan rooted at node \(n\). Identify boundary children as any node \(n_c \in p\) such that there exists a parent node \(n \in sg\) with \(n_c \in \mathrm{Children}_p(n)\) but \(n_c \notin sg\), where \(\mathrm{Children}_p(n)\) denotes the immediate child nodes of \(n\) in plan \(p\). Since \(EC_{sp}(n_{sgr})\) includes the contributions of all children of \(n_{sgr}\), subtracting the contributions \(EC_{sp}(n_c)\) for each boundary child \(n_c\) yields the contribution of \(sg\). Formally:
{
\setlength{\abovedisplayskip}{2pt}
\setlength{\belowdisplayskip}{2pt}
\setlength{\abovedisplayshortskip}{2pt}
\setlength{\belowdisplayshortskip}{2pt}
\begin{equation}
EC(sg) \;=\; \max (0,\ EC_{sp}(n_{sgr}) \;-\; \sum_{\,n_c \in \partial(sg)} EC_{sp}(n_c))
\label{eq:subgraph_contribution_inferring}
\end{equation}
where \(\partial(sg) = \{\,n_c \mid \exists\,n \in sg,\;n_c \in \mathrm{Children}_p(n),\;n_c \notin sg\}\). The \(\max(0,\cdot)\) ensures non-negativity and mitigates spurious negative values caused by estimation errors. This process is also shown in Algorithm~\ref{alg:compute_subgraph_EC}. By combining this technique with our learning-based explainer \(\psi\), it enables contribution estimation for arbitrary subgraphs within a plan and thereby fully addresses Challenge~\ref{challenge_subgraph_contribution}.}

\begin{algorithm}[!t]
\caption{Calculation of Subgraph Contributions from Subplan Explanations}
\label{alg:compute_subgraph_EC}
\begin{algorithmic}[1]
\small
\Require 
Query plan \(p\); Estimated subplan contributions \(EC_{sp}(n)\) for every node \(n \in p\); Any subgraph \(sg \subseteq p\)  
\Ensure Estimated contribution \(EC(sg)\) of subgraph \(sg\)
\Function{ComputeSubgraphEC}{$p, sg, EC_{sp}$}
  \State \(n_{sgr} \gets\) node in \(sg\) with minimal depth  \Comment{The node in \(sg\) closest to \(p\)'s root}
  \State \(\partial(sg) \gets \{\,n_c \mid \exists\,n \in sg,\; n_c \in \mathrm{Children}_p(n),\; n_c \notin sg\}\) \Comment{Boundary children of \(sg\)}
  \State \(EC(sg) \gets EC_{sp}(n_{sgr})\)
  \For{each boundary child \(n_c \in \partial(sg)\)}
    \State \(EC(sg) \gets EC(sg) - EC_{sp}(n_c)\)
  \EndFor
  \State \(EC(sg) \gets \max (0,\ EC(sg))\)
  \State \Return \(EC(sg)\)
\EndFunction
\end{algorithmic}
\end{algorithm}

\subsubsection{Explanation-Based Subplan Pattern Hints}

We introduce a technique to generate Subplan Pattern Hints (SPPHints) as \(H_{\pi}\) (Eq.~\ref{eq_plan_generation_hints}) based on plan explanation results across the workload and apply them via pg\_hint\_plan~\cite{pghintplan}, guiding the optimizer to avoid costly operators or adopt cheaper alternatives for specific subplan patterns in plan generation. The subplan pattern is defined as:

\begin{definition}[Subplan Pattern (SPP)]
Let \(p\) be a query plan and let \(n\in p\) be a node with \(k\) children \(n_{c}\). For each child \(n_{ci}\) \((i=1,\dots,k)\), \(R(n_{ci})\) denotes the sequence of all relations that appear at the leaves in the subplan rooted at \(n_{ci}\), ordered by postorder traversal. The subplan pattern \(SPP\) at \(n\) is defined as \( ((R(n_{c1})),\dots,(R(n_{ck})))\) if \(k \ge 1\), or \((R(n))\) if \(n\) is a leaf node.
\end{definition}

\begin{figure}[t]
    \centering
    \includegraphics[width=0.975\linewidth]{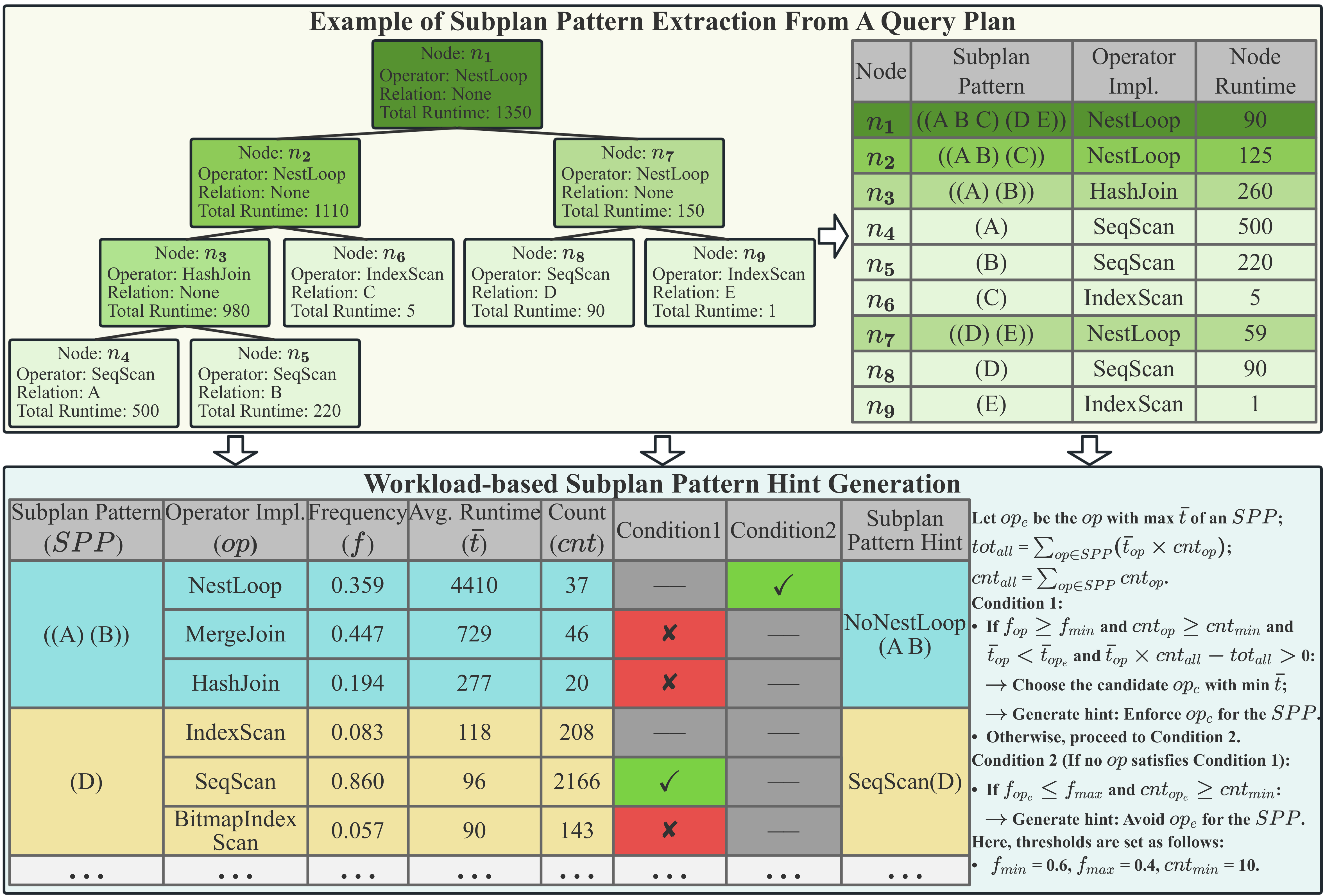}
    \caption{Example of subplan pattern extraction from a query plan and hint generation using workload-level explanation results}
    \label{fig:subplan_pattern_example}
\end{figure}

Given a workload \(W\), for each node \(n\) in plan \(p \in W\), we record its \(SPP_n\) and the tuple \((op, t)\), where \(op\) is the operator implementation at \(n\) and \(t\) is \(n\)'s runtime inferred from explanation results. Collecting all such tuples from \(W\) for a pattern \(SPP\) yields its subplan pattern instances
\(
SPPI(SPP) = \bigl\{(op_1,\,t_1),\, (op_2,\,t_2),\,\dots \bigr\}
\).
Given an \(SPPI(SPP)\) containing \(l\) \(op\) types, we group it by \(op\) types to form \(l\) subplan pattern instance groups \(SPPIG_j = \{\,op_j,\,f_j,\,cnt_j,\,\bar{t}_j\}\) for \(j=1,\dots,l\), where \(op_j\) is the \(j\)-th \(op\), \(cnt_j\) is \(op_j\)'s occurrence count in \(SPPI(SPP)\), \(f_j = cnt_j / \sum_{i=1}^l cnt_i\) is the group's relative frequency within \(SPPI(SPP)\), and \(\bar{t}_j\) is the average of all \(t\) in the group.

\(SPP\) captures each node \(n\)’s contextual information via the relation processing order. Under every \(SPP\), \(SPPIG\) records each observed \(op\)'s relative frequency \(f\), count \(cnt\), and average explained execution time \(\bar{t}\) over the entire explained workload. Figure~\ref{fig:subplan_pattern_example} illustrates an example of SPP extraction from an explained query plan and the algorithm for SPPHints generation. All \(SPP\)s and their \(SPPIG\)s are collected across the workload.
To reduce the impact of outliers, three thresholds are introduced: \(f_{\min}\) and \(f_{\max}\) bound the acceptable range of an \(SPPIG\)'s relative frequency \(f\), and \(cnt_{\min}\) is the minimum \(cnt\) required for consideration. The algorithm defines two conditions:

\textit{Condition~1 (enforce).} For an \(SPP\), consider candidates \(SPPIG_j\) that satisfy \(f_j \geq f_{\min}\) and \(cnt_j \geq cnt_{\min}\). If multiple candidates pass, select \(SPPIG_{\mathrm{sel}}\) with the smallest \(\bar{t}\), denoting as \(\bar{t}_{\mathrm{sel}}\). Choosing the smallest \(\bar{t}\) selects the relatively cheapest operator implementation \(op_c\) at the workload level. We then replace each \(SPPIG\)’s \(\bar{t}\) in the current \(SPPI\) with \(\bar{t}_{sel}\) to verify whether the total runtime improves. If it does, Condition 2 is skipped, and a hint is generated to enforce using \(op_c\) whenever the same \(SPP\) reappears during future plan generation.

\textit{Condition~2 (disable).} If no \(SPPIG\) meets Condition~1, select \(SPPIG_e\) with the largest \(\bar{t}\). If \(f_e \leq f_{\max}\) and \(cnt_e \geq cnt_{\min}\), generate a disable hint that avoids the expensive \(op_e\) for this \(SPP\).

This algorithm instantiates \(\pi\), and its safeguards ensure that costly \(op\)s are avoided only if efficient and frequent alternatives exist. Given the negligible overhead of generating unexecuted plans, the workload can be augmented with additional such plans and their explanations to refine \(SPP\) statistics and produce more precise and effective hints.

\begin{example}
Figure~\ref{fig:subplan_pattern_example} illustrates \(SPP\) extraction and the hint generation process for two patterns. For \(SPP\ = \bigl((A)\,(B)\bigr)\), three operator implementations appear but neither MergeJoin nor HashJoin satisfies Condition 1 (\(f\ge0.6\) and \(cnt\geq10\)), thus we proceed to Condition 2. The most expensive NestLoop as \(op_e\) meets \(f\le0.4\) and \(cnt\geq10\), so we generate a hint to avoid NestLoop on \(\bigl((A)\,(B)\bigr)\). For \(SPP = (D)\), since SeqScan satisfies Condition 1 and yields a positive total runtime improvement, we generate a hint to enforce SeqScan on \((D)\) and skip Condition 2. Subsequent queries apply these hints via pg\_hint\_plan, preventing the optimizer from selecting costly operator implementations. \qed
\end{example}

SPPHints generated from workload-wide explanations form the \(H_{\pi}\) in Problem~\ref{problem_plan_generation_hints}. For each query, a subset of \(H_{\pi}\) is extracted by matching the query’s relations to each hint’s subplan pattern. For example, selecting all hints associated with subsets of \(\{A, B, C\}\) when the query involves \(A\), \(B\), and \(C\).
During plan generation, this subset is applied via pg\_hint\_plan to select more efficient operator implementations. If no hint in \(H_{\pi}\) matches, the query is likely to have a low subplan cost at the workload level, or the operator choice has negligible impact, so no hints are applied. Since SPPHints arise as a byproduct of LCM training and testing, and are updated by rules, maintaining \(H_{\pi}\) incurs low overhead. When the workload changes, hints can be refreshed from test-time explanations without retraining, and existing hints can be applied to unseen queries that share the same \(SPP\)s if the data distribution does not change substantially.

As an application of the obtained explanations, this technique converts them into SPPHints tailored to the current workload and improves the plan generation safely as the solution to Problem~\ref{problem_plan_generation_hints}. These hints reduce the incidence of catastrophic operations among candidate plans, elevate the quality of the plan pool, and thereby indirectly enhance the robustness of plan selection and query performance. We consider this to be only a preliminary exploitation of the explainability technique, with more promising avenues to be explored.

\subsection{An Uncertainty-Aware Learning-to-Rank Cost Estimator for Robust Plan Selection}
\label{learning-to-Rank_uncertainty_quantification}
To address Problem~\ref{problem_robust_plan_selection}, we propose a robust learning-to-rank cost estimator that adaptively integrates uncertainty into cost estimation. Rather than relying on predefined rules, it employs a ranking loss with pairwise plan comparisons to learn how to adaptively combine estimated cost and uncertainty for plan selection. By training on the comparisons, the model reliably identifies cheaper plans while accounting for uncertainty, thereby improving the robustness of plan selection.

\subsubsection{Uncertainty Quantification}

Real-world query plans often exhibit variability due to data uncertainty~\cite{roq}, which in ML can stem from noise in inputs and labels. During cost estimation, uncertainty can arise from various complex factors, including fluctuations in execution time due to changes in the execution environment and the variability in the plan representations. In this work, we focus on this uncertainty and develop a technique to quantify and utilize it.

A neural network can be designed to predict the parameters of the normal distribution~\cite{nix1994estimating}, allowing it to predict not only the conditional expected value, but also the conditional variance of the target given the input and training data. This functionality is achieved by integrating a secondary output branch into the original learning-based cost estimator that is tasked with variance prediction, as shown in Figure~\ref{fig:learning_to_rank_robust_cost_model}. The optimization of this estimator involves minimizing the Gaussian negative log-likelihood, as demonstrated in the following uncertainty loss function:
{
\setlength{\abovedisplayskip}{3pt}
\setlength{\belowdisplayskip}{3pt}
\setlength{\abovedisplayshortskip}{3pt}
\setlength{\belowdisplayshortskip}{3pt}
\begin{equation}
\label{uncertaintyloss}
    \mathcal{L}_{\text{Uncertainty}} = \frac{1}{N} \sum_{i=1}^{N} \left( \frac{\ln \sigma_{p_i}^2}{2} + \frac{(y_{p_i} - \mu_{p_i})^2}{2\sigma_{p_i}^2} \right)
\end{equation}
where, given $N$ plans, for the $i$-th plan embedding as input, $\mu_{p_i}$ is predicted by the first branch of the estimator and represents the expected value of the estimated cost, $\sigma_{p_i}^2$ is predicted by the second branch and reflects the conditional variance as the data uncertainty, and $y_{p_i}$ is the label that represents the actual cost of the input plan. By minimizing this loss function, we can obtain both the estimated cost and its conditional variance for each plan, enabling effective uncertainty quantification without assuming bounded residuals for subsequent plan selection.}

\subsubsection{Uncertainty-Aware Learning-to-Rank via Pairwise Plan Comparisons}
\begin{figure}[t!]
    \centering
    \includegraphics[width=\linewidth]{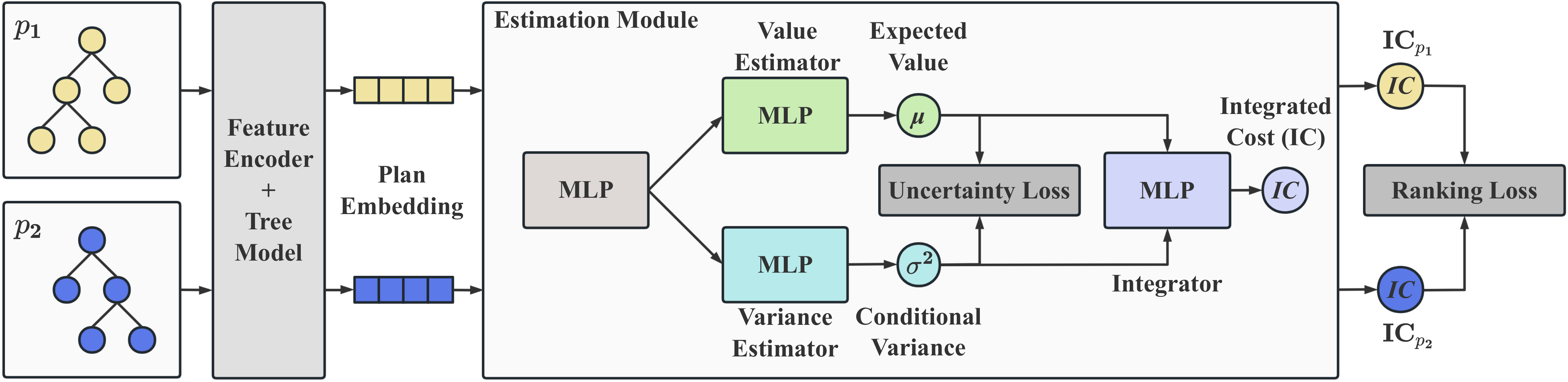}
    \caption{Architecture of the uncertainty-aware learning-to-rank cost estimator}    \label{fig:learning_to_rank_robust_cost_model}
\end{figure}

As discussed in Section~\ref{section_problem_statement}, existing uncertainty-aware cost estimators utilize their obtained uncertainty and estimated costs in a predefined balance strategy, which is independent of the estimator training phase, preventing self-improvement based on plan selection outcomes. To address this, we propose a novel learning-based cost estimator architecture that uses a ranking loss function with plan pairs as inputs, allowing the estimator to adaptively integrate uncertainty and cost estimates, as shown in Figure~\ref{fig:learning_to_rank_robust_cost_model}.

\textbf{Learning-based Integration.} We integrate the estimated expected value $\mu_p$ and variance $\sigma^2_p$ from the estimator module to compute each plan $p$’s integrated cost $IC_p$. Specifically:
\begin{equation}
    IC_p=b(\mu_p,\sigma_p^2)=\omega\bigl([\mu_p,\sigma_p^2]^\top\bigr)
\label{eq_integration_strategy}
\end{equation}
where $\omega$ is a lightweight multi-layer perceptron (MLP). This approach allows the estimator to adapt the integration process to specific workload characteristics.

\textbf{Pairwise Plan Comparison.} Beyond learning from cost prediction accuracy, our cost estimator also learns from pairwise plan comparisons with labels indicating which one is better in each pair, enabling the integrator to be trained jointly. Given a query plan set $P$ of size $N$, a pair of plans $\{p_i, p_j\} \subseteq P$, with their integrated cost $(IC_{p_i}, IC_{p_j})$ and actual cost $(y_{p_i}, y_{p_j})$, a specially designed ranking loss function enables the estimator to learn from the plan comparison results:
\begin{equation}
\label{rankingloss}
\mathcal{L}_{\text{Ranking}} = \sum_{i=1}^{N} \sum_{j=i+1}^{N}
\begin{cases}
\exp(-y_{p_{ij}} \cdot (IC_{p_i} - IC_{p_j}) + \Delta), & \text{if } y_{p_{ij}} \cdot (IC_{p_i} - IC_{p_j}) < 0 \ \wedge \ y_{p_{ij}} \neq 0 \\
0, & \text{otherwise}
\end{cases}
\end{equation}
where \(y_{p_{ij}}\)=1 if \(y_{p_i}>y_{p_j}\), -1 if \({y_{p_i}<y_{p_j}}\), and 0 otherwise (ties are ignored). The \(\Delta\) is a hyperparameter that scales the penalty assigned to misranked plan pairs.
Accordingly, the overall loss of our cost estimator is defined as:
\begin{equation}
\label{robust_cost_model_loss}
    \mathcal{L}_{\text{RobustRank}} = \mathcal{L}_{\text{Uncertainty}}(Eq.~\ref{uncertaintyloss}) + \mathcal{L}_{\text{Ranking}}(Eq.~\ref{rankingloss})
\end{equation}

Our learning-to-rank cost estimator addresses Problem~\ref{problem_robust_plan_selection} by first providing two specialized output branches: one branch serves as \( m \) to estimate cost and another as \( u \) to quantify uncertainty. We then introduce a learning-based integration function (Eq.~\ref{eq_integration_strategy}) as \( b \) (Eq.~\ref{eq:problem_robust_plan_selection}) that adaptively balances the two outputs. By incorporating pairwise plan comparisons into training, for a plan \(p\), the estimator automatically combines \( m(p) \) and \( u(p) \) into a single value to rank candidate plans, precisely capturing the core requirement of plan selection: comparing and choosing the most efficient and robust plan under inherent uncertainty. Consequently, our cost estimator not only strengthens overall robustness by learning from plan comparisons during training, but also supports flexible utilization of quantified uncertainty in plan selection, thereby addressing Challenge~\ref{challenge_robust_plan_selection}.

\begin{figure}[t]
\centering
\includegraphics[width=0.72\textwidth]{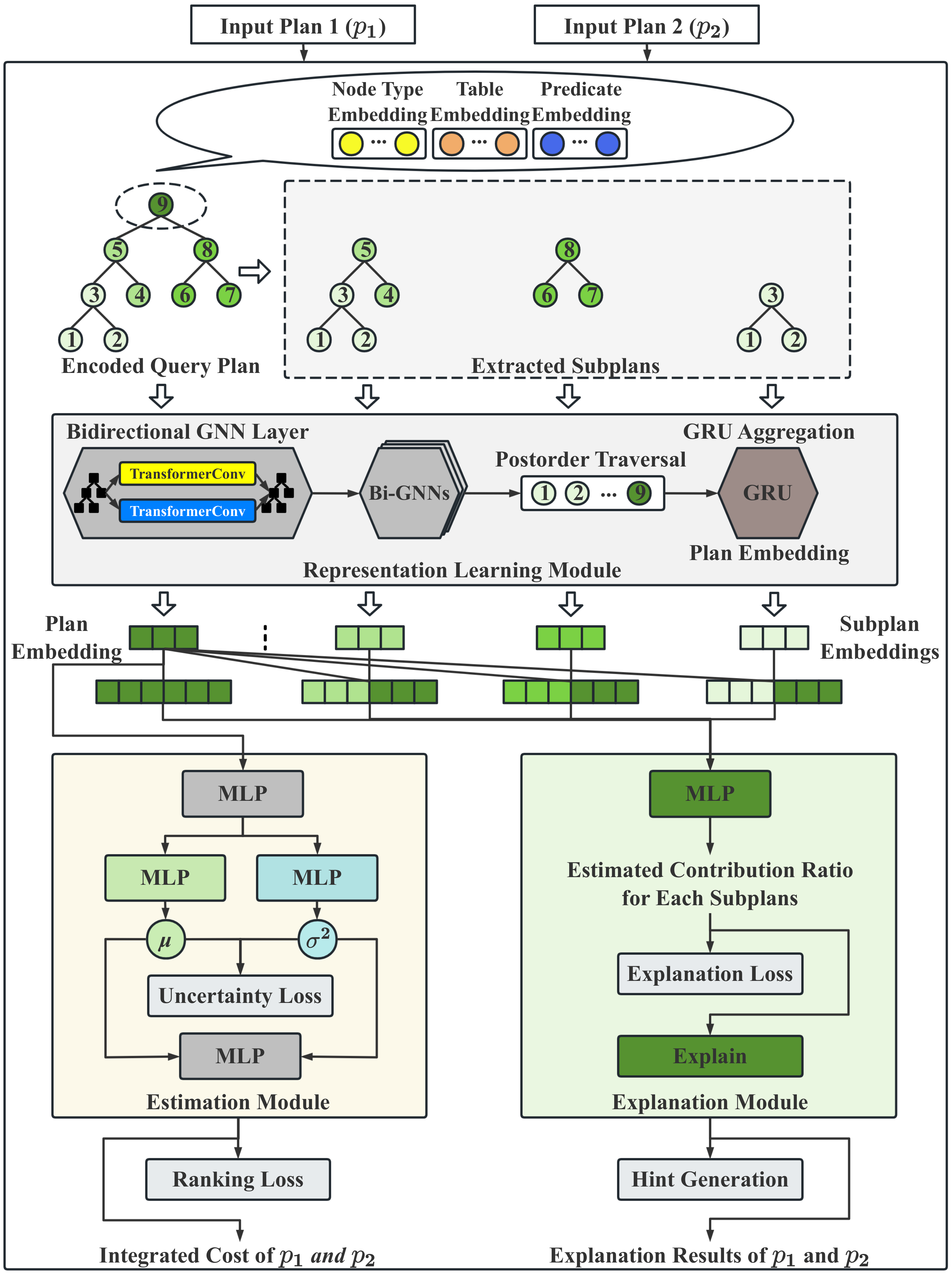}
\caption{The complete architecture of Reqo} \label{figure_model_architecture}
\end{figure}

\section{Model Architecture}
\label{section_model_architecture}
By integrating the three techniques in Section~\ref{section_model_overview}, we propose Reqo, an LCM that enhances cost estimation performance, explainability, and plan selection robustness. Figure~\ref{figure_model_architecture} shows its architecture, consisting of a feature encoder and three modules (representation learning, estimation, and explanation). The details of each component and the training process are described below.

\textbf{Query Plan Node Feature Encoding.}
\label{section_plan_feature_encoding}
A query plan details the operators used, their implementations, execution order, and involved relations. To convert this complex information into fixed-length node features, we propose a plan encoder inspired by RTOS~\cite{rtos}.
Each node feature comprises four parts: a node type embedding from one-hot operator types encoding with a fully connected (FC) layer; a table embedding from one-hot encoding of the node's involved relations with an FC layer; a predicate embedding formed by dividing numerical and character column predicates separately into eight cases based on predicate operators. For numerical columns, predicate values are normalized by the corresponding column range and placed into the column-specific matrix based on the predicate operator type. For non-numeric columns, word2vec~\cite{word2vec} is applied to convert character-type predicate values into embeddings, which are used to populate the column matrix as for numerical columns. Max pooling aggregates each table’s column embeddings into the table's embeddings, and concatenating all table embeddings yields the predicate embedding; and the normalized PostgreSQL's estimated cardinality and cost. Concatenating the four parts yields the complete node feature. When data changes, the minimum and maximum values of numeric columns need to be re-extracted, and the word2vec model for non-numeric columns to be retrained.

\textbf{Representation Learning Module.}
\label{section_representation_learning_module}
The representation learning module generates plan-level embeddings from the encoded query plan tree using our proposed tree model, which consists of Bi-GNN layers and a GRU aggregator (Section~\ref{section_tree_model}). Each Bi-GNN layer splits the input plan into two single-directional tree graphs (with opposite edge directions) that are processed by independent TransformerConv~\cite{transformerconv} layers. The resulting node features are then weighted and recombined into an output tree. After the Bi-GNN layers, the tree is traversed in postorder and processed by a GRU aggregator~\cite{gruaggr}, which aggregates the learned plan tree into a representation.

\textbf{Estimation Module.}
\label{section_estimation_module}
The estimation module (Section~\ref{learning-to-Rank_uncertainty_quantification}) takes the input plan representation and transforms it through an MLP to produce a shared feature vector, which is then fed into two parallel MLP branches. The first branch uses a Sigmoid activation function to produce a normalized expected runtime, while the second employs a SoftPlus for non-negative variance that represents uncertainty. These outputs are integrated by a lightweight MLP to yield the integrated cost that captures the trade-off between estimated cost and uncertainty for plan comparison and selection.

\textbf{Explanation Module.}
\label{section_explanation_module}
The explanation module (Section~\ref{section_explainability_technique}) extracts subplans from the encoded query plan tree. These subplans are processed by the representation module to obtain subplan-level embeddings. Each subplan embedding is then concatenated with the entire plan embedding and fed into the explainer, which consists of an MLP and a Sigmoid activation function. The explainer predicts the contribution ratio of each subplan toward the predicted execution time of the entire plan, which can be used to generate plan generation hints based on subplan patterns. 

\textbf{Model Training and Testing.}
\label{section_model_training}
Reqo is trained on query plan pairs, using the actual execution times of plans and subplans as labels. The explanation module is optional. When it is disabled, Reqo is trained using only \(\mathcal{L}_{\mathrm{RobustRank}}\) (Eq.~\ref{robust_cost_model_loss}). When enabled, Reqo is trained by minimizing the sum of \(\mathcal{L}_{\mathrm{RobustRank}}\) and \(\mathcal{L}_{\mathrm{Explanation}}\) (Eq.~\ref{explanationloss}), as shown in Eq.~\ref{overallloss}:
\begin{equation}
\begin{split}
\label{overallloss}
    \mathcal{L}_{\text{Reqo}} = \mathcal{L}_{\text{RobustRank}}(Eq.~\ref{robust_cost_model_loss})
    +\ \mathcal{L}_{\text{Explanation}}(Eq.~\ref{explanationloss})
\end{split}
\end{equation}

\noindent During testing, Reqo does not require inputs to be paired. Candidate plans are directly ranked based on the integrated cost produced by the estimation module, thereby reducing inference time. 

\section{Experimental Study}
\label{section_experimental_study}
We experimentally evaluate Reqo against state-of-the-art cost models across diverse workloads and evaluation metrics.
Experimental findings show that Reqo consistently outperforms existing approaches, demonstrating its effectiveness in real-world database environments.

\subsection{Experimental Setup}
\label{subsection_experimental_setup}
All experiments run on a Linux server (16-core Intel Silver 4216 CPU, 64GB RAM, NVIDIA V100 GPU). PostgreSQL 15.1 (configuration tuned with PGTune~\cite{pgtune}) is used to compile and execute queries for workload generation, and pg\_hint\_plan 1.5.2 to apply plan generation hints. The prototype is implemented in Python 3.12 using PyTorch~\cite{pytorch}, with hyperparameters tuned via Ray Tune~\cite{raytune}. Adam~\cite{adam} is used as the optimizer during training, with dropout and early stopping applied to prevent overfitting. All experimental results are averaged over 10-fold cross-validation.

\label{dataset}
\textbf{Benchmarks.} We evaluate cost models on six workloads:

The \textbf{\textit{TPC-H benchmark}}~\cite{tpch} evaluates database performance on complex business queries. We use TPC-H 3.0.1 to generate a 10\,GB database with 8 tables and 61 columns. From its 22 query templates, we produce 1.1k queries by instantiating each template with varying predicate constants.

The \textbf{\textit{IMDB dataset}} used by the \textbf{\textit{JOB-light\&full workloads}}~\cite{imdb} contains 21 tables. JOB-light has 70 real-world query templates, and JOB-full contains 113 more complex queries. Using the method described above, we generate 2.6k JOB-light queries from its templates and 1.1k JOB-full queries by treating the 113 queries as templates.

The \textbf{\textit{STATS dataset and STATS-CEB workload}}~\cite{stats} include 8 tables from the Stats Stack Exchange network with more complex data distributions than IMDB. STATS-CEB provides 146 queries with varying join sizes and types. We generate a workload of 3k queries by treating the 146 queries as templates and instantiating them with randomly sampled predicate constants.

The \textbf{\textit{TPC-DS benchmark}}~\cite{tpcds} is an industrial-standard benchmark for evaluating database performance. We use TPC-DS 3.2 to generate a 10\,GB database with 24 tables and 425 columns. However, the built-in templates in such benchmarks provide limited structural diversity. Even when predicate constants vary, template queries retain the same structure and often produce similar execution plans, thereby reducing uncertainty. In contrast, randomly generated queries produce unseen structures and greater variability, providing a more challenging evaluation of robustness. Therefore, we use a random query generator to produce 8k queries. The queries are generated by parameters such as join number, join types (e.g., inner, outer, anti) and the number of predicates. All joins are constructed on common attributes of relations in the database schema, and predicate values are drawn from the database to ensure that every join and predicate yields a non-empty result. Each query is randomly generated within the predefined parameter ranges, including up to 10 joins, up to 3 join predicates per pair of joined tables, and 5 local predicates per table.

The \textbf{\textit{DSB benchmark}}~\cite{dsb} enhances TPC-DS with complex data distribution and challenging query templates. We generate 1.2k queries from its 15 single-block and 22 multi-block templates.

\textbf{Dataset Generation and Preprocessing.} We build our experimental datasets from these workloads. Each query is compiled in PostgreSQL using 13 different GDHints inspired by Bao~\cite{bao}, which impose varying constraints on join and access operators. Each sample in the dataset comprises candidate query plans generated for the same query using these hints. Here, all GDHints are applied by adjusting PostgreSQL’s planner method configuration, and all SPPHints derived from explanations are applied using pg\_hint\_plan. We use PostgreSQL to execute the generated query plans and take their execution times as labels for cost estimation and plan selection. We also collect execution times of each subplan from the execution engine as explanation labels. To reduce the skewness of the execution time values and ensure alignment with the output range of the Sigmoid activation function, we apply a natural-log transformation followed by min-max scaling, mapping each actual execution time \( y \) to \([0, 1]\) using the training data’s minimum and maximum. Unless otherwise indicated, in each workload, all LCMs are trained and evaluated on the same dataset using a 10-fold cross-validation with a 9:1 split between training and test sets.

\subsection{Experimental Methodology}
\label{section_experimental_methodology}
\textbf{Comparison.} We compare Reqo against PostgreSQL and four recent works: Bao~\cite{bao}, Zero-Shot~\cite{zeroshot}, Lero~\cite{lero}, and Roq~\cite{roq}. Bao is chosen for its efficiency and advanced performance, Zero-Shot for its database-agnostic design, Lero for its learning-to-rank mechanism, and Roq for its approach to quantifying uncertainty for robust plan selection. These comparisons let us assess improvements across different aspects against state-of-the-art mechanisms.

\textbf{\textit{PostgreSQL}} serves as the baseline, representing the performance of commercial query optimizers. We use PostgreSQL's estimated cost to evaluate its performance for plan selection and explainability.

\textbf{\textit{Bao}} is a learned query optimizer that enhances classical optimizers by applying hints and reinforcement learning. We compare with its cost model, which predicts execution time by processing the vectorized plan tree through TCNN~\cite{treecnn} and MLP.

\textbf{\textit{Zero-Shot}} is an LCM that adopts a pretraining-based paradigm and can be trained on multiple workloads. Unlike the other baselines, Zero-Shot is trained on the combined training sets of all six workloads and evaluated separately on the test set of each workload.

\textbf{\textit{Lero}} is a learning-to-rank query optimizer. Similar to Bao in plan encoding, Lero trains its cost model on pairwise plan comparisons as a binary classification task rather than predicting numerical values. Thus, it is excluded from our cost estimation accuracy comparison.

\textbf{\textit{Roq}} is a robust risk-aware query optimization framework with a GNN-based query-level encoder and a plan-level encoder similar to Bao. Its cost model estimates execution time and uncertainty, and applies them via fixed plan selection strategies to improve robustness.

\textbf{\textit{Reqo}} is our proposed model and is divided into modules for an ablation study. The base model (base: bi-GNN\&GRU, single-branch MLP with MSE loss, no explanation) serves as the baseline. We isolate each component by substituting Bi-GNN with single-directional GNNs (sd-GNN\&GRU) or undirected GNNs (und-GNN\&GRU), and by replacing the GRU with global add-pooling (sd-GNN\&Addpool), to demonstrate the effectiveness of the bidirectional architecture and GRU aggregation. We extend the base with uncertainty quantification (base\&unc.) by applying the dual-branch MLP and training it solely with \(\mathcal{L}_{\text{Uncertainty}}\) (Eq.~\ref{uncertaintyloss}), without using the quantified uncertainty for plan selection to isolate the impact of the loss. Next, we integrate estimated cost and uncertainty using a constant (base\&unc.\&$I_{\mathrm{fixed}}$, where the integrated cost is obtained based on a fixed rule as $C=\mu + I_{\mathrm{fixed}} \times \sigma^2$) 
to evaluate whether the ranking-based approach  (base\&unc.\&$I_{\mathrm{learned}}$\&rank, i.e.\ Reqo w/o expl.) enhances the robustness, where $I_{\mathrm{learned}}$ is the learning-based integration referred to Eq.~\ref{eq_integration_strategy}. Additionally, we include a variant with the explanation module (Reqo w/ expl.) to assess its impact. Throughout, "Reqo" refers to the model that contains all modules.

\textbf{Evaluation Metrics.} We employ eight evaluation metrics:

\textit{\textbf{1. Q-Error:}} We evaluate cost estimation accuracy using Q-Error~\cite{moerkotte_preventing_2009}, defined as \(Q\text{-}Error = \max(y_{\text{et}}, y_{\text{at}}) / \min(y_{\text{et}}, y_{\text{at}})\), where \(y_{\text{et}}\) and \(y_{\text{at}}\) are estimated and actual execution times on the original scale, obtained by inverting the min--max scaling and log transform used during training.

\textit{\textbf{2. Spearman’s Correlation:}} We use Spearman’s rank correlation coefficient to measure the relationship between estimated and actual execution time, with values closer to 1 indicating a stronger correlation. Unlike Pearson’s correlation coefficient, it is less sensitive to outliers and scale differences, making it suitable for measurements that vary greatly in magnitude.

\textit{\textbf{3. Total Runtime Ratio:}} This ratio is computed by dividing the sum of actual execution times for cost model-selected plans by the sum of actual execution times for optimal plans across all queries, offering an evaluation of overall plan selection performance. In each query’s generated candidate plan set, the cost model-selected plan is the candidate with the minimum estimated execution time predicted by the model, and the optimal plan is the candidate with the minimum actual execution time in the same candidate set.

\textit{\textbf{4. Plan Suboptimality:}} For a set of candidate execution plans \( \mathcal{P} \) for the same query, we rank plans by their actual execution times \( T\) and identify the optimal plan \( p_o \) with the shortest execution time. The suboptimality of a plan \( p_i \in \mathcal{P} \) is obtained by \(T_{p_i}/T_{p_o}\).
This metric ranges from \( [1, \infty) \) and reflects the model's ability to select optimal plans. Analyzing the distribution, especially the worst cases, helps assess the model's robustness in plan selection~\cite{dutt_plan_2014, roq}.

\textit{\textbf{5. Overhead:}} For each LCM, we measure training data generation time to convert all executed query plans in a workload into a learnable format, model training time (with the same batch size), model size, and average inference latency to select the optimal plan from encoded candidate plans for a query during testing. We also report the workload end-to-end execution time, defined as the sum (over all queries in the workload) of data generation time, optimization inference overhead, and the actual execution time of the candidate plan selected by the model during testing.

\textit{\textbf{6. Explanation Top-\(K\) Node Accuracy:}} This metric assesses the model's accuracy in identifying the plan nodes contributing most to the cost prediction. The metric ranks nodes by estimated local contributions and checks whether the cost model selected top-\(K\) most influential nodes match the actual top-\(K\) most influential nodes in correct order, returning 1 if they coincide and 0 otherwise.

\textit{\textbf{7. Explanation Top-\(K\) Node Contribution Ratio:}} This metric evaluates the cost model’s ability to identify the most influential nodes by comparing the sum of actual contributions of the top-\(K\) cost model-selected nodes \(\{n_{\mathrm{pred}\_1},\dots,n_{\mathrm{pred}\_K}\}\) to that of the actual top-\(K\) nodes \(\{n_{\mathrm{actual}\_1},\dots,n_{\mathrm{actual}\_K}\}\). Formally:
{
\setlength{\abovedisplayskip}{2pt}
\setlength{\belowdisplayskip}{3pt}
\setlength{\abovedisplayshortskip}{2pt}
\setlength{\belowdisplayshortskip}{3pt}
\begin{equation}
\mathrm{Expl.\ Top\text{-}K\ Node\ Ctrb.\ Ratio}
=
\frac{\sum_{k=1}^K AC\bigl(n_{\mathrm{pred}\_k}\bigr)}
     {\sum_{k=1}^K AC\bigl(n_{\mathrm{actual}\_k}\bigr)}.
\end{equation}
Unlike metric 6, this ratio captures cases when the model-selected nodes contribute significantly, even if they are not among the top-\(K\), providing a more comprehensive evaluation of explainability.}

\textit{\textbf{8. Subplan Contribution Ratio (SCR)-RMSE:}} This metric evaluates how well the model estimates each subplan’s relative contribution to its corresponding entire plan's cost across the workload. For a plan \(p_i \in P\) with subplans \(\{sp_{ik}\}_{k=1}^{K_i}\) and \(|P|=N\), let \(ET\) and \(T\) denote the estimated and actual execution times, respectively, and define the root mean squared error (RMSE) as:
{
\setlength{\abovedisplayskip}{2pt}
\setlength{\belowdisplayskip}{2pt}
\setlength{\abovedisplayshortskip}{2pt}
\setlength{\belowdisplayshortskip}{2pt}
\begin{equation}
\mathrm{SCR\text{-}RMSE}
=
\sqrt{
\frac{1}{N}\sum_{i=1}^{N}\left(
\frac{1}{K_i}\sum_{k=1}^{K_i}
\left(\frac{ET_{sp_{ik}}}{ET_{p_i}} - \frac{T_{sp_{ik}}}{T_{p_i}}\right)^{2}
\right)
}
\end{equation}
}

\subsection{Experimental Results}
\label{section_experimental_results}

\begin{figure}[t]
\centering

\begin{subfigure}[b]{0.333\columnwidth}
  \includegraphics[width=\linewidth]{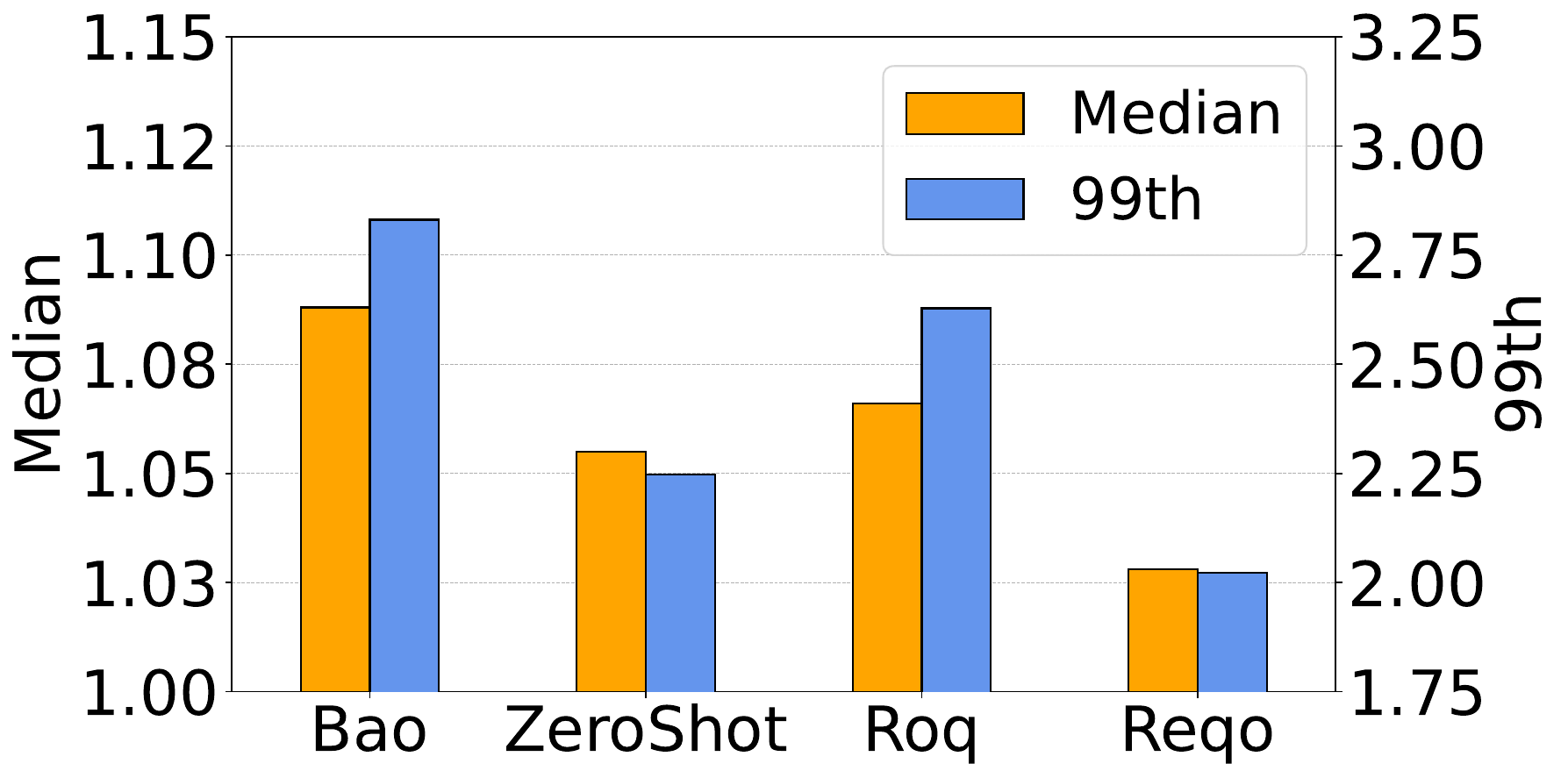}
    \centering
  \caption{TPC-H}
  \label{fig:cost_tpc-h}
\end{subfigure}\hfill
\begin{subfigure}[b]{0.333\columnwidth}
  \includegraphics[width=\linewidth]{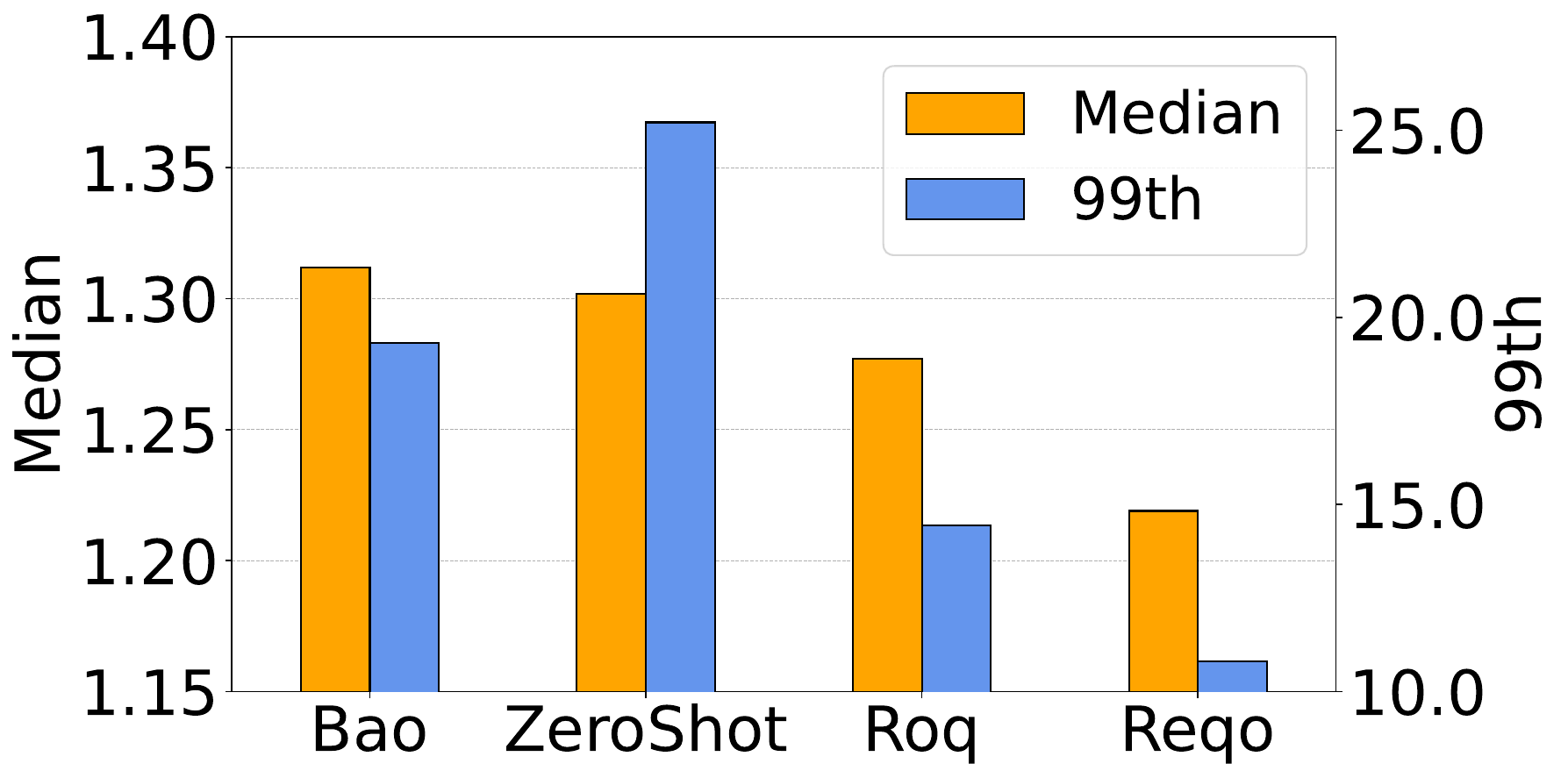}
    \centering
  \caption{JOB-light}
  \label{fig:cost_job-light}
\end{subfigure}\hfill
\begin{subfigure}[b]{0.333\columnwidth}
  \includegraphics[width=\linewidth]{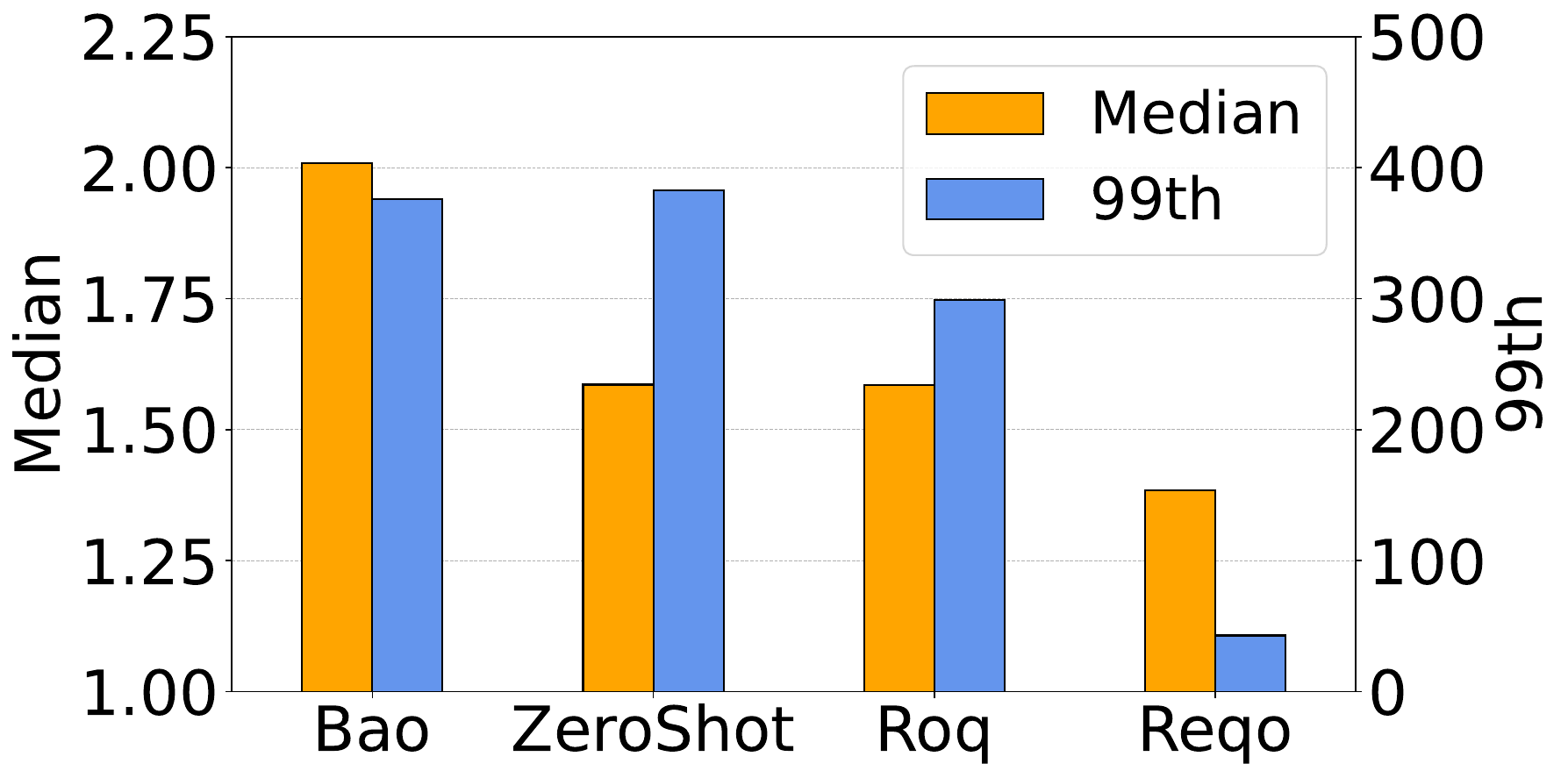}
    \centering
  \caption{JOB-full}
  \label{fig:cost_job-full}
\end{subfigure}

\begin{subfigure}[b]{0.333\columnwidth}
  \includegraphics[width=\linewidth]{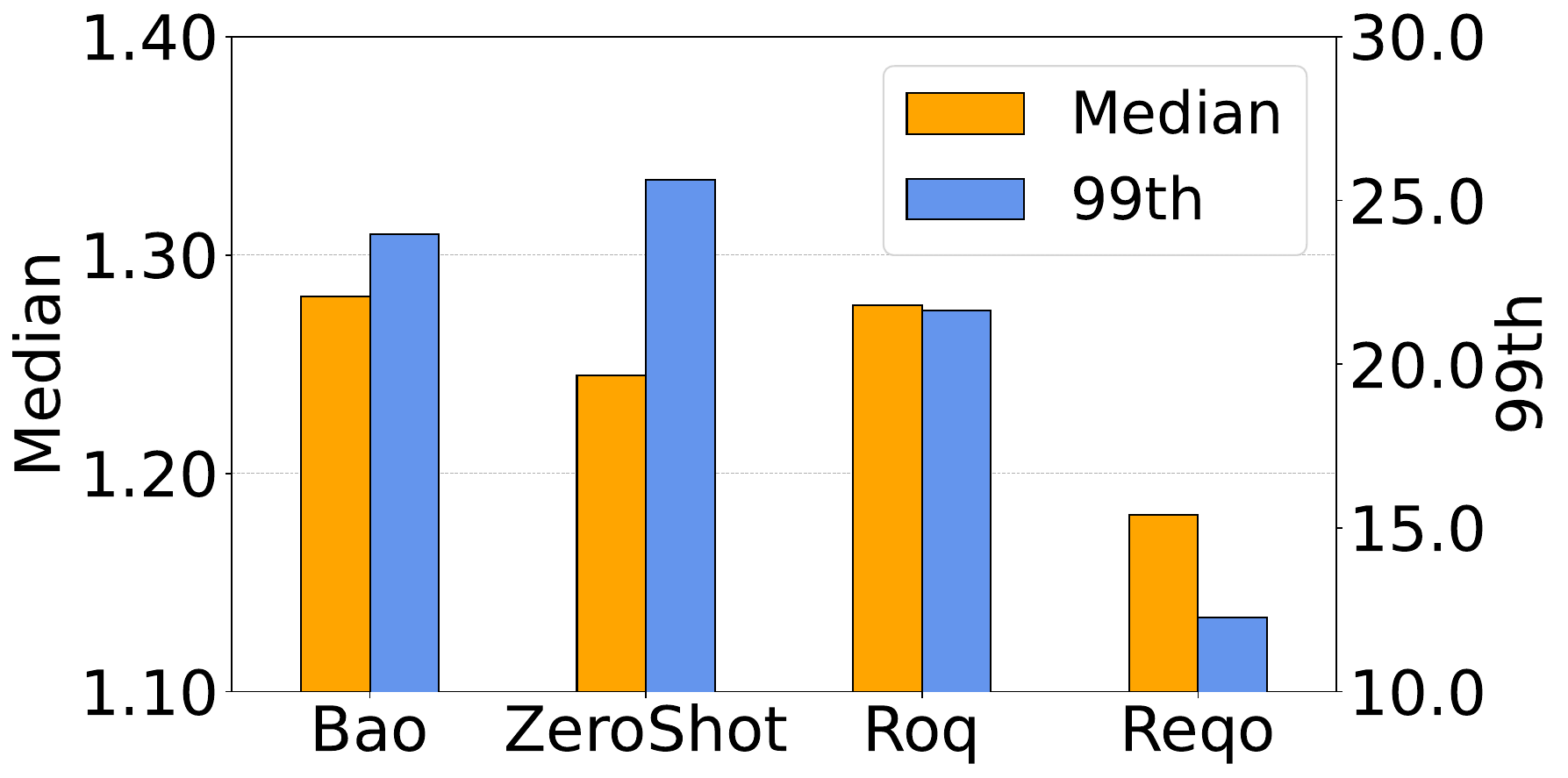}
    \centering
  \caption{STATS}
  \label{fig:cost_stats}
\end{subfigure}\hfill
\begin{subfigure}[b]{0.333\columnwidth}
  \includegraphics[width=\linewidth]{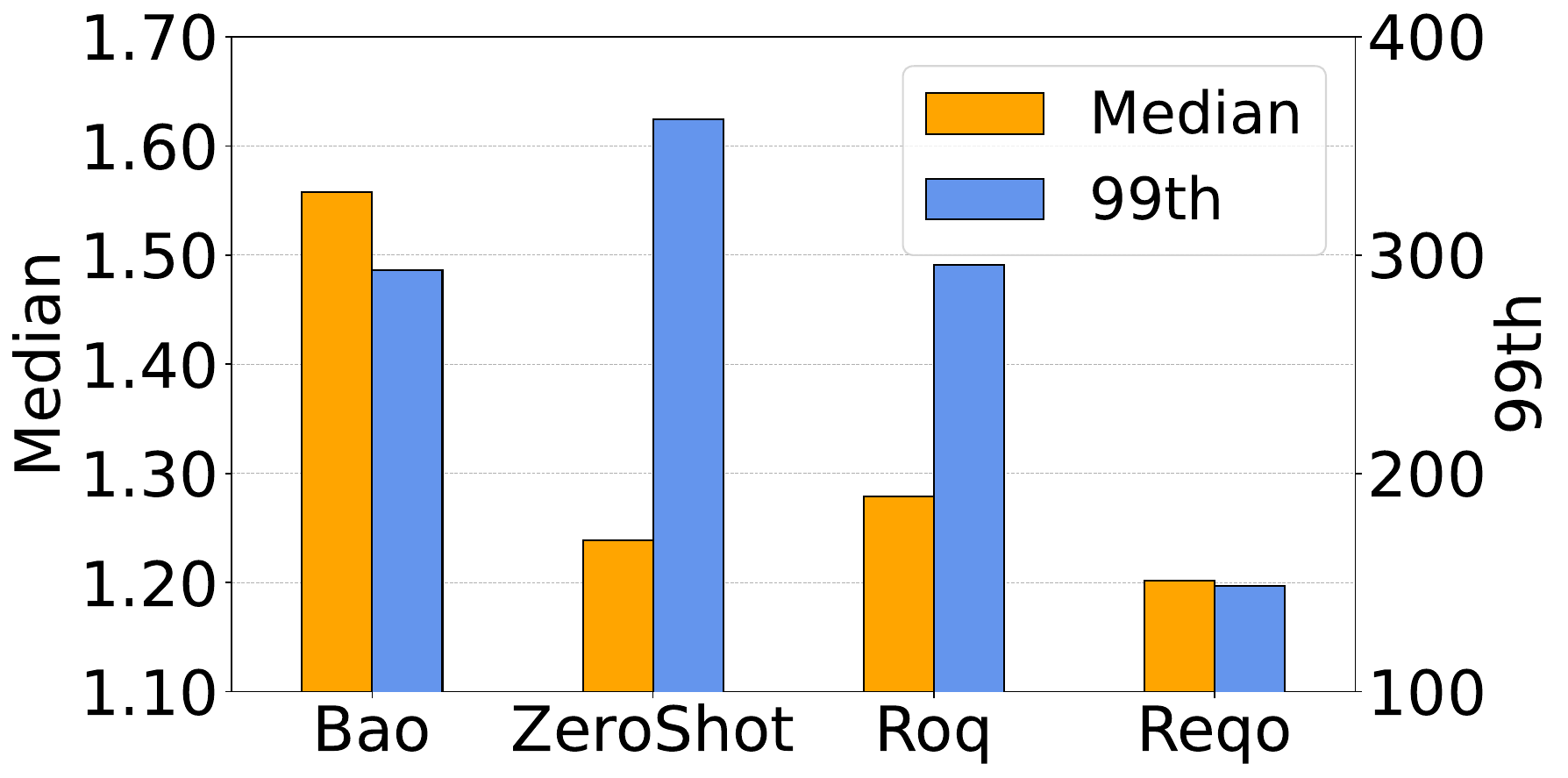}
    \centering
  \caption{DSB}
  \label{fig:cost_dsb}
\end{subfigure}\hfill
\begin{subfigure}[b]{0.333\columnwidth}
  \includegraphics[width=\linewidth]{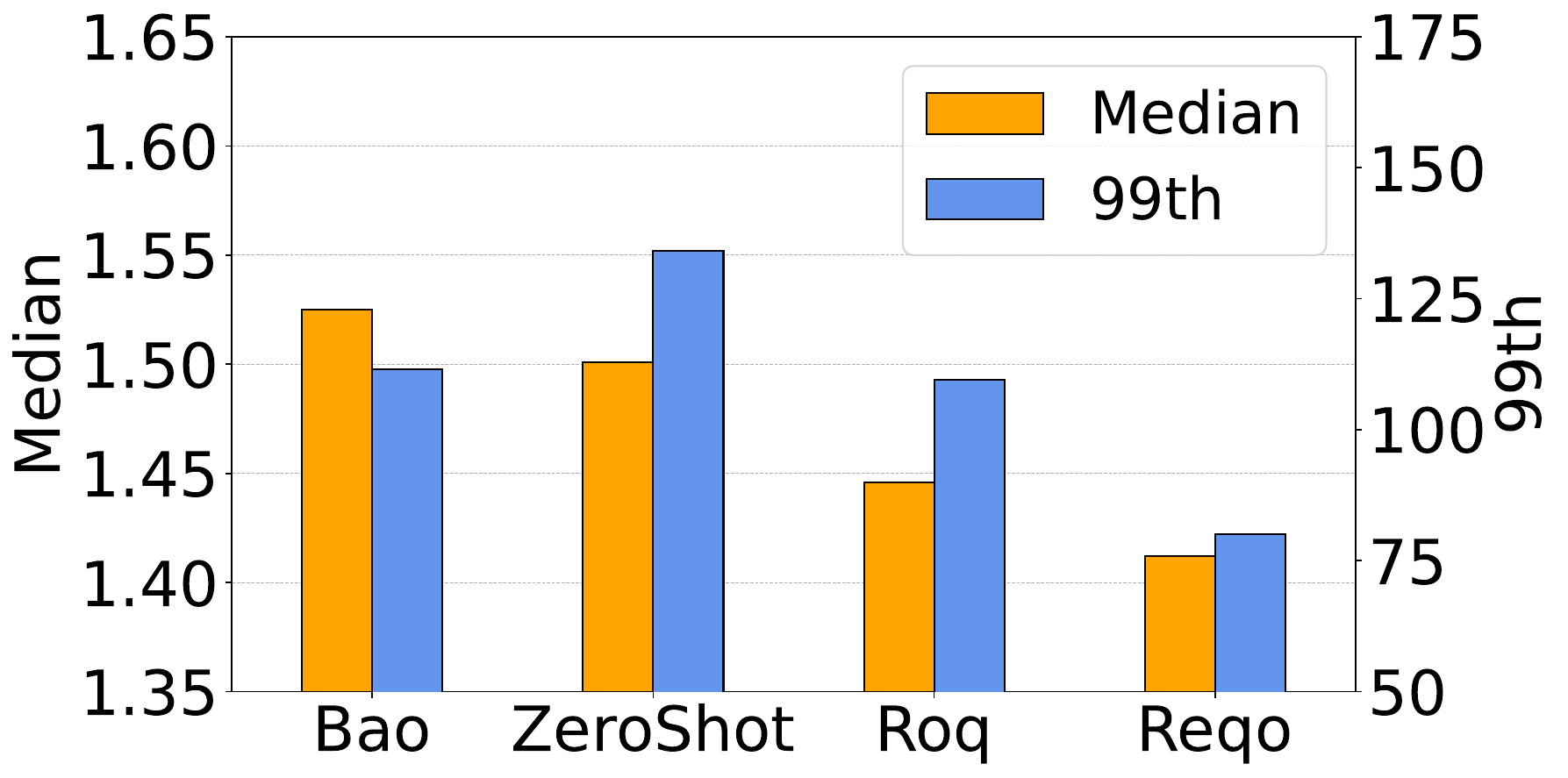}
  \centering
  \caption{TPC-DS}
  \label{fig:cost_tpc-ds}
\end{subfigure}

\caption{Cost estimation performance (Q-Error)}
\label{fig:cost-estimation-performance}
\end{figure}

\begin{figure}[t]
    \centering
    \includegraphics[width=0.7\linewidth]{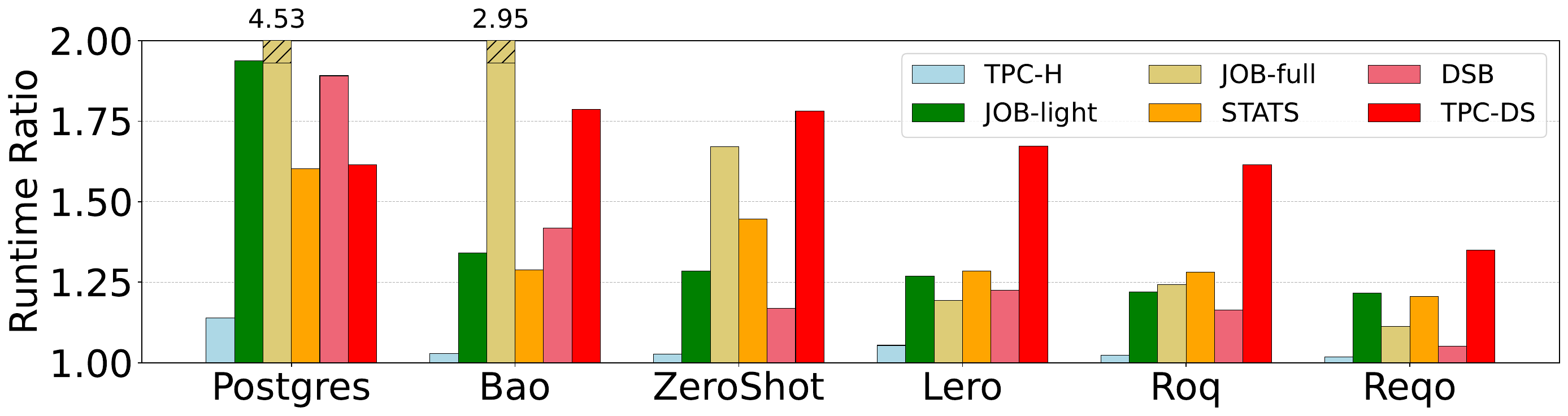}
    \caption{Total runtime ratio performance (Model-selected to optimal)}
    \label{fig:runtime_performance_all}
\end{figure}

\subsubsection{Comparison for Cost Estimation}
Figure~\ref{fig:cost-estimation-performance} demonstrates that Reqo consistently outperforms Bao, Zero-Shot and Roq across all datasets in terms of Q-Error. Zero-Shot achieves a relatively low median Q-Error compared with the other baselines. Reqo achieves lower Q-Error values across various percentiles, with particularly strong gains in the tail, indicating its superior cost estimation performance and more reliable worst-case predictions. Notably, Reqo maintains its advantage on complex workloads such as JOB-full, TPC-DS and DSB, showcasing its effectiveness in handling challenging scenarios. These results confirm Reqo's advancement over existing models and its effectiveness in both simple and complex query optimization tasks.

\subsubsection{Comparison for Plan Selection}
The runtime results in Figure~\ref{fig:runtime_performance_all} show the cost models' plan selection performance across workloads. Reqo consistently surpasses others, demonstrating substantial performance enhancements. In our more complex TPC-DS workload, the other LCM baselines do not perform as well as PostgreSQL, despite exhibiting better performance in simpler workloads. In contrast, Reqo's tree model provides strong representation capability and, combined with the learning-to-rank mechanism with uncertainty quantification, delivers superior performance. In terms of total runtime ratio, Reqo achieves performance enhancements of 16.6\% over PostgreSQL, 24.6\% over Bao, 25.6\% over Zero-Shot, 20.4\% over Lero, and 18.6\% over Roq on TPC-DS. These results underscore Reqo's effectiveness in handling complex query scenarios, highlighting its superiority in actual runtime improvements rather than gains in average estimation error alone.

\begin{figure}[t]
\centering
\begin{subfigure}[b]{\columnwidth}
  \centering
  \includegraphics[width=0.65\linewidth]{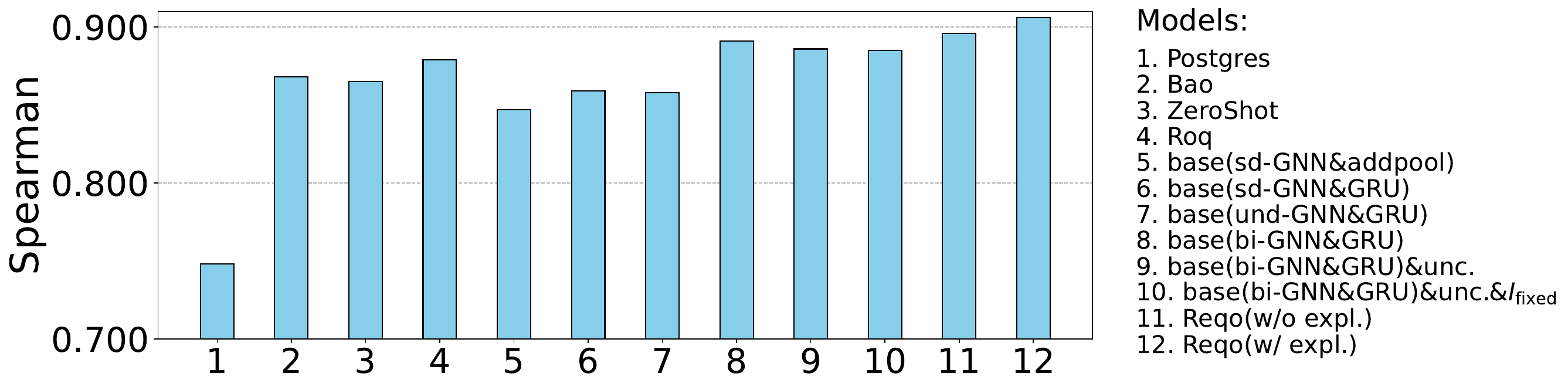}
  \caption{Spearman's Correlation}
  \label{fig:spearman-correlation-tpc-ds}
\end{subfigure}

\begin{subfigure}[b]{\columnwidth}
  \centering
  \includegraphics[width=0.65\linewidth]{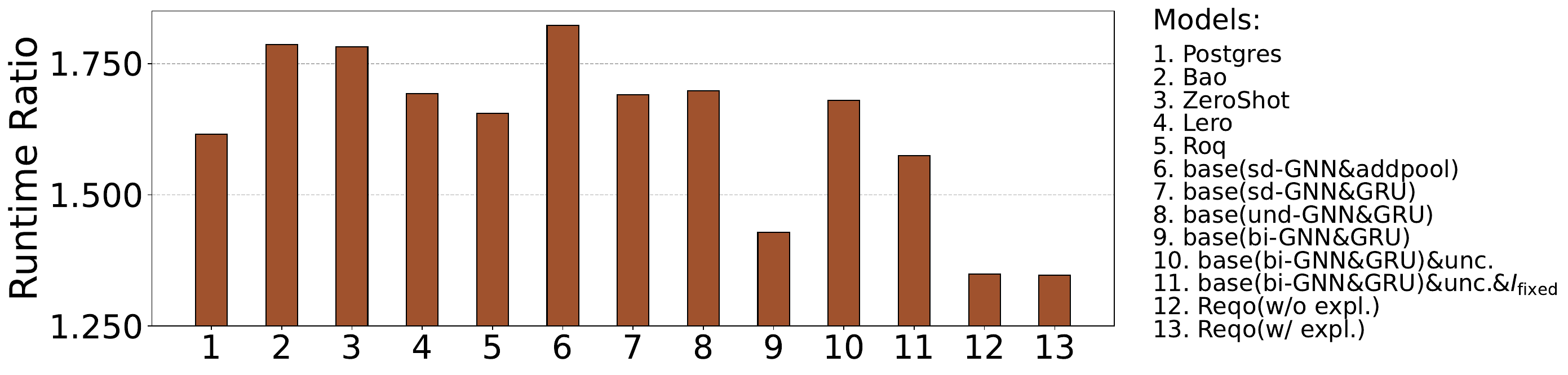}
  \caption{Total Runtime Ratio (Model-selected to optimal)}
  \label{fig:total-runtime-ratio-tpc-ds}
\end{subfigure}

\caption{Ablation study results on TPC-DS in Spearman’s correlation and total runtime ratio}
\label{fig:modules_compare_tpc-ds}
\end{figure}

\subsubsection{Ablation Study}
To assess the impact of Reqo’s proposed techniques on cost estimation and plan selection, we conduct an ablation study on the TPC-DS workload. Figure~\ref{fig:spearman-correlation-tpc-ds} shows variations in Spearman's correlation for different Reqo configurations versus other cost models. All learning-based models significantly outperform PostgreSQL's classical cost model. Base (bi-GNN\&GRU), which leverages our proposed tree model, surpasses Bao, Zero-Shot, and Roq without additional mechanisms, showcasing its powerful representation learning capabilities and more accurate execution runtime estimation. The bidirectional GNN outperforms both single-directional and undirected variants (models 5-7), and replacing global addpooling with a GRU (models 4-5) yields additional gains, confirming the effectiveness of the Bi-GNNs with GRU design over conventional GNNs. Adding uncertainty quantification (base\&unc. and base\&unc.\&$I_{\mathrm{fixed}}$) enhances plan selection robustness by quantifying uncertainty during cost estimation, though it slightly reduces estimation accuracy. Incorporating the learning-to-rank mechanism (Reqo (w/o expl.)) addresses this trade-off and further improves cost estimation performance, ensuring that Reqo maintains strong cost estimation accuracy while enhancing robustness.

To evaluate plan selection performance, Figure~\ref{fig:total-runtime-ratio-tpc-ds} presents the total runtime ratio under the TPC-DS workload. Base (bi-GNN\&GRU) already outperforms PostgreSQL and other learning-based models. However, enabling uncertainty quantification without applying it to plan selection (base\&unc.) leads to reduced performance, exhibiting the trade-off between uncertainty and accuracy. Integrating uncertainty with a fixed rule (base\&unc.\&$I_{\mathrm{fixed}}$) does improve performance, but still not to the level achieved by the base model. Our uncertainty-aware learning-to-rank approach (Reqo (w/o expl.)) further enhances runtime performance, surpassing the base model and confirming the effectiveness of our uncertainty-aware learning-to-rank design for plan selection.

As shown in Figure~\ref{fig:modules_compare_tpc-ds},  the ablation study indicates that integrating the explanation module does not negatively influence the LCM, and slightly enhances Reqo’s cost estimation and robustness performance. These results support the feasibility of applying our explainability technique to LCMs. Additionally, the subplan extraction process for explainability optimizes the utilization of information within the executed query plan, which may further contribute to these gains.

\subsubsection{Comparison for Robustness}
\begin{figure}[t]
\centering
\begin{subfigure}[b]{0.333\columnwidth}
  \includegraphics[width=\linewidth]{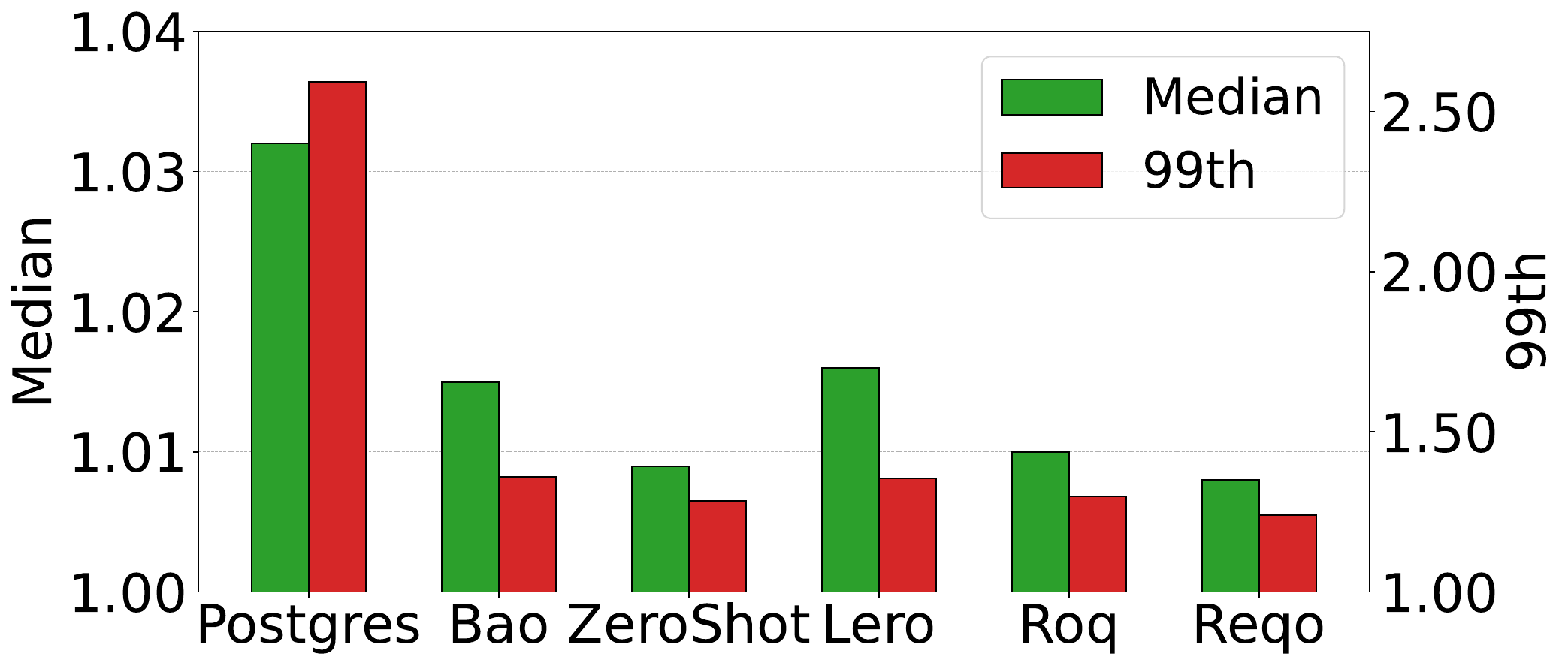}
  \centering
  \caption{TPC-H}
  \label{fig:plan_tpc-h}
\end{subfigure}%
\hfill
\begin{subfigure}[b]{0.333\columnwidth}
  \includegraphics[width=\linewidth]{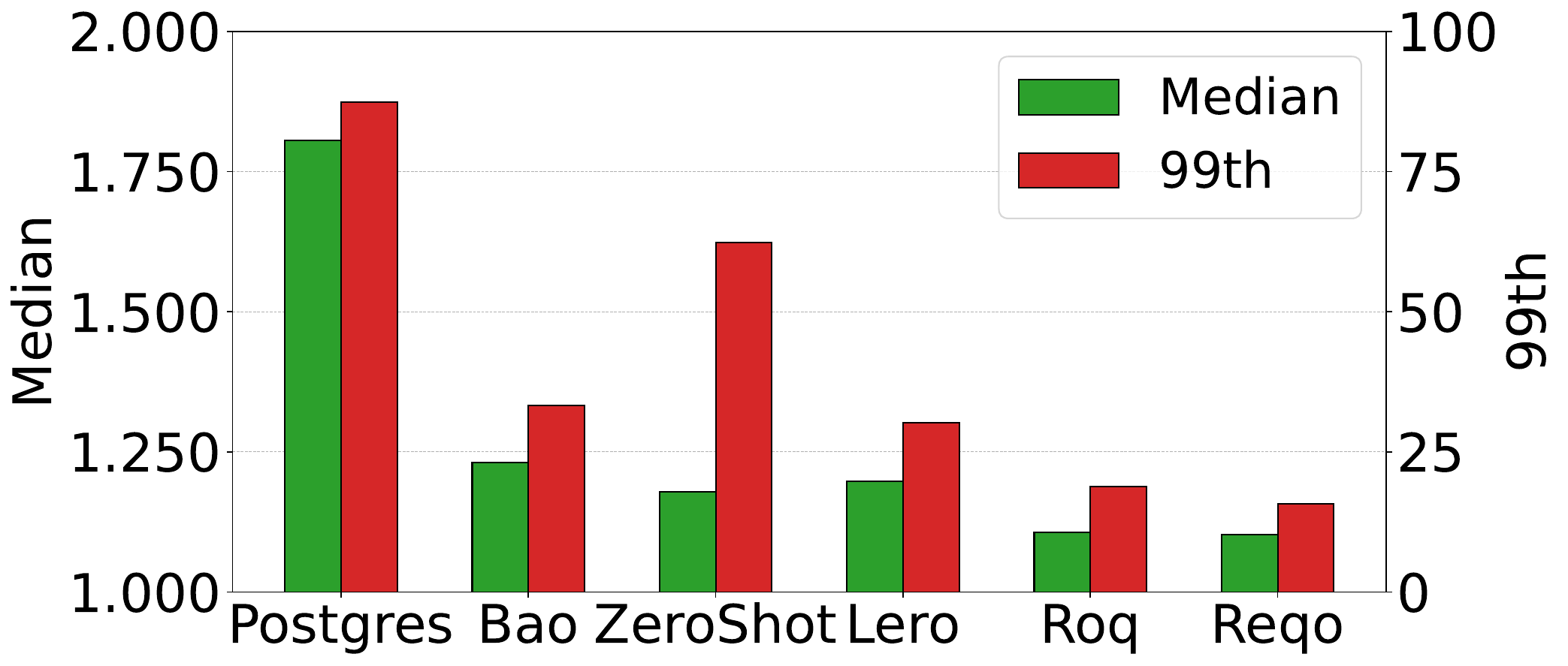}
    \centering
  \caption{JOB-light}
  \label{fig:plan_job-light}
\end{subfigure}%
\hfill
\begin{subfigure}[b]{0.333\columnwidth}
  \includegraphics[width=\linewidth]{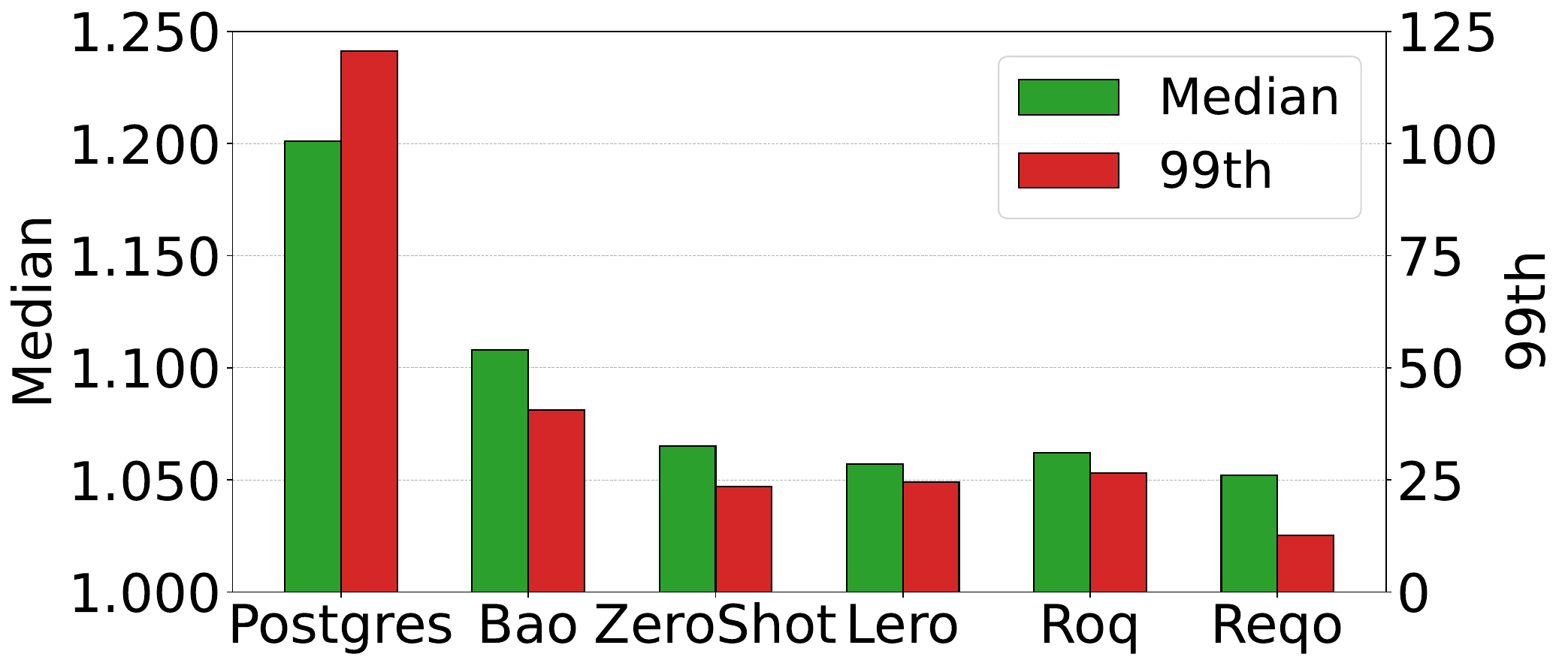}
    \centering
  \caption{JOB-full}
  \label{fig:plan_job-full}
\end{subfigure}

\begin{subfigure}[b]{0.333\columnwidth}
  \includegraphics[width=\linewidth]{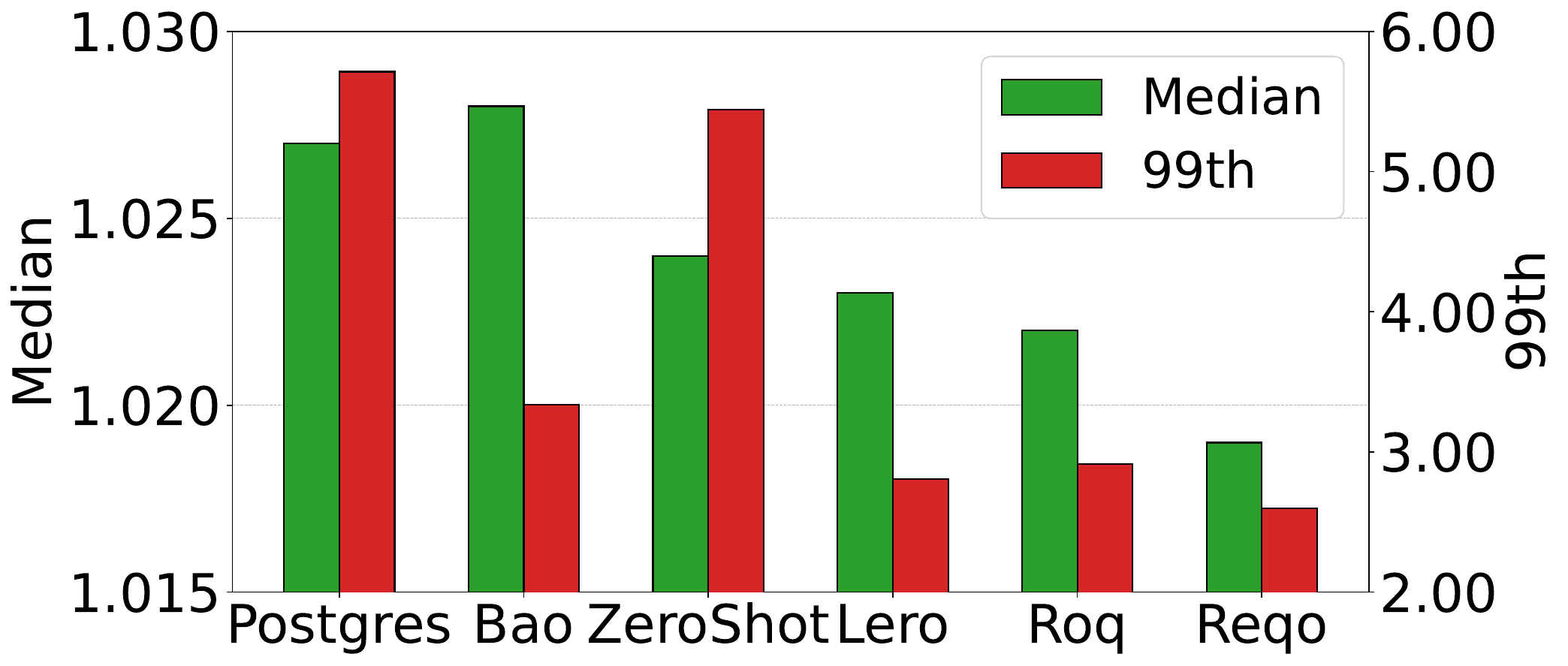}
    \centering
  \caption{STATS}
  \label{fig:plan_stats}
\end{subfigure}%
\hfill
\begin{subfigure}[b]{0.333\columnwidth}
  \includegraphics[width=\linewidth]{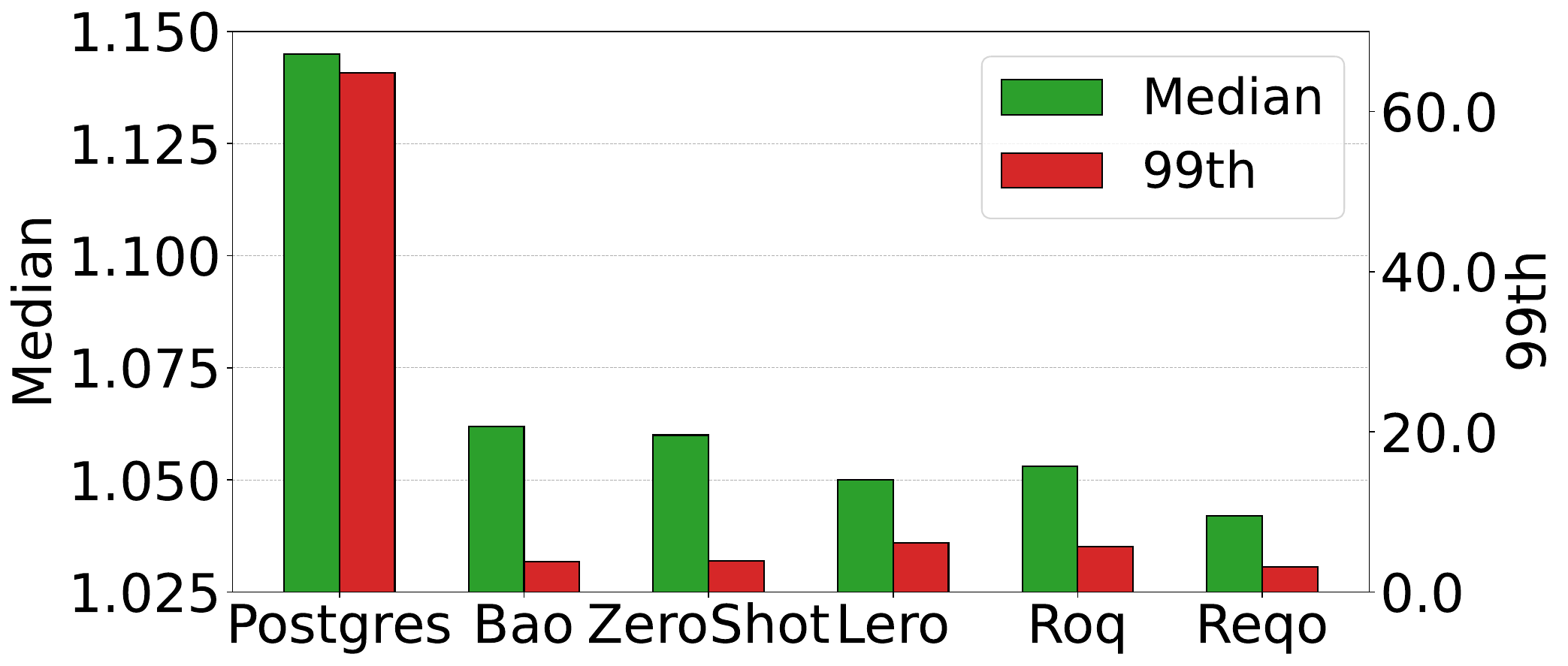}
    \centering
  \caption{DSB}
  \label{fig:plan_dsb}
\end{subfigure}%
\hfill
\begin{subfigure}[b]{0.333\columnwidth}
  \includegraphics[width=\linewidth]{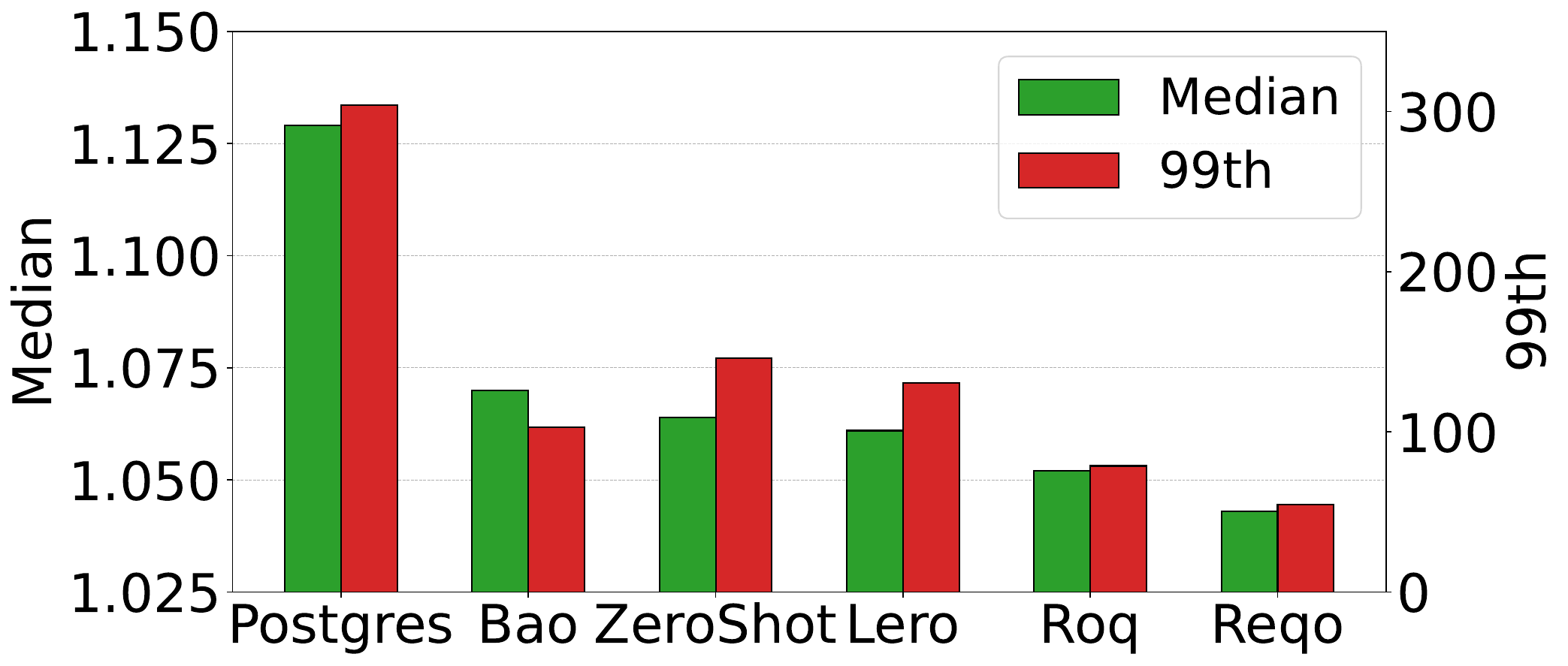}
    \centering
  \caption{TPC-DS}
  \label{fig:plan_tpc-ds}
\end{subfigure}
\caption{Plan suboptimality performance}
\label{fig:plan-suboptimality-performance}
\end{figure}

We evaluate plan selection robustness via plan suboptimality, as shown in Figure~\ref{fig:plan-suboptimality-performance}. For simpler workloads, the gap among models is small, as baselines generally make correct decisions. Nonetheless, Reqo still achieves the most accurate plan selection, particularly excelling at the 99th percentile tail, demonstrating superior robustness under worst-case scenarios. For complex workloads, both Reqo and Roq leverage uncertainty quantification to enhance robustness and outperform other models. Reqo ultimately surpasses Roq through its ranking-based adaptive integration, confirming its superior robustness in plan selection.

\begin{figure}[t]
\centering
\begin{subfigure}[b]{0.249\columnwidth}
  \includegraphics[height=1.95cm,keepaspectratio]{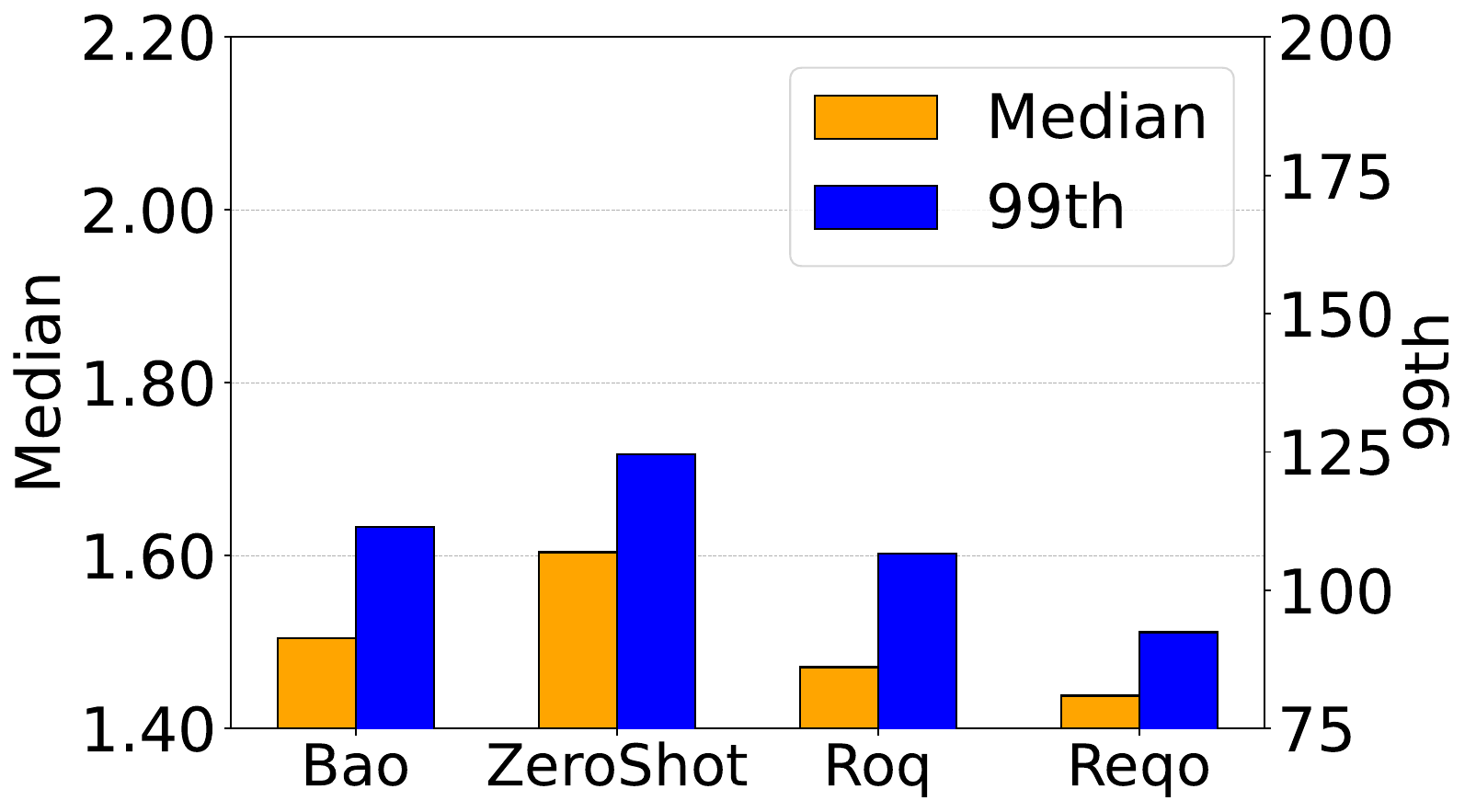}
  \caption{Q-Error (1-5 J)}
  \label{fig:cost_1-5}
\end{subfigure}%
\hfill
\begin{subfigure}[b]{0.249\columnwidth}
  \includegraphics[height=1.95cm,keepaspectratio]{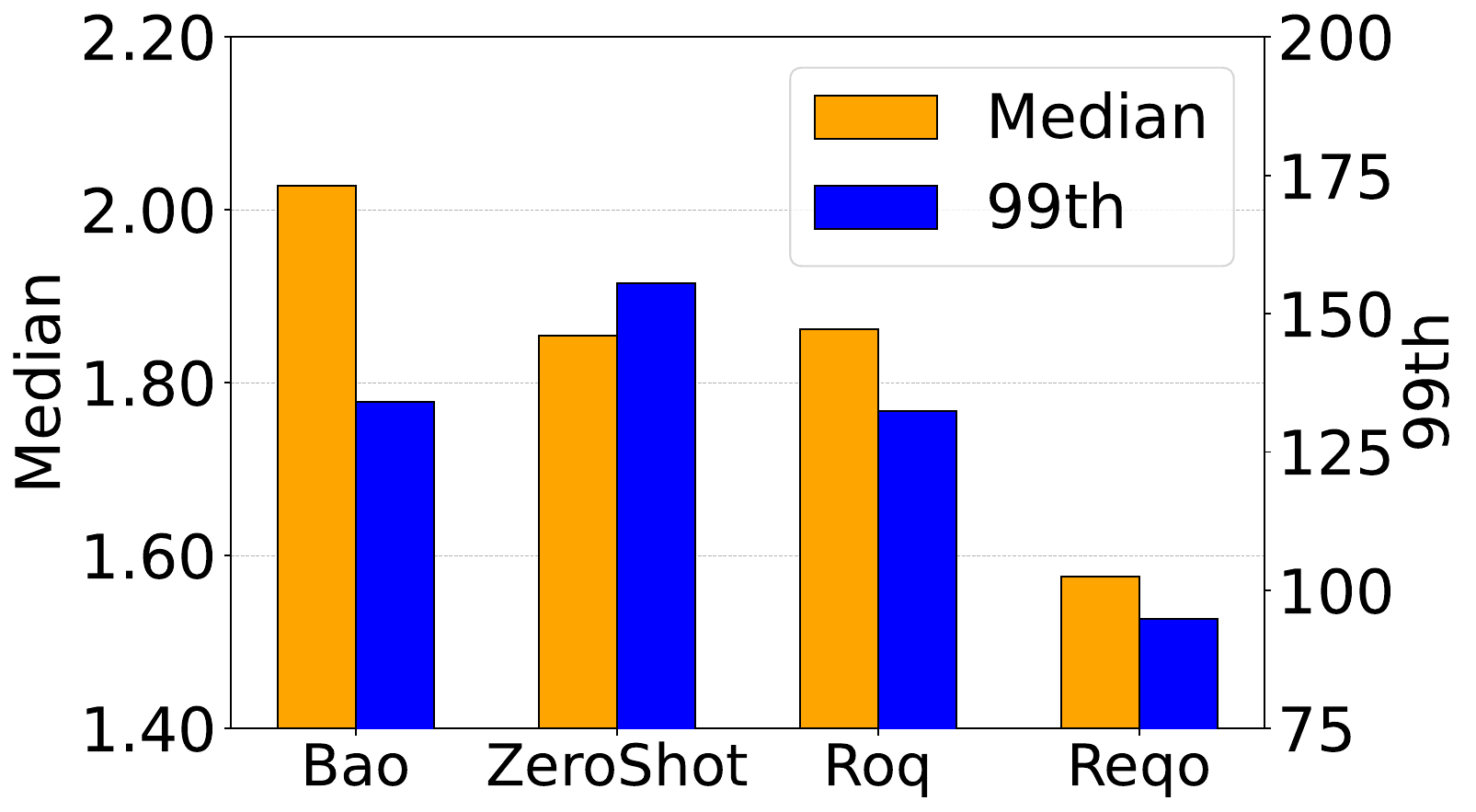}
  \caption{Q-Error (6-10 J)}
  \label{fig:cost_6-10}
\end{subfigure}%
\hfill
\begin{subfigure}[b]{0.249\columnwidth}
  \includegraphics[height=1.95cm,keepaspectratio]{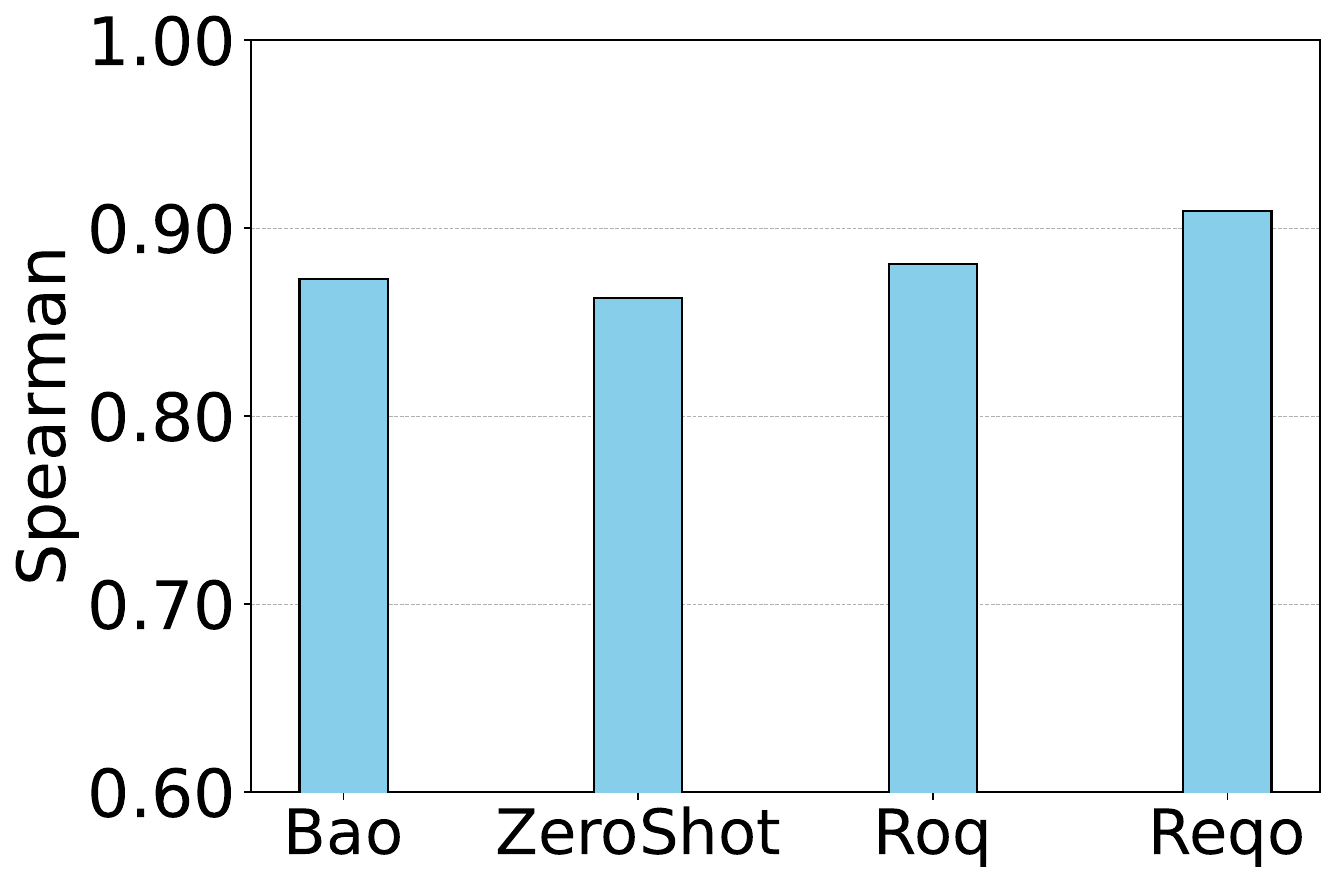}
  \caption{Spearman Corr. (1-5 J)}
  \label{fig:spearman_1-5}
\end{subfigure}%
\hfill
\begin{subfigure}[b]{0.249\columnwidth}
  \includegraphics[height=1.95cm,keepaspectratio]{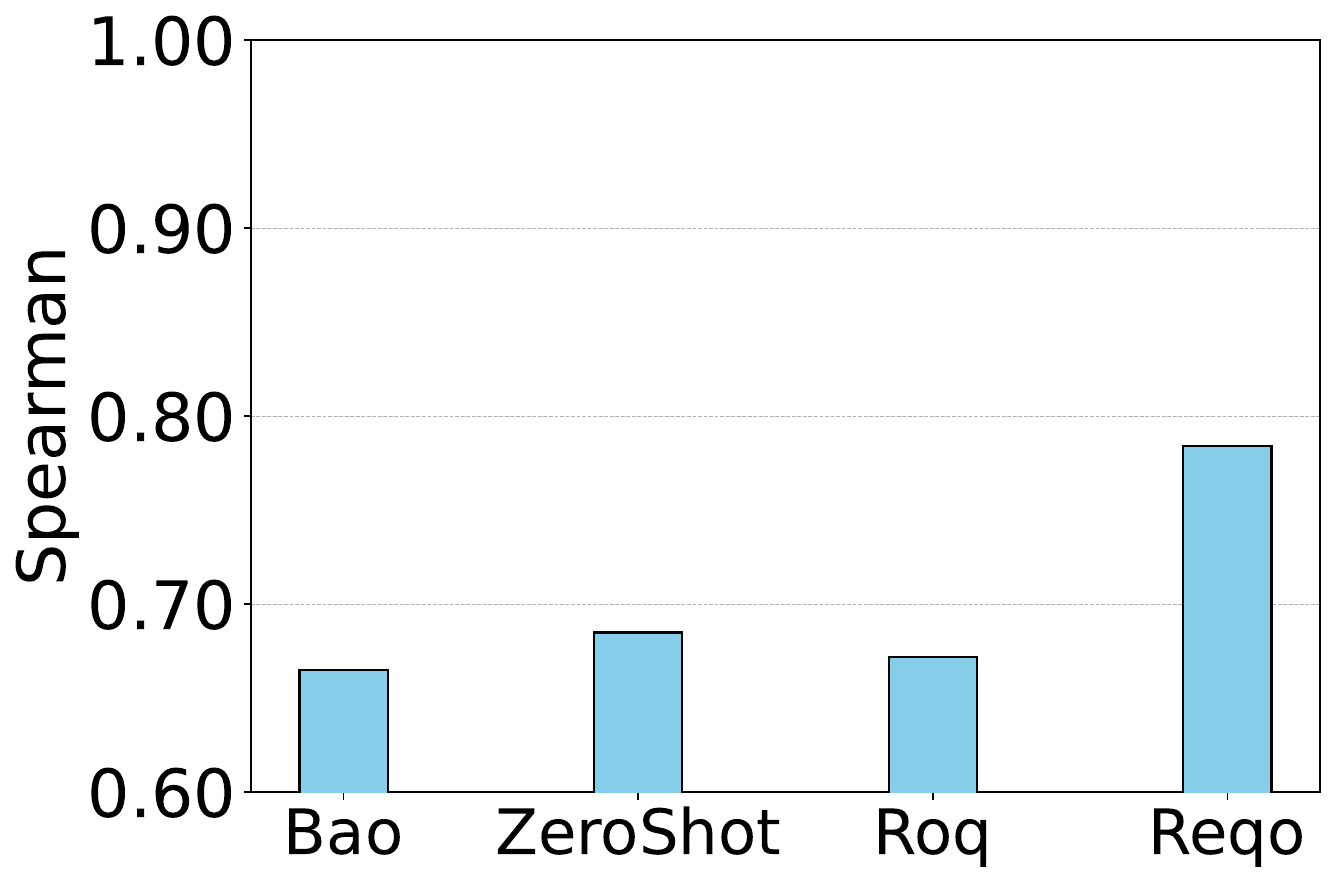}
  \caption{Spearman Corr. (6-10 J)}
  \label{fig:spearman_6-10}
\end{subfigure}

\begin{subfigure}[b]{0.249\columnwidth}
  \includegraphics[height=1.95cm,keepaspectratio]{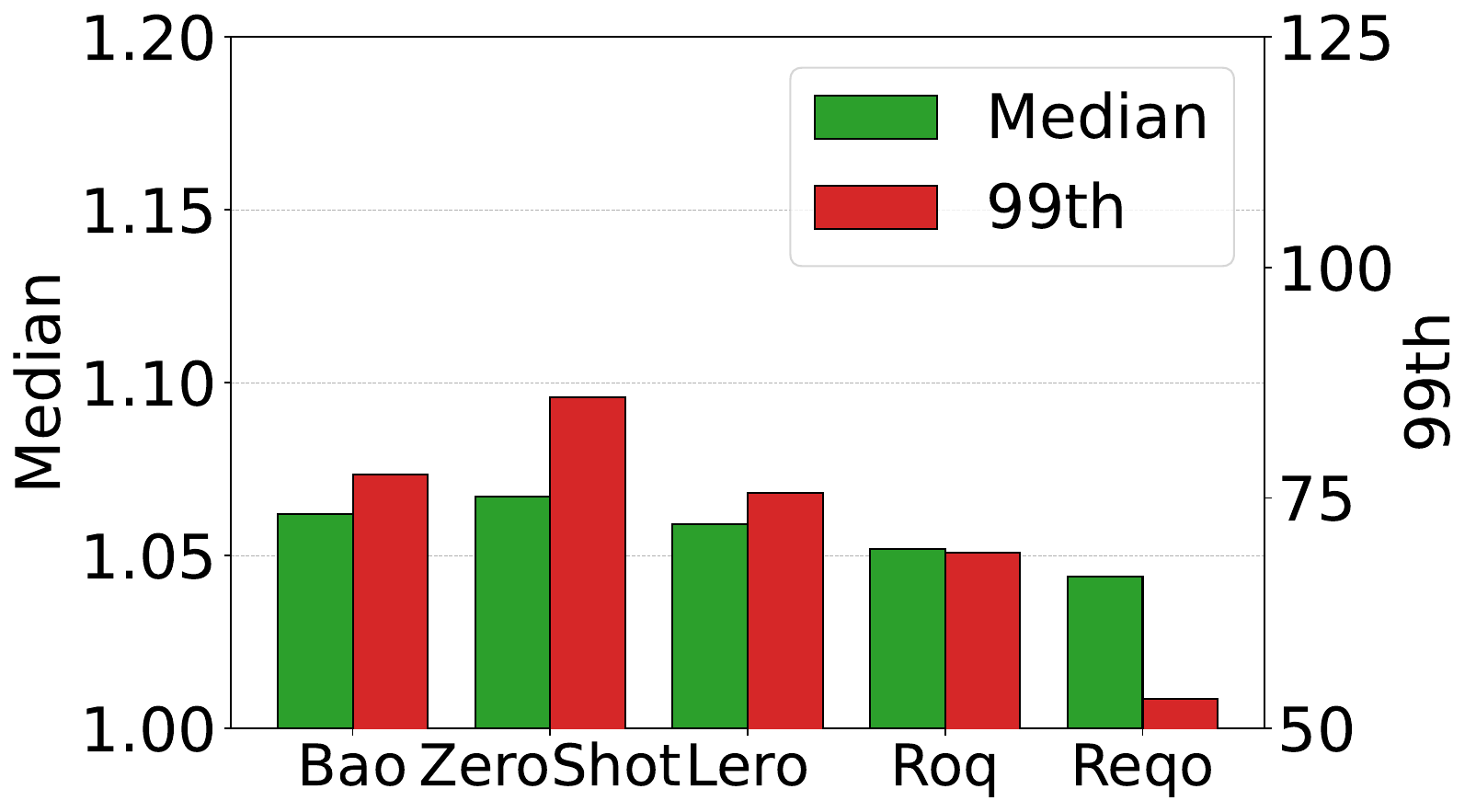}
  \caption{Plan Subopt. (1-5 J)}
  \label{fig:plan_1-5}
\end{subfigure}%
\hfill
\begin{subfigure}[b]{0.249\columnwidth}
  \includegraphics[height=1.95cm,keepaspectratio]{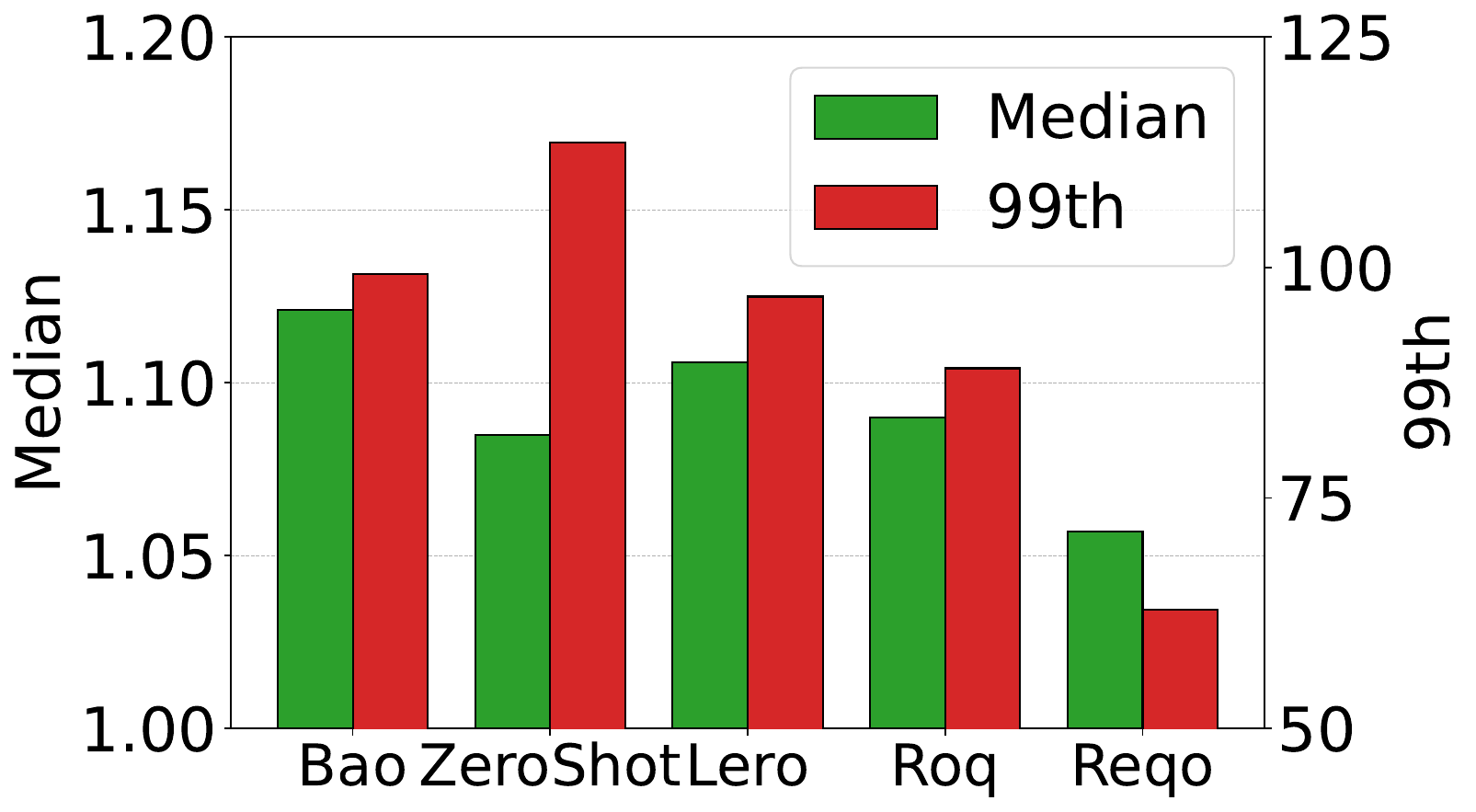}
  \caption{Plan Subopt. (6-10 J)}
  \label{fig:plan_6-10}
\end{subfigure}%
\hfill
\begin{subfigure}[b]{0.249\columnwidth}
  \includegraphics[height=1.95cm,keepaspectratio]{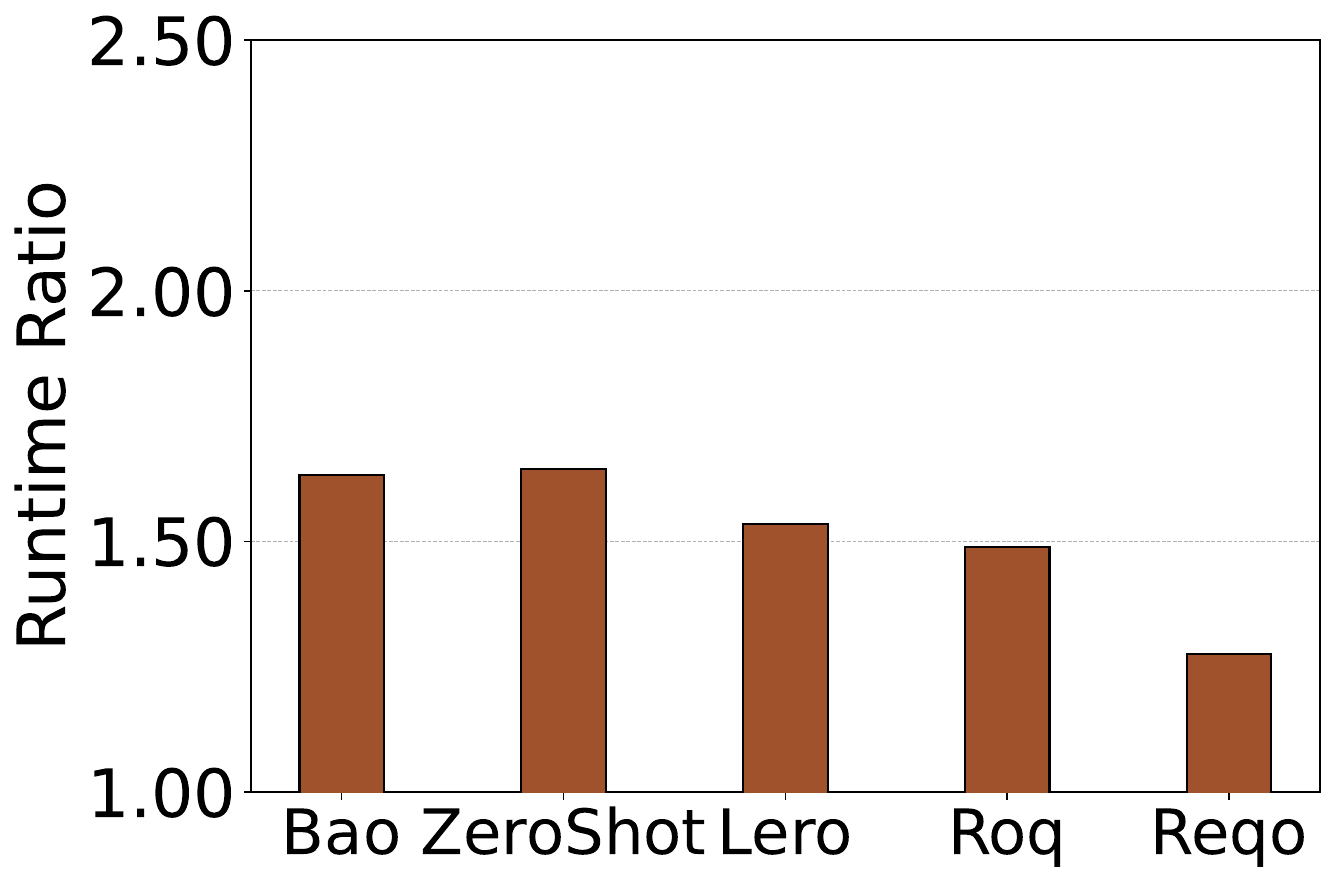}
  \caption{Runtime Ratio (1-5 J)}
  \label{fig:runtime_1-5}
\end{subfigure}%
\hfill
\begin{subfigure}[b]{0.249\columnwidth}
  \includegraphics[height=1.95cm,keepaspectratio]{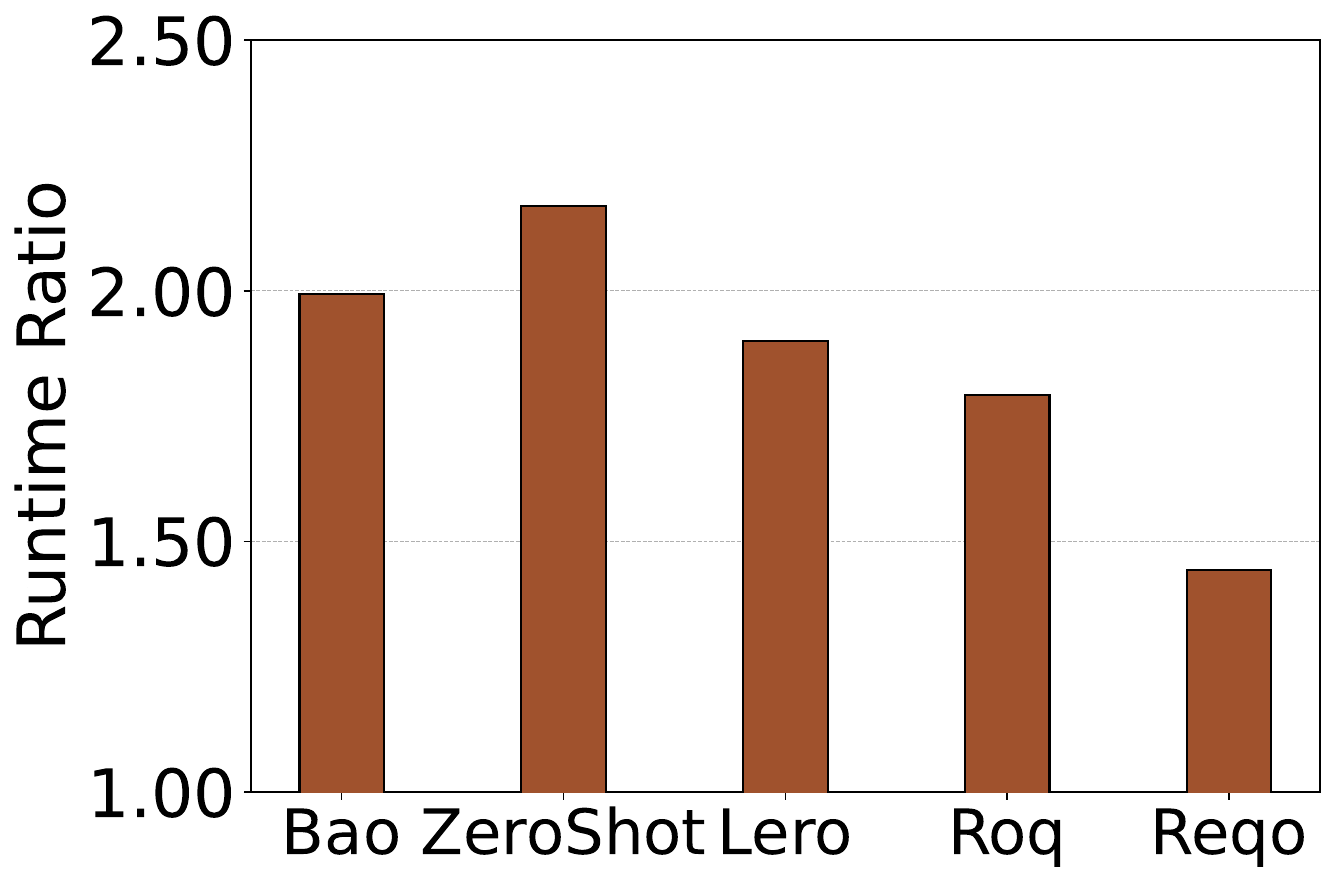}
  \caption{Runtime Ratio (6-10 J)}
  \label{fig:runtime_6-10}
\end{subfigure}

\caption{Workload-shift experiment results on TPC-DS across performance metrics. The models are trained on 4,500 queries with 1-5 joins and evaluated separately on 500 queries with 1-5 joins (in-distribution) and 500 queries with 6-10 joins (shifted workload)}
\label{fig:workload_shift_performance_metrics_tpcds}
\end{figure}

To further assess robustness, we conduct a workload-shift experiment on TPC-DS. Models are trained on queries with 1-5 joins and tested on new queries with 1-5 joins and 6-10 joins separately to examine their adaptability to more complex workloads. For Zero-Shot, these test sets and all TPC-DS queries with more than 5 joins are excluded from its training data. Figure~\ref{fig:workload_shift_performance_metrics_tpcds} shows that Reqo achieves the best results across all metrics and, despite a performance decline in the more complex scenario, Reqo experiences the smallest drop. Moreover, its relative advantage over the baselines becomes more pronounced as complexity increases. Excluding Reqo, Zero-Shot generalizes better to more complex workloads but shows weak tail performance. Reqo is trained on only one workload yet achieves stronger robustness. Combined with Reqo's superior tail-end performance in Q-Error and plan suboptimality, these findings confirm that Reqo demonstrates exceptional robustness and outperforms other state-of-the-art LCMs under challenging conditions.

\begin{figure}[t]
\centering
\begin{subfigure}[b]{0.333\columnwidth}
  \includegraphics[width=\linewidth]{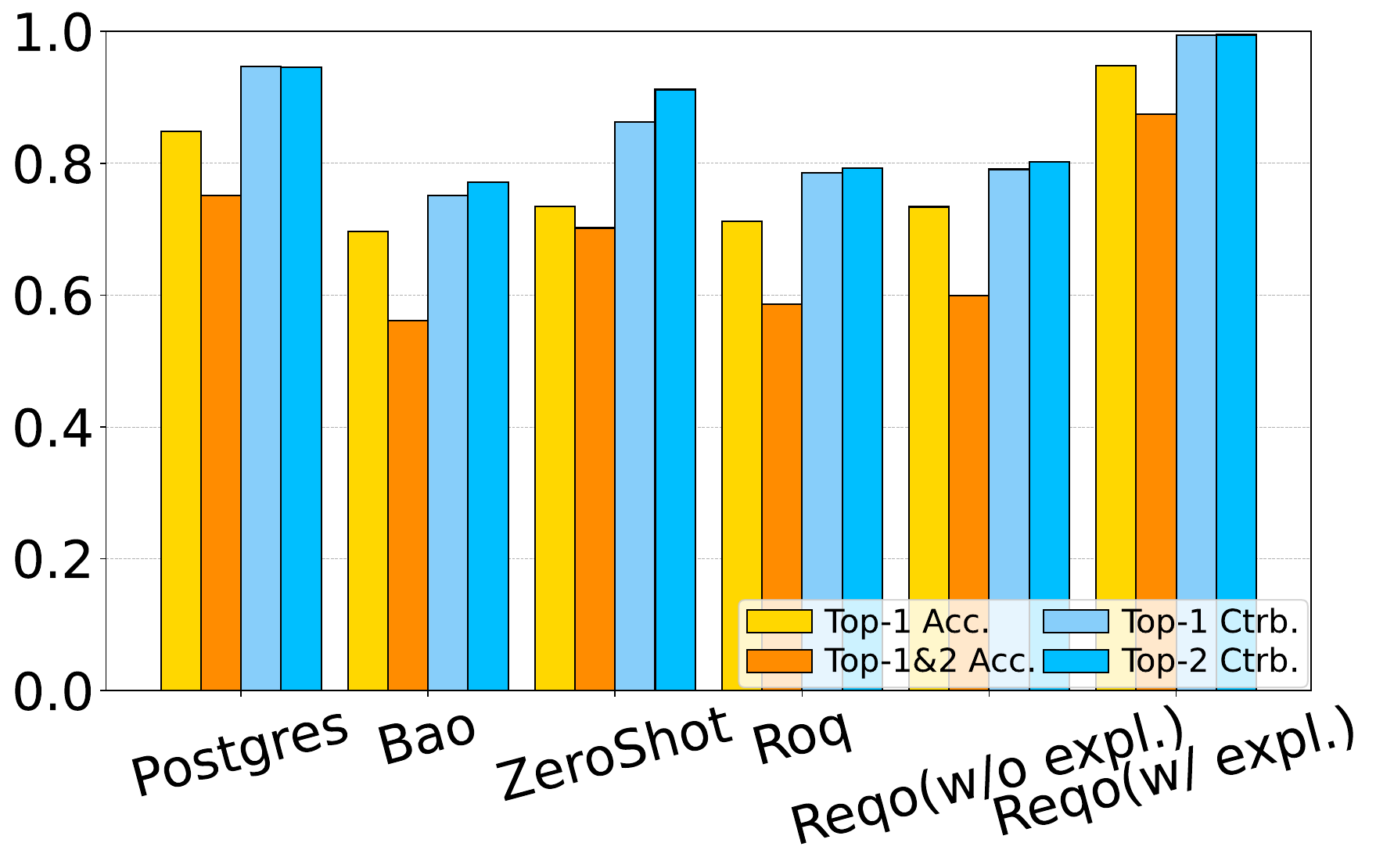}
  \caption{TPC-H}
  \label{fig:explanation_tpc-h}
\end{subfigure}%
\hfill
\begin{subfigure}[b]{0.333\columnwidth}
  \includegraphics[width=\linewidth]{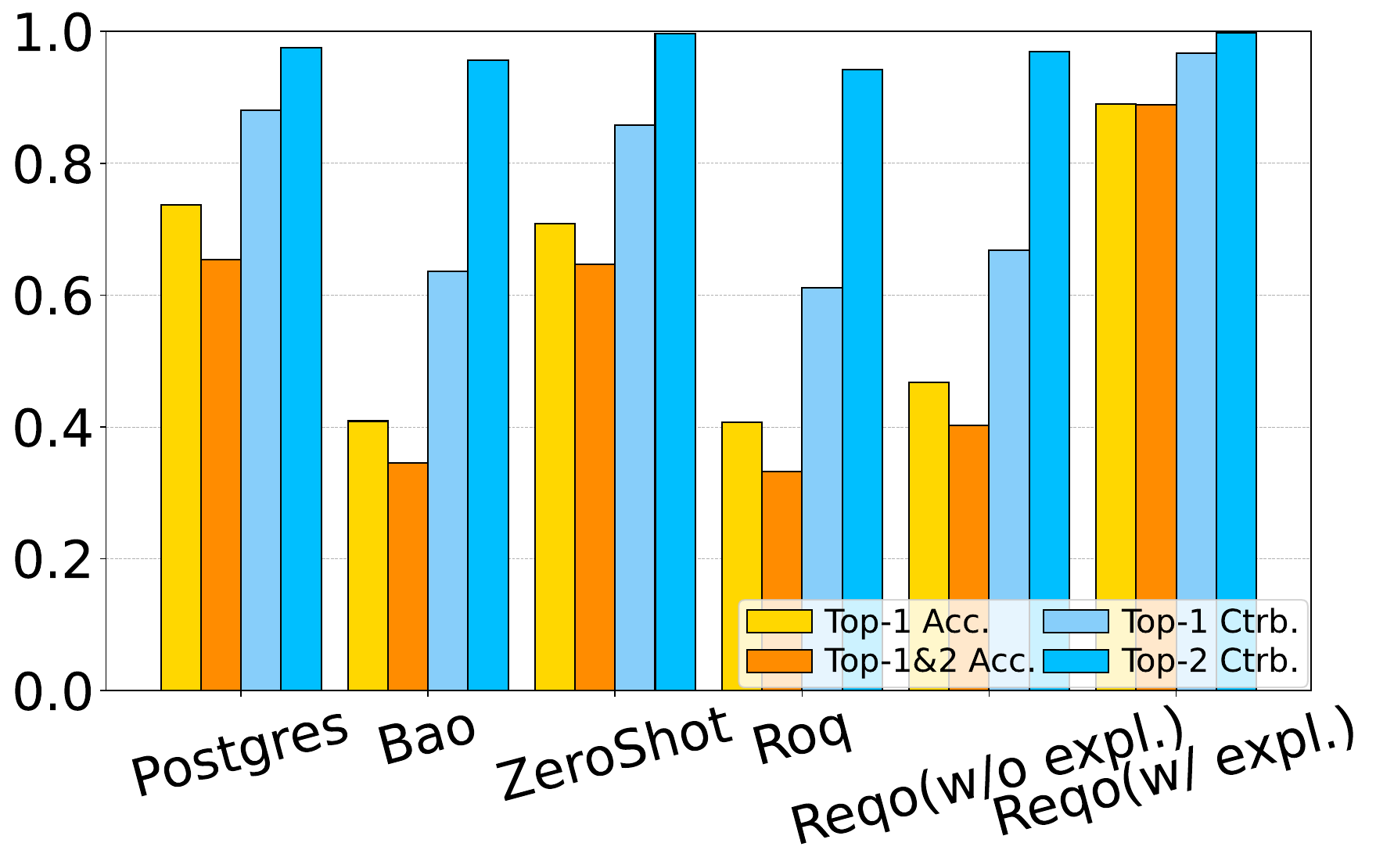}
  \caption{JOB-light}
  \label{fig:explanation_job-light}
\end{subfigure}%
\hfill
\begin{subfigure}[b]{0.333\columnwidth}
  \includegraphics[width=\linewidth]{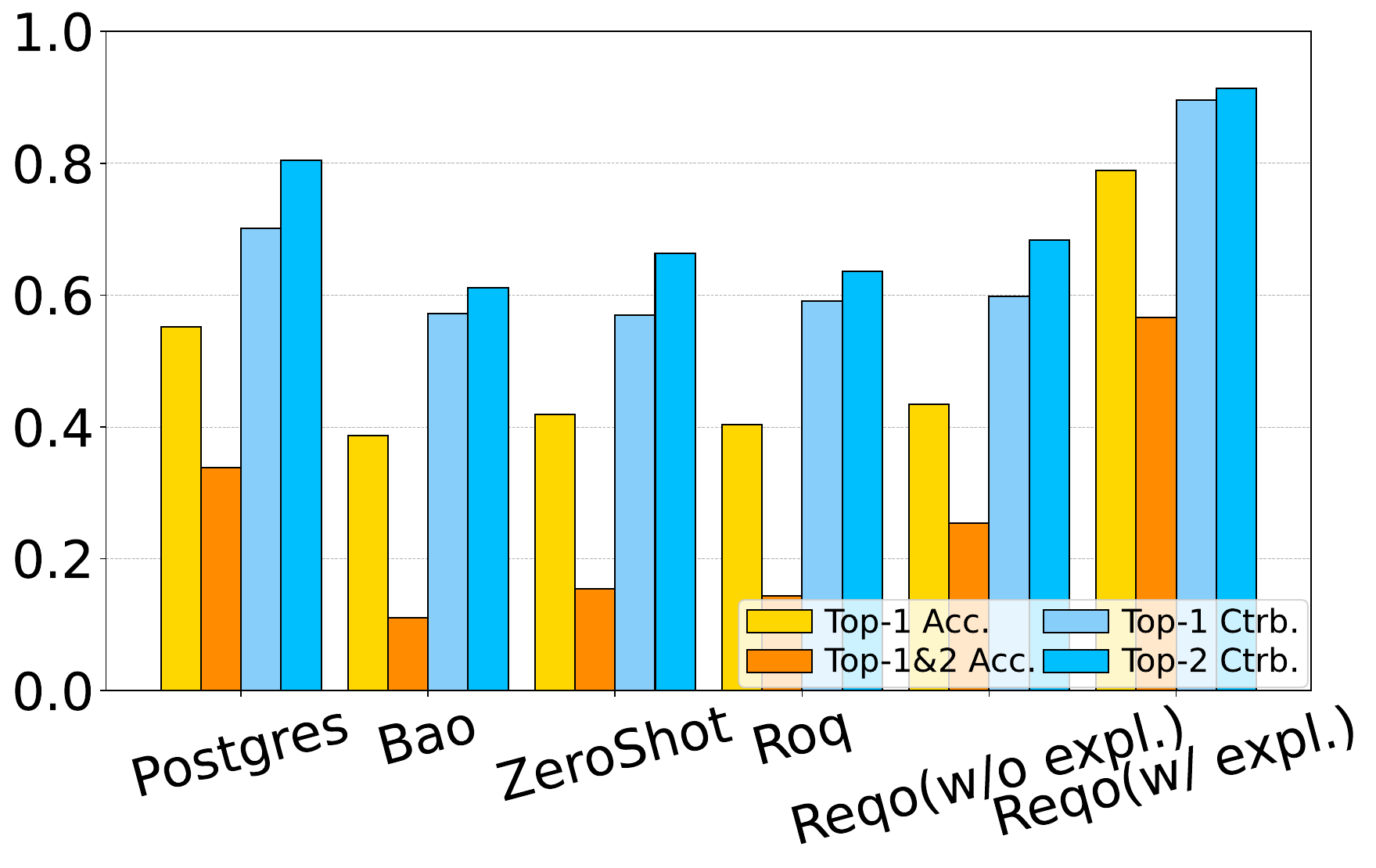}
  \caption{JOB-full}
  \label{fig:explanation_job-full}
\end{subfigure}

\begin{subfigure}[b]{0.333\columnwidth}
  \includegraphics[width=\linewidth]{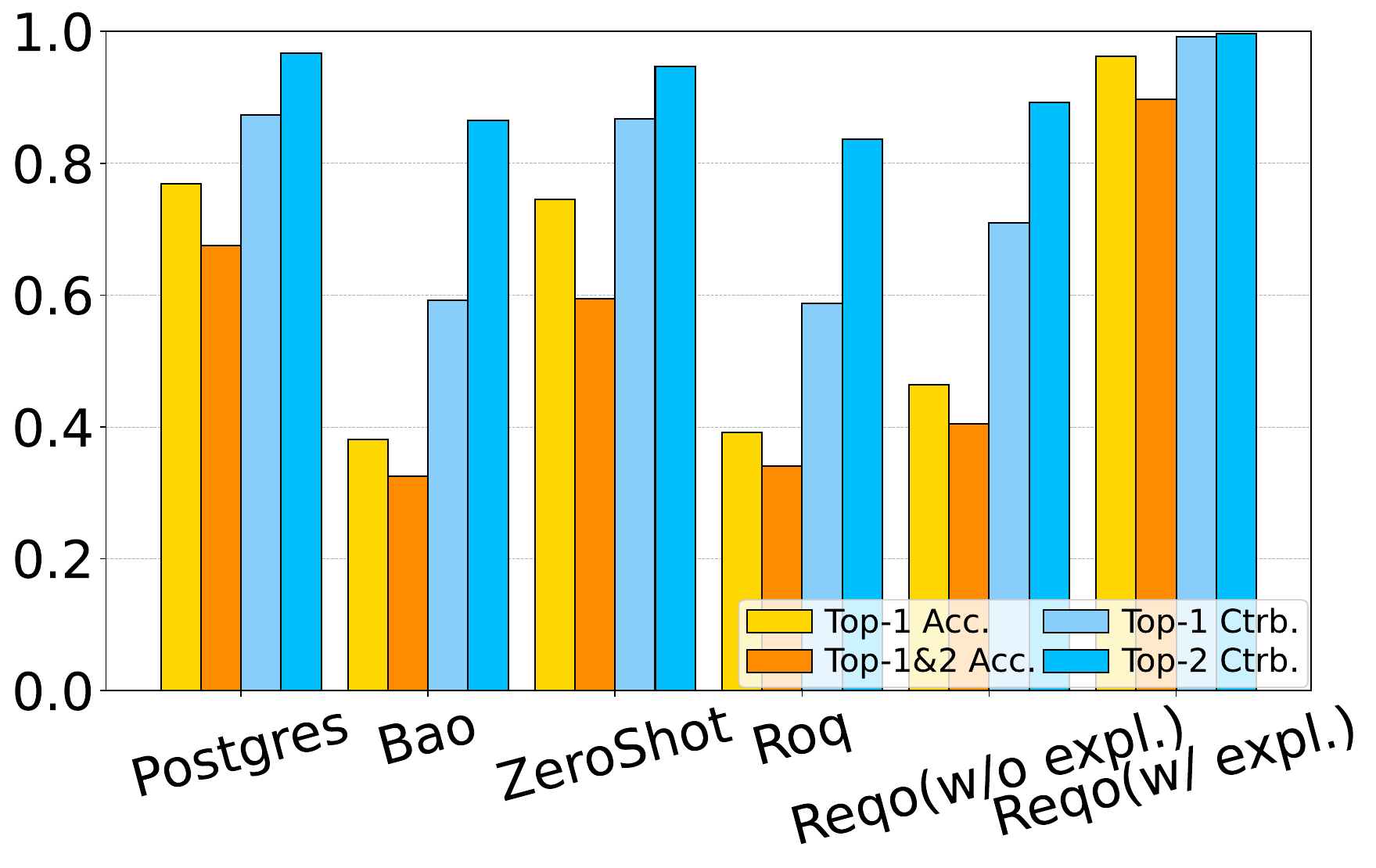}
  \caption{STATS}
  \label{fig:explanation_stats}
\end{subfigure}%
\hfill
\begin{subfigure}[b]{0.333\columnwidth}
  \includegraphics[width=\linewidth]{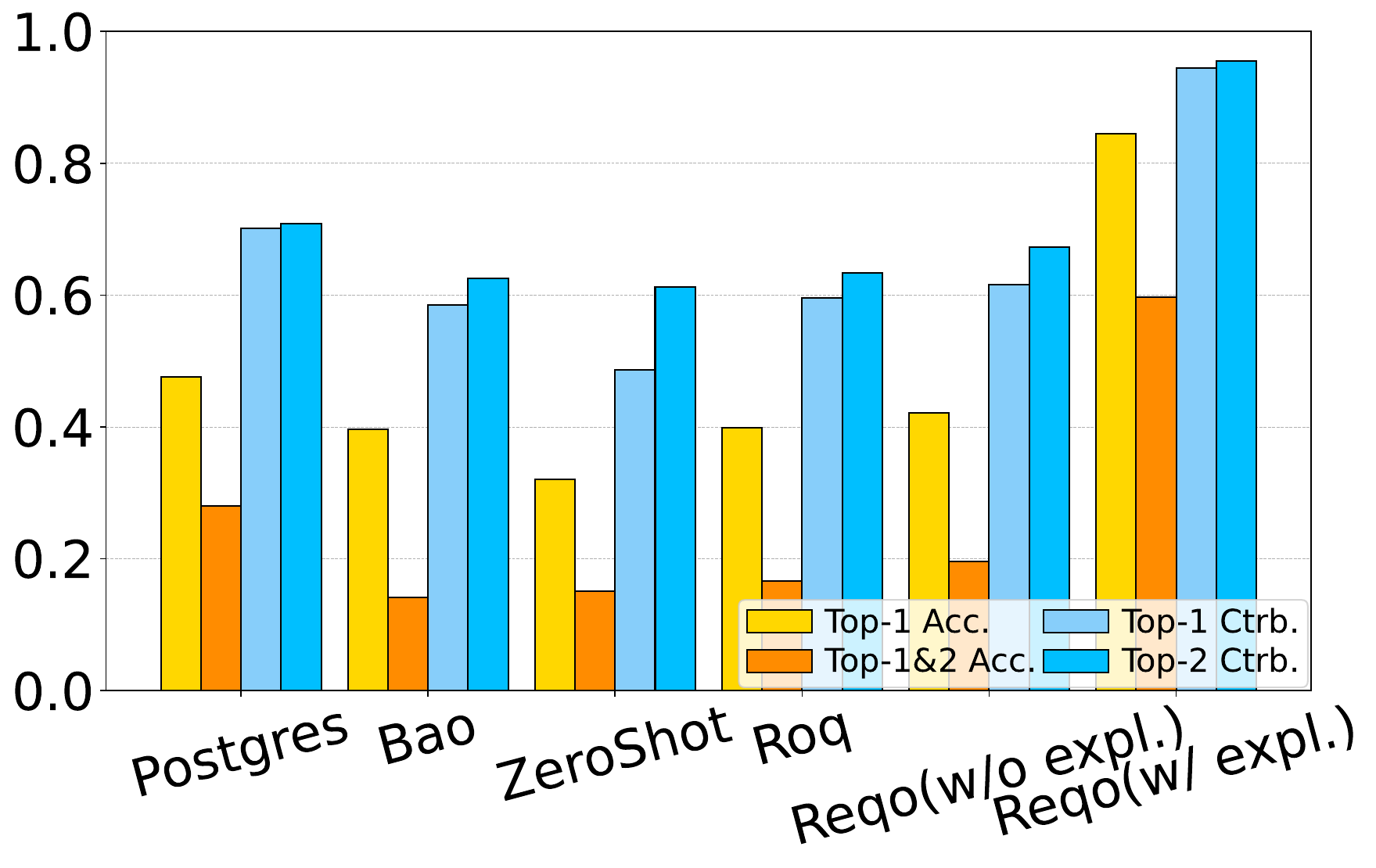}
  \caption{DSB}
  \label{fig:explanation_dsb}
\end{subfigure}%
\hfill
\begin{subfigure}[b]{0.333\columnwidth}
  \includegraphics[width=\linewidth]{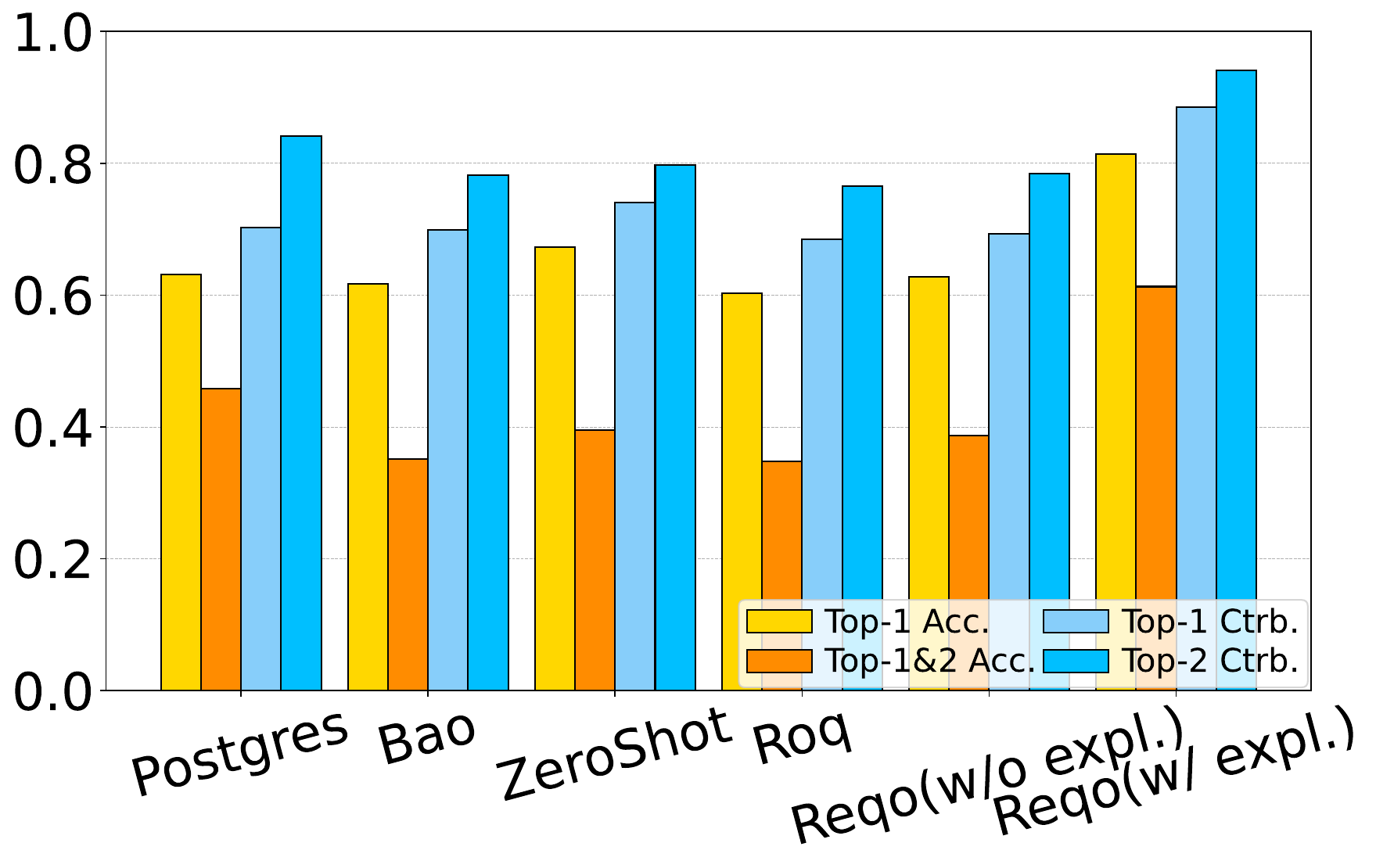}
  \caption{TPC-DS}
  \label{fig:explanation_tpc-ds}
\end{subfigure}
\caption{Explanation performance. Top-1 Acc. measures the accuracy of identifying the most influential node with the largest contribution, and Top-1\&2 Acc. for both the top two most influential nodes in the correct order. Top-1 (or Top-2) Ctrb. Ratio is defined as the sum of actual contributions of the cost model-selected top 1 (or top 2) nodes divided by that of the actual top 1 (or top 2) nodes}
\label{fig:explanation-results}
\end{figure}

\begin{figure}
\centering
\begin{subfigure}[b]{0.333\columnwidth}
  \includegraphics[width=\linewidth]{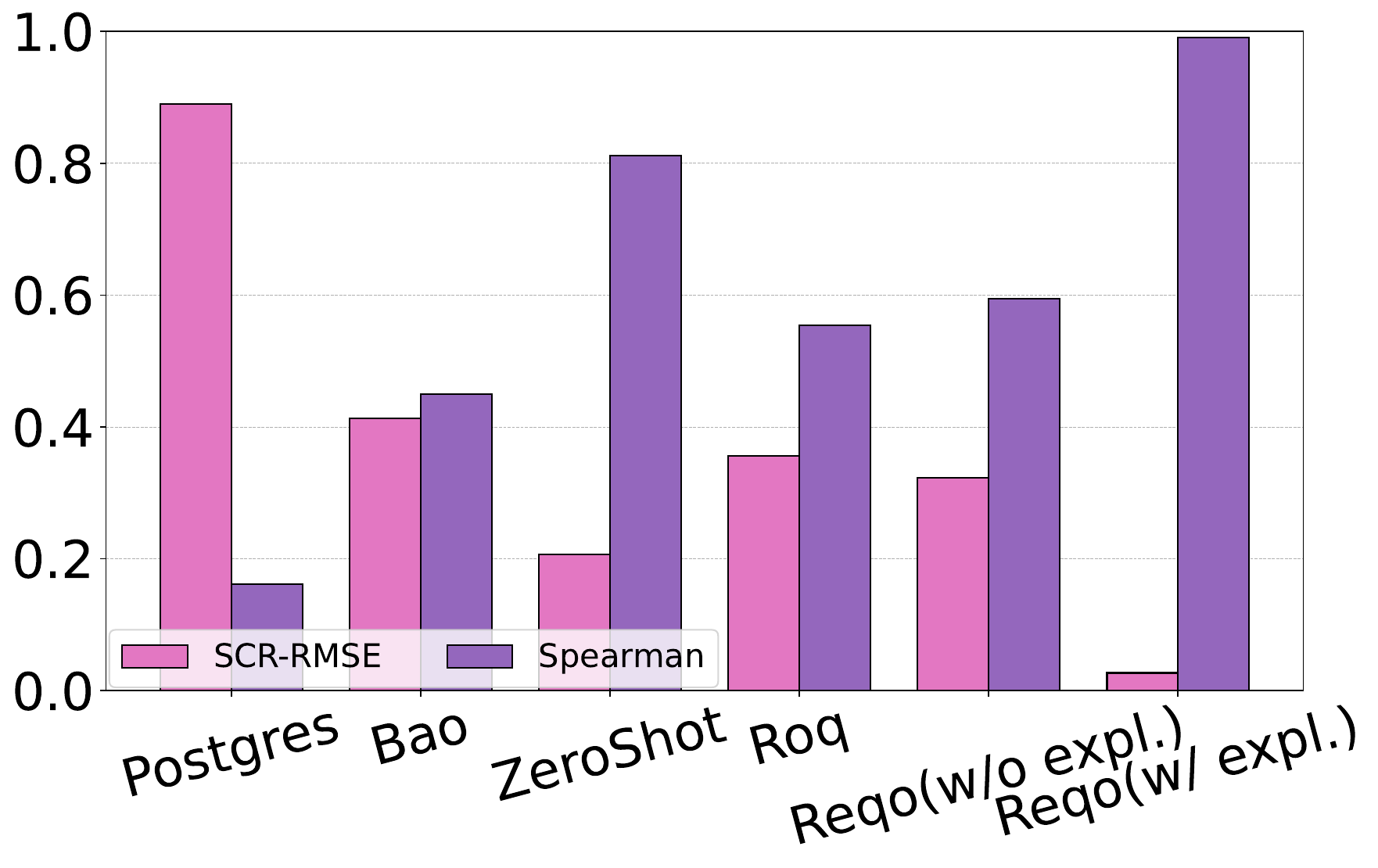}
  \caption{TPC-H}
  \label{fig:subplan_cost_estimation_tpc-h}
\end{subfigure}%
\hfill
\begin{subfigure}[b]{0.333\columnwidth}
  \includegraphics[width=\linewidth]{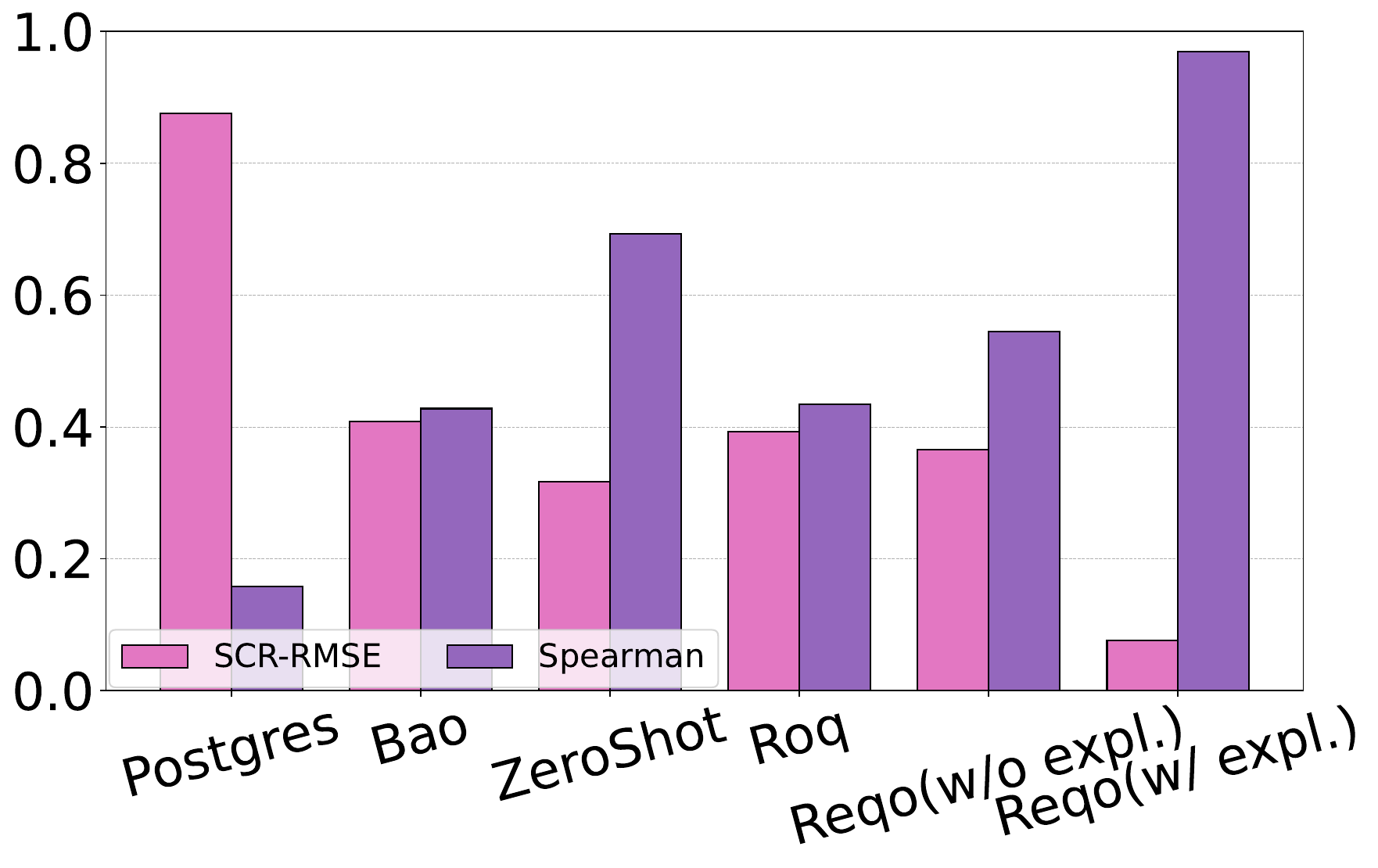}
  \caption{JOB-light}
  \label{fig:subplan_cost_estimation_job-light}
\end{subfigure}%
\hfill
\begin{subfigure}[b]{0.333\columnwidth}
  \includegraphics[width=\linewidth]{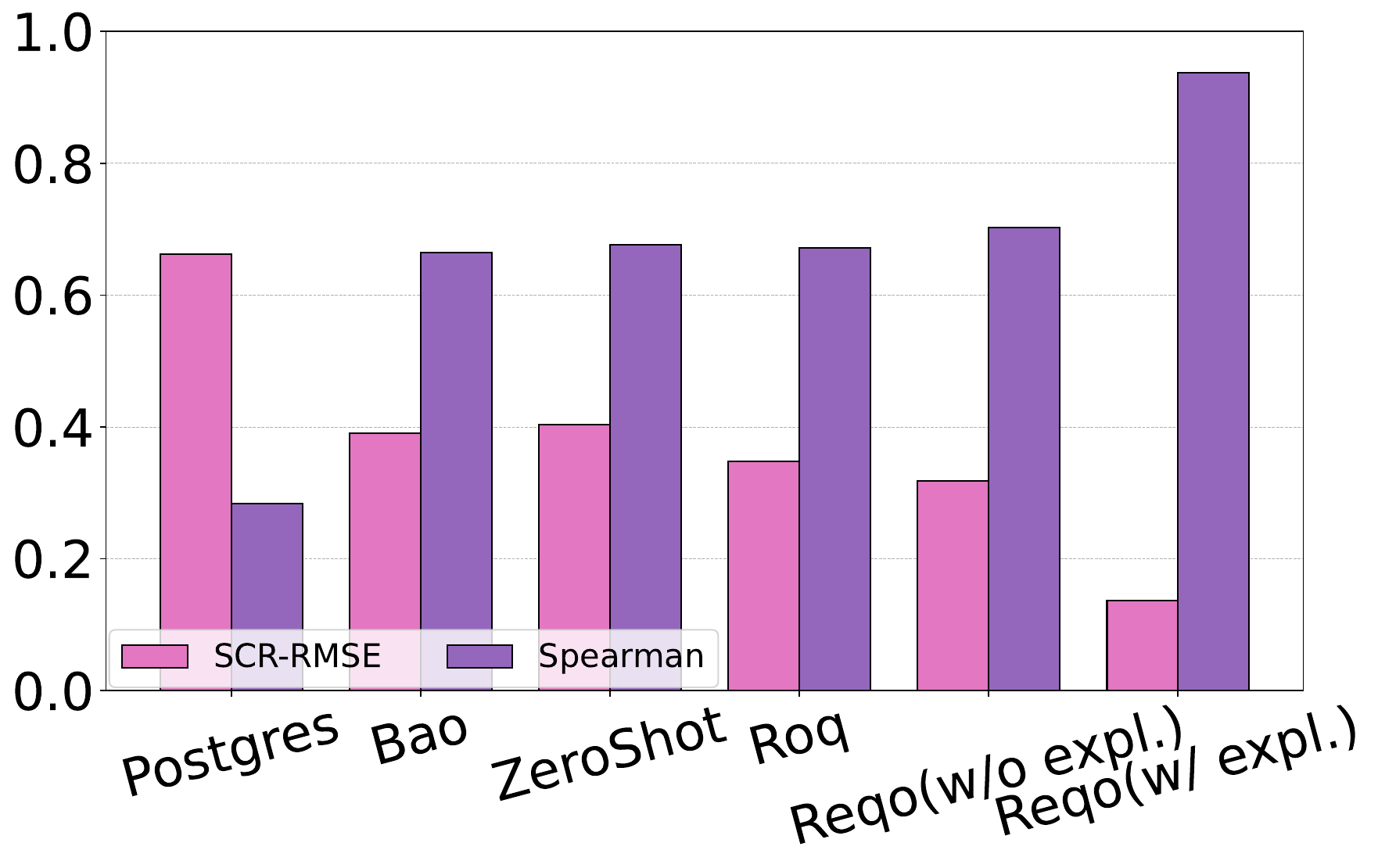}
  \caption{JOB-full}
  \label{fig:subplan_cost_estimation_job-full}
\end{subfigure}

\begin{subfigure}[b]{0.333\columnwidth}
  \includegraphics[width=\linewidth]{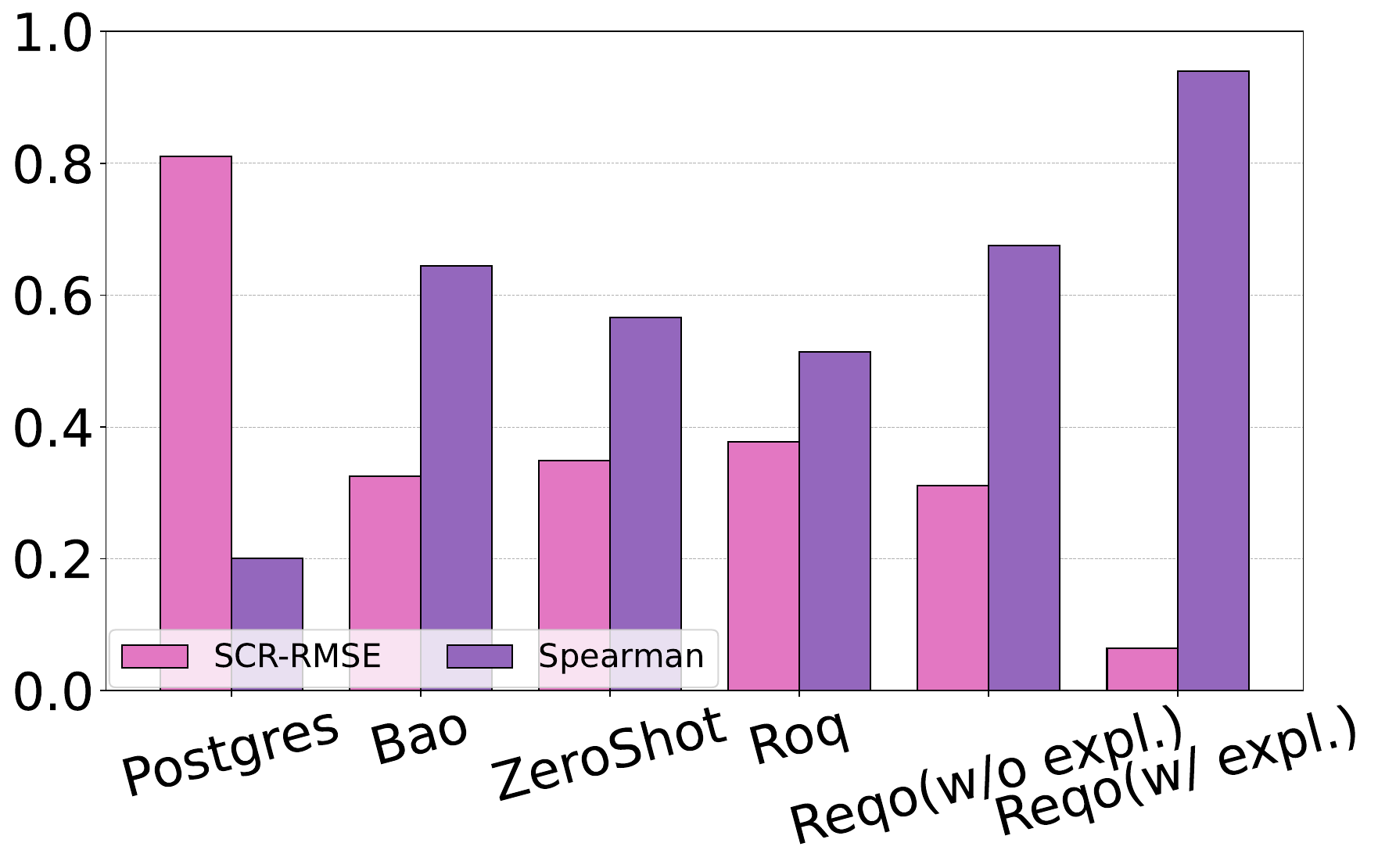}
  \caption{STATS}
  \label{fig:subplan_cost_estimation_stats}
\end{subfigure}%
\hfill
\begin{subfigure}[b]{0.333\columnwidth}
  \includegraphics[width=\linewidth]{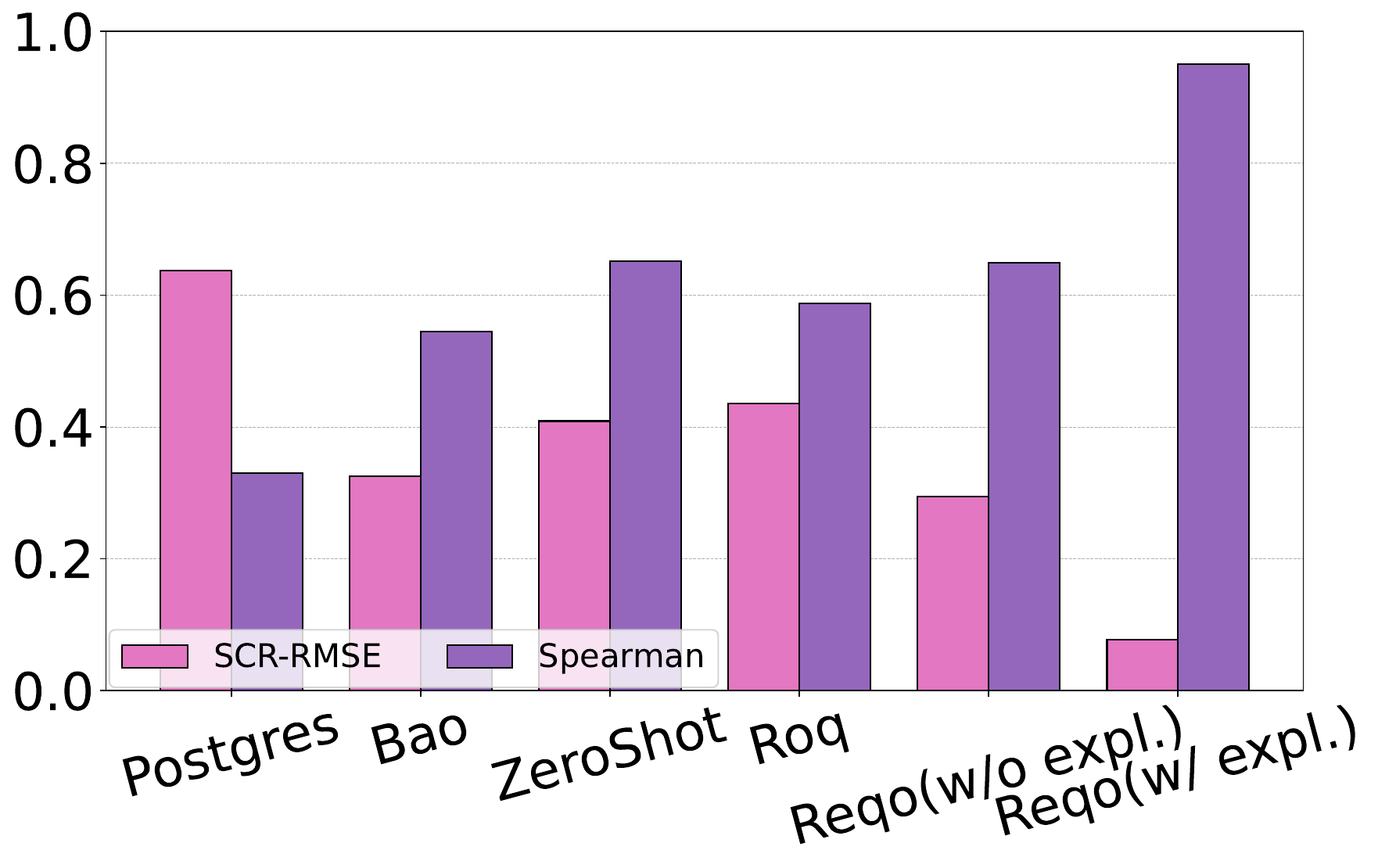}
  \caption{DSB}
  \label{fig:subplan_cost_estimation_dsb}
\end{subfigure}%
\hfill
\begin{subfigure}[b]{0.333\columnwidth}
  \includegraphics[width=\linewidth]{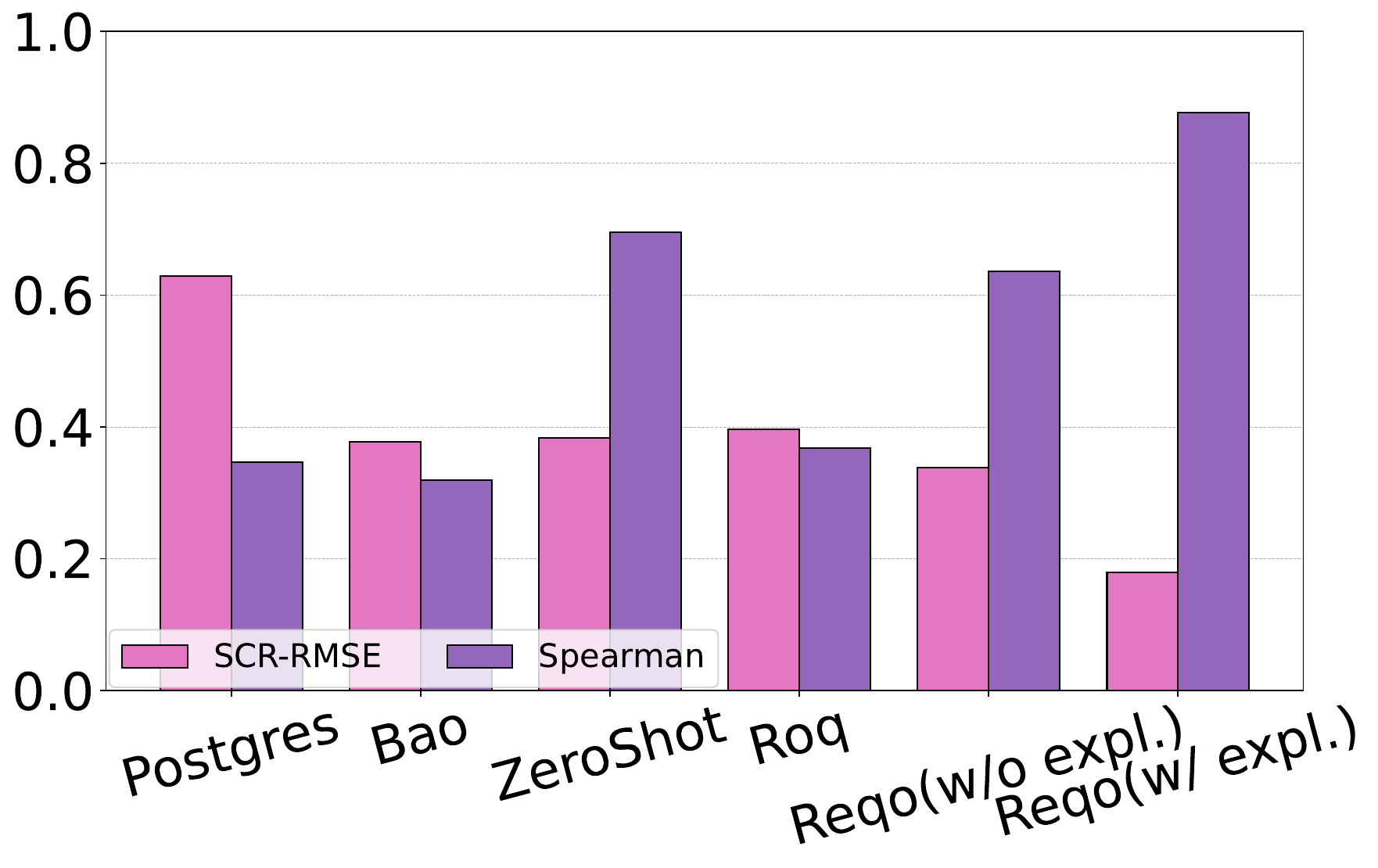}
  \caption{TPC-DS}
  \label{fig:subplan_cost_estimation_tpc-ds}
\end{subfigure}
\caption{Subplan cost estimation performance in SCR-RMSE and Spearman's Correlation}
\label{fig:subplan_cost_estimation_results}
\end{figure}

\subsubsection{Comparison for Explainability}
To evaluate the explainability of LCMs that lack native support (except Reqo (w/ expl.)), subplans from each plan are extracted and fed directly to these models to obtain subplan cost estimates. Algorithm~\ref{alg:compute_subgraph_EC} is then applied to infer each node’s explanation from these subplan cost estimates rather than their contributions. Explanation performance is evaluated using metrics 6 and 7. Figure~\ref{fig:explanation-results} shows that all LCM baselines, including Reqo (w/o expl.), underperform at explaining the contributions of specific nodes. In contrast, PostgreSQL’s classical optimizer accumulates detailed cost statistics for each node in a bottom-up manner, making its decision process transparent and yielding higher explanation accuracy than the other baselines. Without our proposed explainability technique, all LCM baselines underperform PostgreSQL on all explanation metrics. The gap is especially pronounced in Top-1\&2 accuracy, where the models must identify both the two most influential nodes in correct order, indicating that these models cannot precisely isolate and rank node contributions. This limitation undermines trust in learning-based query optimization and underscores the necessity of integrated explainability in LCMs.

However, due to limitations of classical optimizers, PostgreSQL’s cost estimates are not accurate and provide limited explanation performance. Experiments show that Reqo with the explainability module outperforms all baselines across all metrics. In simpler workloads, it achieves a nearly perfect Top-2 contribution ratio, accurately identifying the most influential nodes. Even in JOB-full, DSB and TPC-DS, Reqo’s advantage becomes more pronounced over all baselines, with Top-1\&2 node accuracy up over 20\% and Top-2 node contribution ratio up over 10\% versus PostgreSQL. Therefore, our explainability technique demonstrably improves LCM transparency and yields explanation performance superior to the classical cost model across diverse workloads.

Figure~\ref{fig:subplan_cost_estimation_results} shows pure subplan cost estimation performance and helps explain why LCMs exhibit poor explanation accuracy. Unlike other LCM baselines, Reqo (w/ expl.) derives each subplan’s estimated runtime by multiplying its predicted subplan contribution ratio by the estimated runtime of the corresponding entire plan. We compare these estimates with actual runtimes for all subplans in the test set and report SCR-RMSE and Spearman’s correlation. The results show that although LCMs estimate subplan runtime more accurately than PostgreSQL, their accuracy on subplans is substantially lower than on entire plans.
In contrast, Reqo maintains substantial improvements on both metrics, validating our observation that directly predicting subplan cost without contextual information is insufficient. Although these LCM baselines outperform PostgreSQL in subplan cost estimation, they perform worse at explaining specific nodes, as shown in Figure~\ref{fig:explanation-results}, confirming that they do not preserve monotonic estimates across nested subplans within a plan. This lack of monotonicity, together with their limited accuracy, undermines node-level explanation accuracy. Reqo’s explanation loss forces it to learn subplan contribution ratios and promotes monotonicity, enabling finer discrimination among subplans within the same plan, more accurate relative cost estimates between nested subplans, and consequently more precise explanation inference.

\begin{table}[t]
  \caption{Training/inference overheads of LCMs}
  \label{tab:overheads}
  \centering
  \footnotesize
  \setlength{\tabcolsep}{0.8pt}
  \renewcommand{\arraystretch}{1.05}
  \begin{threeparttable}
  {
  \begin{tabular}{
    >{\raggedright\arraybackslash}p{2.5cm}
    S[table-format=3.2, round-mode=places, round-precision=2, table-column-width=1.825cm]  
    S[table-format=3.2, round-mode=places, round-precision=2, table-column-width=1.825cm]  
    S[table-format=3.2, round-mode=places, round-precision=2, table-column-width=1.825cm]  
    S[table-format=2.2, round-mode=places, round-precision=2, table-column-width=1.825cm]  
    S[table-format=2.2, round-mode=places, round-precision=2, table-column-width=1.825cm]  
    S[table-format=1.2, round-mode=places, round-precision=2, table-column-width=1.825cm]  
  }
    \toprule
    \multicolumn{1}{c}{Model} &
    \multicolumn{1}{c}{\makecell{Training data\\generation [s]}} &
    \multicolumn{1}{c}{\makecell{Training overh.\\/ epoch [s]}} &
    \multicolumn{1}{c}{\makecell{Training overh.\\total [min]}} &
    \multicolumn{1}{c}{\makecell{Model size\\{[MB]}}} &
    \multicolumn{1}{c}{\makecell{Inference overh.\\/ query [ms]}} &
    \multicolumn{1}{c}{\makecell{End-to-end\\time [h]}} \\
    \midrule
    Bao               &  13.925 &   1.602 &   0.899 &  0.513 &  2.403 & 6.384 \\
    Zero-Shot          & 153.436 &  19.330 &  29.516 & 15.405 & 40.899 & 5.31 \\
    Lero              &  14.695 & 340.787 & 170.393 &  1.190 & 50.156 & 5.525 \\
    Roq               &  21.420 &   1.737 &   2.818 &  1.451 & 13.513 & 5.243 \\
    Reqo (w/o expl.)  &  46.205 &   5.991 &   4.270 &  9.244 & 11.168 & 4.740 \\
    Reqo (w/ expl.)   & 109.510 &   8.138 &   11.385 & 10.919 & 16.489 & 4.854 \\
    \bottomrule
  \end{tabular}
  }
  \begin{tablenotes}[para,flushleft]
    \footnotesize
    Note: The actual execution time of the per-query optimal plans in the DSB workload sums to 4.49 h.
  \end{tablenotes}
  \end{threeparttable}
\end{table}

\subsubsection{Comparison for Overhead}
We measure each LCM’s overhead on the DSB workload used in prior experiments, as shown in Table~\ref{tab:overheads}. For Zero-Shot, the reported overhead reflects training it on this single workload. Although Reqo’s more complex architecture incurs more overhead, its improved robustness yields a clearly shorter end-to-end execution time across the workload despite longer data generation and inference time. For inference latency, Reqo is mid-range. Unlike Lero, another learning-to-rank approach that relies solely on pairwise comparisons and does not yield numerical cost estimates, it often requires multiple rounds to find the optimal plan, thereby increasing inference time. Reqo outputs a single integrated cost that enables one-pass ranking and reduces latency. Enabling the explanation module adds training overhead and slightly increases inference latency. Since the module is optional, it can be disabled for maximum throughput or enabled for detailed insights, allowing Reqo to accommodate diverse use cases.

\begin{figure}[t]
    \centering
    \includegraphics[width=0.75\linewidth]{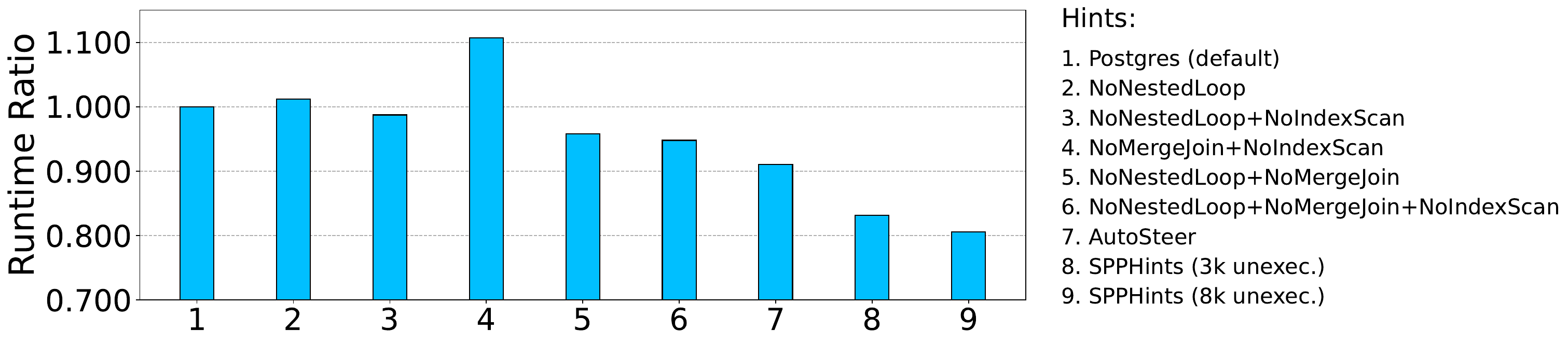}
    \caption{Total runtime ratio for PostgreSQL default, GDHints, and SPPHints on 3k new TPC-DS queries}
    \label{fig:SPPHints}
\end{figure}

\subsubsection{Evaluation for Subplan Pattern Hints}

To evaluate our SPPHints, we train Reqo with a TPC-DS workload of executed plans generated from 3k queries. We generate 8k additional queries to obtain unexecuted plans, use the trained Reqo model to estimate their explanations and produce SPPHints. We set the parameters $f_{max}$ to 0.4, $f_{min}$ to 0.6, $cnt_{min}$ to 3\% of the total number of query plans in the workload, 
and generate 1k new TPC-DS queries for validation and 3k for testing. These parameters are chosen via parameter search by optimizing validation-set query performance using SPPHints generated by the technique illustrated in Figure~\ref{fig:subplan_pattern_example} with these parameters.
We use the default PostgreSQL plans as the baseline, and compare against plans produced with five GDHint sets from our prior experiments, AutoSteer hints~\cite{autosteer} and our SPPHints. AutoSteer extends Bao~\cite{bao} with automatic hint set discovery. It is trained on the same queries as Reqo and employs a TCNN trained on plans generated during hint exploration. During testing, we apply AutoSteer to explore hint sets and use TCNN predictions to select the optimal hint set for each query.

The effectiveness of generated hints is evaluated by the ratio of the total runtime for 3k new test queries under each hint set to the total runtime using the default PostgreSQL plans, as shown in Figure~\ref{fig:SPPHints}. Although the baselines also provide improvements, our SPPHints generated from explanations of 3k unexecuted plans reduce the total runtime by 17\% compared to PostgreSQL plans. SPPHints from 8k unexecuted plans deliver an additional 3\% benefit, surpassing AutoSteer by 8.5\% and demonstrating the effectiveness of our approach.
The experimental results indicate that our explainability technique not only clarifies the decision process of the learning-based cost model but also produces explanations that can be leveraged for targeted query optimization, thereby further improving query performance. Beyond the SPPHints scenario, we believe the explanations can be extended to many other facets of query optimization, which remain open for future exploration.

\section{Related Work}
\label{section_related_work}
This section reviews learning-based tree models, robustness and explainability techniques for LCMs, limitations of current LCMs, and hint-based query optimization approaches.

\textbf{Learning-based Tree Models.} Some learning-based optimizers~\cite{sun2019end, avgdl} adopt RNN-based models such as LSTM~\cite{lstm} as tree models, but these approaches require flattening trees into sequences, which can lose structural information. Saturn~\cite{saturn} and QueryFormer~\cite{queryformer} apply self-attention to enhance plan representation but still work on sequences. Tree-specific models like Tree-LSTM~\cite{treelstm} and Tree-CNN~\cite{treecnn} preserve structural relationships by processing plans directly as a tree but underperform on deep query plans~\cite{bigg}. Our proposed tree model addresses these issues by combining Bi-GNNs with GRU, demonstrating superior plan representation and significant performance improvements in cost estimation and downstream tasks over existing tree models.

\textbf{Learning-based Robustness Techniques for LCMs.} Most LCMs aim for accurate cost estimates but do not explicitly target robustness or leverage inherent uncertainty. Prior work such as~\cite{neo, bao} shows partial robustness to estimation errors or bounds worst-case outcomes, yet lacks systematic uncertainty quantification.
Zero-Shot~\cite{zeroshot} uses a pretraining-based learning paradigm that trains on multiple workloads and provides relatively accurate cost estimation on unseen databases, indicating its robustness.
Recent research addresses this gap by predicting variance alongside cost estimates. Studies such as~\cite{liu2021fauce, zhao2022lightweight, zhao2021uncertainty, yu2022cost, roq} employ Gaussian negative log-likelihood~\cite{nix1994estimating} and ~\cite{doshi2023kepler} introduce spectral-normalized neural Gaussian processes~\cite{liu2023simple}, while others~\cite{liu2021fauce, zhao2021uncertainty, chen2023leon, roq} use Bayesian or approximate probabilistic neural networks such as Monte Carlo Dropout. These approaches discard high-uncertainty plans or revert to classical methods when uncertainty is high. However, they apply uncertainty to plan selection but keep selection independent from model training and depend on fixed rules such as uncertainty thresholds, which limit adaptability and prevent automatic refinement of these rules based on workload characteristics. Reqo addresses these limitations by automatically learning to integrate cost estimates and uncertainty from plan comparison without manual tuning or preset rules. In contrast to other learning-to-rank cost models, such as Lero~\cite{lero}, which do not produce numeric cost prediction and must perform multiple rounds of pairwise plan comparisons to select the optimal plan, Reqo benefits from pairwise comparisons only during training but directly outputs a single integrated cost-and-uncertainty value for inference, eliminating repeated comparisons and reducing inference overhead.

\textbf{Explainability Techniques for LCMs.} To our knowledge, Reqo is the first technique to provide accurate node-level explanations for LCM predictions. Flow-Loss~\cite{flowloss} proposes a flow-aware loss for cardinality estimation that identifies high-flow subplans sensitive to plan cardinality errors. Its goal is to improve cardinality estimation accuracy for the entire plan rather than explainability. It does not provide faithful and fine-grained explanations of LCM predictions.

\textbf{Limitations of Current LCMs.} A recent study~\cite{howgoodarelcm} systematically evaluates state-of-the-art LCMs for query optimization, summarizes their limitations, and provides recommendations for future LCM design. Reqo follows these recommendations. Specifically, Reqo uses physical plans as input rather than relying solely on SQL strings, and it is evaluated with metrics such as total runtime ratio and plan suboptimality to capture actual runtime improvement beyond cost estimation accuracy. To mitigate training data bias, Reqo proposes an uncertainty-aware design and supports executing each query under diverse hints to enrich the training data. As suggested, Reqo incorporates expert knowledge from the classical optimizer by encoding PostgreSQL estimated cardinalities and costs as part of plan node features. Despite greater architectural complexity, Reqo achieves lower estimation error and also improves the robustness and explainability of learning-based cost estimation. Overall, Reqo aligns with the recommended practices.

\textbf{Hint-based Query Optimization.} Recent hint-based query optimization approaches have explored various strategies. Bao~\cite{bao} models fixed hint sets as arms in a contextual bandit and uses Thompson sampling~\cite{thompson_sampling} to balance exploration and exploitation when steering the optimizer. AutoSteer~\cite{autosteer} extends Bao by automating hint set discovery via query-span approximation and greedy search, pruning candidates with a TCNN for runtime hint selection within the contextual bandit framework. FASTgres~\cite{fastgres} exhaustively evaluates all boolean hint combinations offline and trains classifiers to predict optimal hints. HERO~\cite{hero} conducts a budget-limited local search over operator and parallelism flags and trains a context-aware ensemble for inference. In contrast, Reqo generates hints directly from its explainer outputs, avoiding exhaustive executions for hint evaluation and additional training while still producing high-quality hints.

\section{Conclusion}
\label{section_conclusion}
We introduce Reqo, a comprehensive LCM that integrates three innovations: a novel tree model (Bi-GNN with GRU aggregation), an uncertainty-aware learning-to-rank cost estimator and a subplan-based explainability technique. While achieving top-tier cost estimation accuracy, Reqo adaptively integrates cost estimates with uncertainties to improve plan selection robustness. Our explainability technique enhances transparency and prediction monotonicity of learned cost estimation, making Reqo the first LCM capable of explaining its estimates. We further leverage these explanations to generate SPPHints that guide plan generation and improve candidate plan quality. Extensive experiments demonstrate Reqo’s superiority in cost estimation accuracy, robustness, explainability, and hint generation, consistently outperforming state-of-the-art approaches across all three stages.

\bibliographystyle{ACM-Reference-Format}
\bibliography{references}

@inproceedings{bigg,
author = {Chang, Baoming and Kamali, Amin and Kantere, Verena},
title = {A Novel Technique for Query Plan Representation Based on Graph Neural Nets},
year = {2024},
isbn = {978-3-031-68322-0},
publisher = {Springer-Verlag},
address = {Berlin, Heidelberg},
url = {https://doi.org/10.1007/978-3-031-68323-7_25},
doi = {10.1007/978-3-031-68323-7_25},
booktitle = {Big Data Analytics and Knowledge Discovery: 26th International Conference, DaWaK 2024, Naples, Italy, August 26–28, 2024, Proceedings},
pages = {299–314},
numpages = {16},
location = {Naples, Italy}
}

@inproceedings{dirgnn,
  title={Edge directionality improves learning on heterophilic graphs},
  author={Rossi, Emanuele and Charpentier, Bertrand and Di Giovanni, Francesco and Frasca, Fabrizio and G{\"u}nnemann, Stephan and Bronstein, Michael M},
  booktitle={Learning on graphs conference},
  pages={25--1},
  year={2024},
  organization={PMLR}
}

@article{gruaggr,
  title={Graph neural networks with adaptive readouts},
  author={Buterez, David and Janet, Jon Paul and Kiddle, Steven J and Oglic, Dino and Li{\`o}, Pietro},
  journal={Advances in Neural Information Processing Systems},
  volume={35},
  pages={19746--19758},
  year={2022}
}

@inproceedings{treelstm,
    title = "Improved Semantic Representations From Tree-Structured Long Short-Term Memory Networks",
    author = "Tai, Kai Sheng  and
      Socher, Richard  and
      Manning, Christopher D.",
    editor = "Zong, Chengqing  and
      Strube, Michael",
    booktitle = "Proceedings of the 53rd Annual Meeting of the Association for Computational Linguistics and the 7th International Joint Conference on Natural Language Processing (Volume 1: Long Papers)",
    month = jul,
    year = "2015",
    address = "Beijing, China",
    publisher = "Association for Computational Linguistics",
    url = "https://aclanthology.org/P15-1150/",
    doi = "10.3115/v1/P15-1150",
    pages = "1556--1566"
}

@inproceedings{treecnn,
author = {Mou, Lili and Li, Ge and Zhang, Lu and Wang, Tao and Jin, Zhi},
title = {Convolutional neural networks over tree structures for programming language processing},
year = {2016},
publisher = {AAAI Press},
booktitle = {Proceedings of the Thirtieth AAAI Conference on Artificial Intelligence},
pages = {1287–1293},
numpages = {7},
location = {Phoenix, Arizona},
series = {AAAI'16}
}

@article{roq,
author = {Kamali, Amin and Kantere, Verena and Zuzarte, Calisto and Corvinelli, Vincent},
title = {Robust Plan Evaluation Based on Approximate Probabilistic Machine Learning},
year = {2025},
issue_date = {April 2025},
publisher = {VLDB Endowment},
volume = {18},
number = {8},
issn = {2150-8097},
url = {https://doi.org/10.14778/3742728.3742753},
doi = {10.14778/3742728.3742753},
journal = {Proceedings of the VLDB Endowment},
month = sep,
pages = {2626–2638},
numpages = {13}
}

@article{neo,
author = {Marcus, Ryan and Negi, Parimarjan and Mao, Hongzi and Zhang, Chi and Alizadeh, Mohammad and Kraska, Tim and Papaemmanouil, Olga and Tatbul, Nesime},
title = {Neo: a learned query optimizer},
year = {2019},
issue_date = {July 2019},
publisher = {VLDB Endowment},
volume = {12},
number = {11},
issn = {2150-8097},
url = {https://doi.org/10.14778/3342263.3342644},
doi = {10.14778/3342263.3342644},
journal = {Proceedings of the VLDB Endowment},
month = jul,
pages = {1705–1718},
numpages = {14}
}

@inproceedings{bao,
author = {Marcus, Ryan and Negi, Parimarjan and Mao, Hongzi and Tatbul, Nesime and Alizadeh, Mohammad and Kraska, Tim},
title = {Bao: Making Learned Query Optimization Practical},
year = {2021},
isbn = {9781450383431},
publisher = {Association for Computing Machinery},
address = {New York, NY, USA},
url = {https://doi.org/10.1145/3448016.3452838},
doi = {10.1145/3448016.3452838},
booktitle = {Proceedings of the 2021 International Conference on Management of Data},
pages = {1275–1288},
numpages = {14},
location = {Virtual Event, China},
series = {SIGMOD '21}
}

@article{doshi2023kepler,
author = {Doshi, Lyric and Zhuang, Vincent and Jain, Gaurav and Marcus, Ryan and Huang, Haoyu and Altinb\"{u}ken, Deniz and Brevdo, Eugene and Fraser, Campbell},
title = {Kepler: Robust Learning for Parametric Query Optimization},
year = {2023},
issue_date = {May 2023},
publisher = {Association for Computing Machinery},
address = {New York, NY, USA},
volume = {1},
number = {1},
url = {https://doi.org/10.1145/3588963},
doi = {10.1145/3588963},
journal = {Proceedings of the ACM on Management of Data},
month = may,
articleno = {109},
numpages = {25}
}

@inproceedings{zhao2021uncertainty,
author = {Zhao, Kangfei and Yu, Jeffrey Xu and He, Zongyan and Li, Rui and Zhang, Hao},
title = {Lightweight and Accurate Cardinality Estimation by Neural Network Gaussian Process},
year = {2022},
isbn = {9781450392495},
publisher = {Association for Computing Machinery},
address = {New York, NY, USA},
url = {https://doi.org/10.1145/3514221.3526156},
doi = {10.1145/3514221.3526156},
booktitle = {Proceedings of the 2022 International Conference on Management of Data},
pages = {973–987},
numpages = {15},
location = {Philadelphia, PA, USA},
series = {SIGMOD '22}
}

@article{liu2021fauce,
author = {Liu, Jie and Dong, Wenqian and Zhou, Qingqing and Li, Dong},
title = {Fauce: fast and accurate deep ensembles with uncertainty for cardinality estimation},
year = {2021},
issue_date = {July 2021},
publisher = {VLDB Endowment},
volume = {14},
number = {11},
issn = {2150-8097},
url = {https://doi.org/10.14778/3476249.3476254},
doi = {10.14778/3476249.3476254},
journal = {Proceedings of the VLDB Endowment},
month = jul,
pages = {1950–1963},
numpages = {14}
}

@article{chen2023leon,
author = {Chen, Xu and Chen, Haitian and Liang, Zibo and Liu, Shuncheng and Wang, Jinghong and Zeng, Kai and Su, Han and Zheng, Kai},
title = {LEON: A New Framework for ML-Aided Query Optimization},
year = {2023},
issue_date = {May 2023},
publisher = {VLDB Endowment},
volume = {16},
number = {9},
issn = {2150-8097},
url = {https://doi.org/10.14778/3598581.3598597},
doi = {10.14778/3598581.3598597},
journal = {Proceedings of the VLDB Endowment},
month = may,
pages = {2261–2273},
numpages = {13}
}

@inproceedings{transformerconv,
  title={Masked Label Prediction: Unified Message Passing Model for Semi-Supervised Classification},
  author={Shi, Yunsheng and Huang, Zhengjie and Feng, Shikun and Zhong, Hui and Wang, Wenjing and Sun, Yu},
  booktitle={Proceedings of the Thirtieth International Joint Conference on Artificial Intelligence},
  pages={1548--1554},
  year={2021},
  organization={International Joint Conferences on Artificial Intelligence Organization}
}

@inproceedings{selfattention,
author = {Vaswani, Ashish and Shazeer, Noam and Parmar, Niki and Uszkoreit, Jakob and Jones, Llion and Gomez, Aidan N. and Kaiser, \L{}ukasz and Polosukhin, Illia},
title = {Attention is all you need},
year = {2017},
isbn = {9781510860964},
publisher = {Curran Associates Inc.},
address = {Red Hook, NY, USA},
booktitle = {Proceedings of the 31st International Conference on Neural Information Processing Systems},
pages = {6000–6010},
numpages = {11},
location = {Long Beach, California, USA},
series = {NIPS'17}
}

@inproceedings{saturn,
author = {Liu, Shuncheng and Chen, Xu and Zhao, Yan and Chen, Jin and Zhou, Rui and Zheng, Kai},
title = {Efficient Learning with Pseudo Labels for Query Cost Estimation},
year = {2022},
isbn = {9781450392365},
publisher = {Association for Computing Machinery},
address = {New York, NY, USA},
url = {https://doi.org/10.1145/3511808.3557305},
doi = {10.1145/3511808.3557305},
booktitle = {Proceedings of the 31st ACM International Conference on Information \& Knowledge Management},
pages = {1309–1318},
numpages = {10},
location = {Atlanta, GA, USA},
series = {CIKM '22}
}

@article{kraskov2004estimating,
  title={Estimating mutual information},
  author={Kraskov, Alexander and St{\"o}gbauer, Harald and Grassberger, Peter},
  journal={Physical Review E—Statistical, Nonlinear, and Soft Matter Physics},
  volume={69},
  number={6},
  pages={066138},
  year={2004},
  publisher={APS}
}

@article{pytorch,
  title={Pytorch: An imperative style, high-performance deep learning library},
  author={Paszke, Adam and Gross, Sam and Massa, Francisco and Lerer, Adam and Bradbury, James and Chanan, Gregory and Killeen, Trevor and Lin, Zeming and Gimelshein, Natalia and Antiga, Luca and others},
  journal={Advances in neural information processing systems},
  volume={32},
  year={2019}
}

@article{word2vec,
  title={Efficient estimation of word representations in vector space},
  author={Mikolov, Tomas and Chen, Kai and Corrado, Greg and Dean, Jeffrey},
  journal={arXiv preprint arXiv:1301.3781},
  year={2013}
}

@inproceedings{rtos,
  author={Yu, Xiang and Li, Guoliang and Chai, Chengliang and Tang, Nan},
  booktitle={2020 IEEE 36th International Conference on Data Engineering (ICDE)}, 
  title={Reinforcement Learning with Tree-LSTM for Join Order Selection}, 
  year={2020},
  volume={},
  number={},
  pages={1297-1308},
  doi={10.1109/ICDE48307.2020.00116}
}

@article{lero,
author = {Zhu, Rong and Chen, Wei and Ding, Bolin and Chen, Xingguang and Pfadler, Andreas and Wu, Ziniu and Zhou, Jingren},
title = {Lero: A Learning-to-Rank Query Optimizer},
year = {2023},
issue_date = {February 2023},
publisher = {VLDB Endowment},
volume = {16},
number = {6},
issn = {2150-8097},
url = {https://doi.org/10.14778/3583140.3583160},
doi = {10.14778/3583140.3583160},
journal = {Proceedings of the VLDB Endowment},
month = feb,
pages = {1466–1479},
numpages = {14}
}

@inproceedings{tpcds,
author = {Poess, Meikel and Nambiar, Raghunath Othayoth and Walrath, David},
title = {Why you should run TPC-DS: a workload analysis},
year = {2007},
isbn = {9781595936493},
publisher = {VLDB Endowment},
booktitle = {Proceedings of the 33rd International Conference on Very Large Data Bases},
pages = {1138–1149},
numpages = {12},
location = {Vienna, Austria},
series = {VLDB '07}
}

@article{imdb,
author = {Leis, Viktor and Gubichev, Andrey and Mirchev, Atanas and Boncz, Peter and Kemper, Alfons and Neumann, Thomas},
title = {How good are query optimizers, really?},
year = {2015},
issue_date = {November 2015},
publisher = {VLDB Endowment},
volume = {9},
number = {3},
issn = {2150-8097},
url = {https://doi.org/10.14778/2850583.2850594},
doi = {10.14778/2850583.2850594},
journal = {Proceedings of the VLDB Endowment},
month = nov,
pages = {204–215},
numpages = {12}
}

@article{tpch,
author = {Poess, Meikel and Floyd, Chris},
title = {New TPC benchmarks for decision support and web commerce},
year = {2000},
issue_date = {Dec. 2000},
publisher = {Association for Computing Machinery},
address = {New York, NY, USA},
volume = {29},
number = {4},
issn = {0163-5808},
url = {https://doi.org/10.1145/369275.369291},
doi = {10.1145/369275.369291},
journal = {SIGMOD Record},
month = dec,
pages = {64–71},
numpages = {8}
}

@article{stats,
author = {Han, Yuxing and Wu, Ziniu and Wu, Peizhi and Zhu, Rong and Yang, Jingyi and Tan, Liang Wei and Zeng, Kai and Cong, Gao and Qin, Yanzhao and Pfadler, Andreas and Qian, Zhengping and Zhou, Jingren and Li, Jiangneng and Cui, Bin},
title = {Cardinality estimation in DBMS: a comprehensive benchmark evaluation},
year = {2021},
issue_date = {December 2021},
publisher = {VLDB Endowment},
volume = {15},
number = {4},
issn = {2150-8097},
url = {https://doi.org/10.14778/3503585.3503586},
doi = {10.14778/3503585.3503586},
journal = {Proceedings of the VLDB Endowment},
month = dec,
pages = {752–765},
numpages = {14}
}

@article{adam,
  title={Adam: A method for stochastic optimization},
  author={Kingma, Diederik P and Ba, Jimmy},
  journal={arXiv preprint arXiv:1412.6980},
  year={2014}
}

@article{raytune,
    title={Tune: A Research Platform for Distributed Model Selection and Training},
    author={Liaw, Richard and Liang, Eric and Nishihara, Robert
            and Moritz, Philipp and Gonzalez, Joseph E and Stoica, Ion},
    journal={arXiv preprint arXiv:1807.05118},
    year={2018}
}

@article{lstm,
author = {Hochreiter, Sepp and Schmidhuber, J\"{u}rgen},
title = {Long Short-Term Memory},
year = {1997},
issue_date = {November 15, 1997},
publisher = {MIT Press},
address = {Cambridge, MA, USA},
volume = {9},
number = {8},
issn = {0899-7667},
url = {https://doi.org/10.1162/neco.1997.9.8.1735},
doi = {10.1162/neco.1997.9.8.1735},
journal = {Neural Computation},
month = nov,
pages = {1735–1780},
numpages = {46}
}

@article{sun2019end,
author = {Sun, Ji and Li, Guoliang},
title = {An end-to-end learning-based cost estimator},
year = {2019},
issue_date = {November 2019},
publisher = {VLDB Endowment},
volume = {13},
number = {3},
issn = {2150-8097},
url = {https://doi.org/10.14778/3368289.3368296},
doi = {10.14778/3368289.3368296},
journal = {Proceedings of the VLDB Endowment},
month = nov,
pages = {307–319},
numpages = {13}
}

@inproceedings{avgdl,
  author={Yuan, Haitao and Li, Guoliang and Feng, Ling and Sun, Ji and Han, Yue},
  booktitle={2020 IEEE 36th International Conference on Data Engineering (ICDE)}, 
  title={Automatic View Generation with Deep Learning and Reinforcement Learning}, 
  year={2020},
  volume={},
  number={},
  pages={1501-1512},
  doi={10.1109/ICDE48307.2020.00133}
}

@article{queryformer,
author = {Zhao, Yue and Cong, Gao and Shi, Jiachen and Miao, Chunyan},
title = {QueryFormer: a tree transformer model for query plan representation},
year = {2022},
issue_date = {April 2022},
publisher = {VLDB Endowment},
volume = {15},
number = {8},
issn = {2150-8097},
url = {https://doi.org/10.14778/3529337.3529349},
doi = {10.14778/3529337.3529349},
journal = {Proceedings of the VLDB Endowment},
month = apr,
pages = {1658–1670},
numpages = {13}
}

@article{yu2022cost,
author = {Yu, Xiang and Chai, Chengliang and Li, Guoliang and Liu, Jiabin},
title = {Cost-Based or Learning-Based? A Hybrid Query Optimizer for Query Plan Selection},
year = {2022},
issue_date = {September 2022},
publisher = {VLDB Endowment},
volume = {15},
number = {13},
issn = {2150-8097},
url = {https://doi.org/10.14778/3565838.3565846},
doi = {10.14778/3565838.3565846},
journal = {Proceedings of the VLDB Endowment},
month = sep,
pages = {3924–3936},
numpages = {13}
}

@inproceedings{zhao2022lightweight,
author = {Zhao, Kangfei and Yu, Jeffrey Xu and He, Zongyan and Li, Rui and Zhang, Hao},
title = {Lightweight and Accurate Cardinality Estimation by Neural Network Gaussian Process},
year = {2022},
isbn = {9781450392495},
publisher = {Association for Computing Machinery},
address = {New York, NY, USA},
url = {https://doi.org/10.1145/3514221.3526156},
doi = {10.1145/3514221.3526156},
booktitle = {Proceedings of the 2022 International Conference on Management of Data},
pages = {973–987},
numpages = {15},
location = {Philadelphia, PA, USA},
series = {SIGMOD '22}
}

@article{kipf2016semi,
  title={Semi-supervised classification with graph convolutional networks},
  author={Kipf, Thomas N and Welling, Max},
  journal={arXiv preprint arXiv:1609.02907},
  year={2016}
}

@inproceedings{dutt_plan_2014,
author = {Dutt, Anshuman and Haritsa, Jayant R.},
title = {Plan bouquets: query processing without selectivity estimation},
year = {2014},
isbn = {9781450323765},
publisher = {Association for Computing Machinery},
address = {New York, NY, USA},
url = {https://doi.org/10.1145/2588555.2588566},
doi = {10.1145/2588555.2588566},
booktitle = {Proceedings of the 2014 ACM SIGMOD International Conference on Management of Data},
pages = {1039–1050},
numpages = {12},
location = {Snowbird, Utah, USA},
series = {SIGMOD '14}
}

@article{moerkotte_preventing_2009,
author = {Moerkotte, Guido and Neumann, Thomas and Steidl, Gabriele},
title = {Preventing bad plans by bounding the impact of cardinality estimation errors},
year = {2009},
issue_date = {August 2009},
publisher = {VLDB Endowment},
volume = {2},
number = {1},
issn = {2150-8097},
url = {https://doi.org/10.14778/1687627.1687738},
doi = {10.14778/1687627.1687738},
journal = {Proceedings of the VLDB Endowment},
month = aug,
pages = {982–993},
numpages = {12}
}

@inproceedings{nix1994estimating,
  title={Estimating the mean and variance of the target probability distribution},
  author={Nix, D.A. and Weigend, A.S.},
  booktitle={Proceedings of 1994 IEEE International Conference on Neural Networks (ICNN'94)}, 
  title={Estimating the mean and variance of the target probability distribution}, 
  year={1994},
  volume={1},
  number={},
  pages={55-60 vol.1},
  doi={10.1109/ICNN.1994.374138}
}

@article{liu2023simple,
author = {Liu, Jeremiah Zhe and Padhy, Shreyas and Ren, Jie and Lin, Zi and Wen, Yeming and Jerfel, Ghassen and Nado, Zachary and Snoek, Jasper and Tran, Dustin and Lakshminarayanan, Balaji},
title = {A simple approach to improve single-model deep uncertainty via distance-awareness},
year = {2023},
issue_date = {January 2023},
publisher = {JMLR.org},
volume = {24},
number = {1},
issn = {1532-4435},
journal = {Journal of Machine Learning Research},
month = jan,
articleno = {42},
numpages = {63}
}

@article{lipton2018mythos,
author = {Lipton, Zachary C.},
title = {The Mythos of Model Interpretability: In machine learning, the concept of interpretability is both important and slippery.},
year = {2018},
issue_date = {May-June 2018},
publisher = {Association for Computing Machinery},
address = {New York, NY, USA},
volume = {16},
number = {3},
issn = {1542-7730},
url = {https://doi.org/10.1145/3236386.3241340},
doi = {10.1145/3236386.3241340},
journal = {Queue},
month = jun,
pages = {31–57},
numpages = {27}
}

@misc{pghintplan,
  author       = {{pg\_hint\_plan Development Team}},
  title        = {{pg\_hint\_plan}: PostgreSQL query-plan hinting extension},
  version      = {1.5.2},
  year         = {2024},
  url          = {https://github.com/ossc-db/pg_hint_plan}
}

@article{autosteer,
author = {Anneser, Christoph and Tatbul, Nesime and Cohen, David and Xu, Zhenggang and Pandian, Prithviraj and Laptev, Nikolay and Marcus, Ryan},
title = {AutoSteer: Learned Query Optimization for Any SQL Database},
year = {2023},
issue_date = {August 2023},
publisher = {VLDB Endowment},
volume = {16},
number = {12},
issn = {2150-8097},
url = {https://doi.org/10.14778/3611540.3611544},
doi = {10.14778/3611540.3611544},
journal = {Proceedings of the VLDB Endowment},
month = aug,
pages = {3515–3527},
numpages = {13}
}

@article{fastgres,
author = {Woltmann, Lucas and Thiessat, Jerome and Hartmann, Claudio and Habich, Dirk and Lehner, Wolfgang},
title = {FASTgres: Making Learned Query Optimizer Hinting Effective},
year = {2023},
issue_date = {July 2023},
publisher = {VLDB Endowment},
volume = {16},
number = {11},
issn = {2150-8097},
url = {https://doi.org/10.14778/3611479.3611528},
doi = {10.14778/3611479.3611528},
journal = {Proceedings of the VLDB Endowment},
month = jul,
pages = {3310–3322},
numpages = {13}
}

@article{hero,
  title={HERO: Hint-Based Efficient and Reliable Query Optimizer},
  author={Zinchenko, Sergey and Iazov, Sergey},
  journal={arXiv preprint arXiv:2412.02372},
  year={2024}
}

@article{thompson_sampling,
  title={An empirical evaluation of thompson sampling},
  author={Chapelle, Olivier and Li, Lihong},
  journal={Advances in neural information processing systems},
  volume={24},
  year={2011}
}

@article{howgoodarelcm,
author = {Heinrich, Roman and Luthra, Manisha and Wehrstein, Johannes and Kornmayer, Harald and Binnig, Carsten},
title = {How Good are Learned Cost Models, Really? Insights from Query Optimization Tasks},
year = {2025},
issue_date = {June 2025},
publisher = {Association for Computing Machinery},
address = {New York, NY, USA},
volume = {3},
number = {3},
url = {https://doi.org/10.1145/3725309},
doi = {10.1145/3725309},
journal = {Proceedings of the ACM on Management of Data},
month = jun,
articleno = {172},
numpages = {27}
}

@article{zeroshot,
author = {Hilprecht, Benjamin and Binnig, Carsten},
title = {Zero-shot cost models for out-of-the-box learned cost prediction},
year = {2022},
issue_date = {July 2022},
publisher = {VLDB Endowment},
volume = {15},
number = {11},
issn = {2150-8097},
url = {https://doi.org/10.14778/3551793.3551799},
doi = {10.14778/3551793.3551799},
journal = {Proceedings of the VLDB Endowment},
month = jul,
pages = {2361–2374},
numpages = {14}
}

@article{dsb,
author = {Ding, Bailu and Chaudhuri, Surajit and Gehrke, Johannes and Narasayya, Vivek},
title = {DSB: a decision support benchmark for workload-driven and traditional database systems},
year = {2021},
issue_date = {September 2021},
publisher = {VLDB Endowment},
volume = {14},
number = {13},
issn = {2150-8097},
url = {https://doi.org/10.14778/3484224.3484234},
doi = {10.14778/3484224.3484234},
journal = {Proceedings of the VLDB Endowment},
month = sep,
pages = {3376–3388},
numpages = {13}
}

@online{pgtune,
  author  = {Oleksii Vasyliev},
  title   = {PGTune: Tuning PostgreSQL configuration by hardware},
  year    = {2014},
  url     = {https://github.com/le0pard/pgtune}

}

@article{flowloss,
author = {Negi, Parimarjan and Marcus, Ryan and Kipf, Andreas and Mao, Hongzi and Tatbul, Nesime and Kraska, Tim and Alizadeh, Mohammad},
title = {Flow-loss: learning cardinality estimates that matter},
year = {2021},
issue_date = {July 2021},
publisher = {VLDB Endowment},
volume = {14},
number = {11},
issn = {2150-8097},
url = {https://doi.org/10.14778/3476249.3476259},
doi = {10.14778/3476249.3476259},
journal = {Proceedings of the VLDB Endowment},
month = jul,
pages = {2019–2032},
numpages = {14}
}

\end{document}